\documentclass[hidelinks,a4paper,10pt,twocolumn, switch]{article}

\usepackage{fancyhdr}
\fancypagestyle{plain}{%
  \fancyhf{}
\fancyfoot[R]{\thepage}
\fancyfoot[L]{\small \textit{Preprint submitted to Springer}\\ \texttt{arXiv}}
}

\usepackage[affil-it]{authblk}
\makeatletter
\def\@maketitle{%
  \newpage
  \null
  \vskip 2em%
  \begin{center}%
  \let \footnote \thanks
    {\Large\bfseries \@title \par}%
    \vskip 1.5em%
    {\normalsize
      \lineskip .5em%
      \begin{tabular}[t]{c}%
        \@author
      \end{tabular}\par}%
    \vskip 1em%
    {\normalsize \@date}%
  \end{center}%
  \par
  \vskip 1.5em}
\makeatother

\usepackage[a4paper, total={17.5cm,24cm}]{geometry}

\usepackage[font=normal,labelfont=bf]{caption}

\usepackage{sectsty} 

\usepackage{titlesec}
\titlespacing\section{0pt}{12pt plus 3pt minus 3pt}{1pt plus 1pt minus 1pt}
\titlespacing\subsection{0pt}{10pt plus 3pt minus 3pt}{1pt plus 1pt minus 1pt}
\titlespacing\subsubsection{0pt}{8pt plus 3pt minus 3pt}{1pt plus 1pt minus 1pt}

\titleformat{\section}{\normalfont\large\bfseries}{\thesection}{1em}{}

\titleformat{\subsection}{\normalfont\normalsize\bfseries}{\thesubsection}{1em}{}

\titleformat{\subsubsection}{\normalfont\normalsize}{\thesubsubsection}{1em}{}

\titleformat{\paragraph}[runin]{\normalfont\normalsize\itshape}{\theparagraph}{1em}{}

\usepackage{mathtools,upgreek}

\usepackage[utf8]{inputenc}	
\usepackage[T1]{fontenc}	
\usepackage{xcolor}		
\usepackage{booktabs} 		
\usepackage{nicefrac}		
\usepackage{microtype}		
\usepackage{lineno}		
\usepackage{float}			

\usepackage{nicefrac}
\usepackage{paralist}
\usepackage{graphicx}
\usepackage{caption}
\usepackage{subcaption}
\usepackage{amsmath}  
\usepackage{amsfonts} 
\usepackage{pifont} 
\usepackage{lineno}
\usepackage[linktoc=page,
colorlinks=true,	
linkcolor=blue,
citecolor=blue,
breaklinks=true]{hyperref}  

\usepackage{enumitem} 
\usepackage{stmaryrd}  

\usepackage{algorithm}
\usepackage{algpseudocode}
\usepackage{hyperref}

\usepackage{siunitx}
\usepackage{breakcites}
\newcommand{\true}{\checkmark}
\newcommand{\false}{\color{red}\hspace{1pt}\ding{55}}

\usepackage{mathptmx}      
\usepackage{newtxtext,newtxmath}      
\usepackage{tabularx}
\usepackage{listings}
\mathtoolsset{centercolon}

\title{Microstructure Characterization and Reconstruction in Python: MCRpy}

\begin{document}

\author{Paul Seibert, \quad Alexander Ra{\ss}loff, \quad Karl Kalina, \quad Marreddy Ambati, \quad Markus K\"astner
  \thanks{Contact: \texttt{markus.kaestner@tu-dresden.de} }}
\affil{Institute of Solid Mechanics, \\ TU Dresden, Germany}

\date{}
\maketitle

\begin{abstract}
Microstructure characterization and reconstruction (MCR) is an important prerequisite for empowering and accelerating integrated computational materials engineering.
Much progress has been made in MCR recently, however, in absence of a flexible software platform it is difficult to use ideas from other researchers and to develop them further.
To address this issue, this work presents~\emph{MCRpy}\footnote{\url{https://github.com/NEFM-TUDresden/MCRpy}} for easy-to-use, extensible and flexible MCR.
The software platform that can be used as a program with graphical user interface, as a command line tool and as a Python library.
The central idea is that microstructure reconstruction is formulated as a modular and extensible optimization problem.
In this way, \emph{any} descriptors can be used for characterization and \emph{any} loss function combining \emph{any} descriptors can be minimized using \emph{any} optimizer for reconstruction. 
With stochastic optimizers, this leads to variations of the well-known Yeong-Torquato algorithm.
Furthermore, \emph{MCRpy} features automatic differentiation, enabling the utilization of gradient-based optimizers.
In this work, after a brief introduction to the underlying concepts, the capabilities of \emph{MCRpy} are demonstrated by exemplarily applying it to typical MCR tasks.
Finally, it is shown how to extend \emph{MCRpy} by defining a new microstructure descriptor and readily using it for reconstruction without additional implementation effort.

\vspace{5mm}
\noindent
\textbf{Keywords: } Microstructure~\textendash~Characterization~\textendash~Reconstruction~\textendash~Descriptor~\textendash~Software~\textendash~ICME

\end{abstract}

\section{Introduction}
\label{sec:introduction}
Establishing and inverting process-structure-property (PSP) linkages is a central goal in integrated computational materials engineering (ICME) in order to accelerate the development of new materials.
With increasing computational resources and much development in data processing and machine learning, data-centric workflows for microstructure design receive more and more attention~\cite{chen_data-centric_2022}.
These workflows rely on large databases that are created using numerical simulations.
One central aspect to consider in this context is how to choose and create the microstructures to simulate from the extremely big set of possible structures.
To avoid extremely time-consuming and cost-intensive experimental campaigns, an efficient microstructure characterization and reconstruction (MCR) tool is therefore a key ingredient to making large-scale ICME workflows feasible.
A very brief introduction to MCR is given in the following and the reader is kindly referred to~\cite{bostanabad_computational_2018} for an in-depth review.

Microstructure characterization, the first aspect of MCR, is required to handle the stochasticity of the microstructures: 
Two distinct image sections of the same microstructure are similar from a visual and statistical perspective, but completely different in terms of a pixel-based representation. 
Thus, for operations like quantitative comparisons, it is reasonable to map the pixel-based microstructure to a translation-invariant, stationary descriptor~$D$ that allows for these operations. 
Therefore,~$D$ is a reasonable choice for representing structures in PSP linkages. 
Furthermore, it provides a possibility to explore the microstructure space in data-driven materials development workflows.

Microstructure reconstruction, the second aspect of MCR, can be regarded as the inverse operation to microstructure characterization: 
The goal is to find a microstructure such that the corresponding descriptor equals the given value.
Microstructure reconstruction allows to \textit{(i)} create a plausible 3D volume element from a 2D slice like a microscopy image, \textit{(ii)} create a set of similar microstructures given one realization and \textit{(iii)} interpolating between microstructures in terms of descriptors.

These two aspects of MCR, namely characterization and reconstruction, can be treated independently, for example using spatial correlations as descriptors and modern machine learning-based techniques for reconstruction. 
However, automatic ICME workflows for complex materials highly benefit from a principled exploration of the descriptor space, where microstructures are selected for reconstruction, simulation and homogenization in a way that maximizes the expected information gain for the PSP linkage~\cite{khatamsaz_adaptive_2021}.
Therefore, it is important to combine characterization and reconstruction so that given any combination of descriptors and their values, the reconstruction can be triggered from these descriptors.
Furthermore, recent research indicates that there is no single best descriptor for microstructure reconstruction~\cite{seibert_descriptor-based_2022} and for PSP linkages~\cite{liu_how_2022}, but that it is reasonable to choose descriptors based on the structure at hand.
For this purpose, we present~\emph{MCRpy}, a modular and extensible open-source tool that facilitates easy microstructure characterization and reconstruction based on arbitrary descriptors.

Free open-source platforms are a great way of harnessing the advantages of digitization and modern computational infrastructure.
The free accessibility allows researchers to quickly test each others' ideas and to develop them further.
The open-source nature of such a platform enables it to become a collaborative project, considerably leveraging its potential.
Especially in complex scientific disciplines, such collaboration is indispensable.
As an example, consider the field of machine learning, specifically neural networks~\cite{sonnenburg_need_2007}. 
In the beginning of research on neural networks, newcomers had to implement relatively complex procedures like back-propagation before being able to reproduce results from the literature, let alone to develop them further.
Later, easy-to-use open-source libraries like \emph{TensorFlow} and \emph{pyTorch} have greatly lowered the hurdle, allowing more researchers to enter the field easily.
This surely contributed to the rapid progress in the last decades and to the plethora of neural network architectures and applications that is observed today.

The digital infrastructure of the materials science community has grown considerably as a consequence of the materials genome initiative~\cite{de_pablo_materials_2014} and similar projects.
Despite the rapidly growing number of tools for materials innovation in general, MCR specifically is in a comparable position now as machine learning was 20 years ago:
A great variety of methods exists, but in absence of a common platform and interface, every newcomer in the field has to implement fundamental technologies like the lineal path function and the Yeong-Torquato algorithm by hand.
This is a big hurdle and thwarts any rapid progress. 
Thus, the goal of this contribution is to accelerate MCR research by providing~\emph{MCRpy} as an easy-to-use, extensible and flexible software solution that aims at realizing a seamless workflow by providing various interfaces to new and established techniques.

The work starts with Section~\ref{sec:others}, where the current digital infrastructure is reviewed and it is outlined how \emph{MCRpy} integrates into it.
Then, \emph{MCRpy} is presented in Section~\ref{sec:overview}.
Typical application workflows are presented in Section~\ref{sec:application} and finally, a conclusion is drawn in Section~\ref{sec:summary}. 

\section{Current Digital Infrastructure}
\label{sec:others}
After President {Barack Obama} announced the US-American \emph{Materials Genome Initiative}~\cite{de_pablo_materials_2014} that provided substantial funding for accelerated materials development, collaborative projects and digital frameworks were initiated all over the world.
A non-exhaustive list includes the European \emph{NOMAD-CoE}~\cite{noauthor_european_2021} and its platform described in~\cite{ghiringhelli_towards_2016} and the Swiss~\emph{NCCR MARVEL}~\cite{noauthor_computational_2021} with its \emph{AiiDA} platform~\cite{noauthor_automated_2021} described in~\cite{pizzi_aiida_2016}.
The extremely popular and often-cited \emph{pymatgen} library~\cite{ong_python_2013} can be mentioned as an early contribution to open-source materials science software infrastructure.
This trend continues, as can be seen with the recent example \emph{radonpy}~\cite{hayashi_radonpy_2022}.
However, much of this research is focused on deriving material properties from considerations on the atomistic length scale.

On the continuum length scale, the \emph{Python Materials Knowledge System (pyMKS)}~\cite{brough_materials_2017} is a notable open-source framework.
Its efficient FFT-based implementation of the spatial two-point correlation~$S_2$ facilitates easy microstructure characterization.
However, in \emph{pyMKS}, microstructure characterization is limited to~$S_2$ and no further descriptors are available.
Moreover, \emph{pyMKS} does not allow for microstructure reconstruction, only characterization.
A strong focus lies on efficient homogenization~\cite{fast_formulation_2011} and direct coupling to an internal finite element solver,~\emph{SfePy}~\cite{cimrman_sfepy_2014, cimrman_multiscale_2019}.
This is very convenient for simple problems like elasticity.
For advanced techniques like crystal plasticity, external software like the Düsseldorf Advanced Materials Simulation Kit~(\emph{DAMASK})~\cite{roters_damask_2019} can be used.
Furthermore,~\emph{pyMKS} provides an easy interface for dimensionality reduction of the descriptor space and for establishing structure-property-linkages based on the reduced descriptors and the corresponding homogenized properties.
In summary, \emph{pyMKS} acts as an overarching framework to implement ICME workflows. 

For numerical simulation of microstructures, many open-source tools are available, ranging from general and easy-to-use packages like \emph{SfePy}~\cite{cimrman_sfepy_2014, cimrman_multiscale_2019} to special-purpose software like \emph{DAMASK}~\cite{roters_damask_2019}, which comes with a highly optimized Fourier-based crystal plasticity solver. 
Furthermore, current research on FFT-based homogenization~\cite{keshav_fft-based_2022} is making remarkable progress that might lead to an open-source tool soon.
Thus, with \emph{pyMKS} as an overarching framework and numerous tools and progress for numerical simulation and homogenization, an open-source MCR software package can be identified as a final component of ICME workflows.

To the authors' best knowledge, the only widely-used software tool for MCR is \emph{DREAM.3D}~\cite{groeber_dream3d_2014}, a long developed and full-fledged program.
Its roots date back around~20 years to the early works of Michael Groeber and the Carnegie Mellon University microstructure builder.
Despite this long history,~\emph{DREAM.3D} still enables numerous current research activities in materials innovation and ICME, see for example~\cite{azhari_comparison_2022}.
This success is empowered by the many features, robustness, efficiency and easy user interface of \emph{DREAM.3D}, which may be partially attributable to its open-source core.
Thus,~\emph{DREAM.3D} can be highly recommended for the workflows it implements.
However, the internal microstructure representation and the available pipelines in \emph{DREAM.3D} are mainly intended for certain material systems and microstructure descriptors.
The internal data format as well as the provided characterization and reconstruction algorithms are centered around classical descriptors like grain size distribution functions and orientation distribution functions.
This makes \emph{DREAM.3D} excellent at reconstructing geometric inclusions like ellipses and texture as in metallic materials, but multi-phase materials with complex morphology as shown in Figure~\ref{fig:linealpathresults} cannot be realized. 
Furthermore, \emph{DREAM.3D} is written in c++, which is not common among engineering researchers due to its complexity.
In recent research, new microstructure descriptors or reconstruction algorithms are sometimes provided as Python or Matlab code in a GitHub repository, but are hardly ever implemented in c++ as a \emph{DREAM.3D} pipeline.
Even if that was the case, then these descriptors could not be readily used for reconstruction since the~\emph{DREAM.3D} reconstruction pipelines are tailored towards specific descriptors and would need to be re-implemented.
Thus, \emph{DREAM.3D} is an excellent and robust program, but it is mainly suited for specific practical applications and for certain materials.

In contrast, the present work aims at creating a flexible research platform for multiphase materials of high morphological complexity.
Thus, \emph{MCRpy} clearly differs from \emph{DREAM.3D} regarding the targeted audience and the scope of materials systems.
As a Python package, it integrates naturally with numerous tools for numerical simulation, machine learning or materials science workflows
Especially~\emph{pyMKS} can act as an overarching ICME framework, where the present work provides an MCR solution.
In summary, \emph{MCRpy} attempts to fill a striking gap in the ICME software landscape.
A theoretical understanding of \emph{MCRpy} is provided in Section~\ref{sec:overview}, followed by an illustration of typical workflows in Section~\ref{sec:application}.

\section{Overview of MCRpy}
\label{sec:overview}
Microstructure characterization and reconstruction in Python (\emph{MCRpy}) is an open-source software tool accessible under \url{https://github.com/NEFM-TUDresden/MCRpy}. It is released under the \emph{Apache 2.0} license and can be used 
\begin{itemize}
    \item[\textit{(i)}] as a program with graphical user interface (GUI), intended for non-programmers and as an easy introduction to MCR,
    \item[\textit{(ii)}] as a command line tool, intended for automated and large-scale application on high-performance computers without graphical interface, and
    \item[\textit{(iii)}] as a regular PIP-installable Python module, intended for performing advanced and custom operations in the descriptor space.
\end{itemize}

\begin{figure*}[t]
    \centering
    \includegraphics[width=01.0\textwidth]{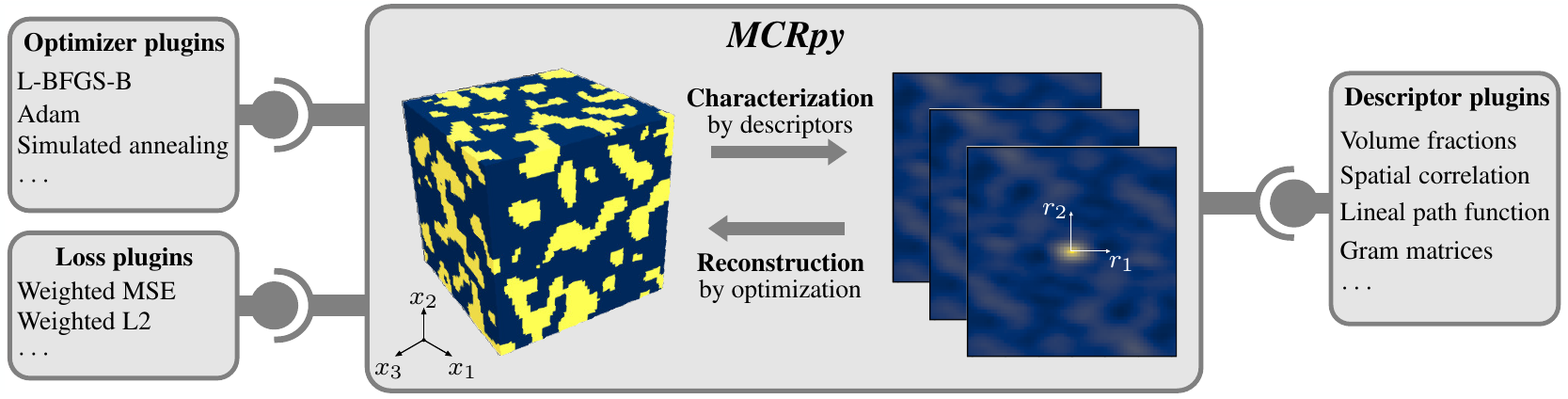}
    \caption{Schematic overview of \emph{MCRpy}: Microstructures can be characterized by descriptors and reconstructed by optimization. Herein, descriptors, losses and optimizers can be provided as flexible plugin modules.\label{fig:schema}}
\end{figure*}

A schematic overview is given in Figure~\ref{fig:schema}:
The main functionalities of \emph{MCRpy}, characterization and reconstruction, are explained in Sections~\ref{sec:characterization} and~\ref{sec:reconstruction} respectively.
Furthermore, additional functions are provided to manipulate the microstructures and descriptors and to visualize data.
A complete set of the available operations is given in Table~\ref{tab:functions}.
The core idea of \emph{MCRpy} is its extensibility in that \emph{any} descriptors can be used for characterization and \emph{any} loss function combining \emph{any} descriptors can be minimized using \emph{any} optimizer for reconstruction. 
This is outlined in Section~\ref{sec:extensibility}.

\begin{table*}[h]
	\centering
	\caption{Functions.}
	\label{tab:functions}
	\begin{tabular}{l | l }
		\toprule
		 function & explanation \\
		\midrule
		\texttt{characterize} & characterize a microstructure, see Section~\ref{sec:characterization} \\
		\texttt{reconstruct} & reconstruct a microstructure given the descriptors, see Section~\ref{sec:reconstruction} \\
		\texttt{match} & characterize and reconstruct immediately, for validation and for 2D-to-3D workflows \\
		\texttt{view} & plot microstructures, descriptors and convergence data interactively or save to a file \\
		\texttt{smooth} & smooth a microstructure \\
		\texttt{merge} & merge different descriptors to prescribe them on orthogonal sections for reconstructing anisotropic structures \\
		\texttt{interpolate} & interpolate between given descriptors \\
	\end{tabular}
\end{table*}

\subsection{Characterization}
\label{sec:characterization}
The characterization function
\begin{equation}
    f_C : M \mapsto \{D_i\}_{i=1}^{n_D}
\end{equation}
assigns a given pixel-based microstructure~$M$ to a set of~$n_D$ corresponding descriptors~$D_i$.
These descriptors quantify the microstructural morphology in a statistical and translation-invariant manner.
Hereby, a microstructure with~$n_\text{p}$ different phases is represented as a set of~$p$ indicator functions
\begin{equation}
    I_p(x) =  \left\{
	\begin{array}{ll}
	1, & \qquad \text{if $x$ in phase $p$} \\
	0, & \qquad \text{else.} \\
	\end{array}
	\right. \;  \\
\end{equation}

For example, the volume fraction~$v_\text{f}$ of a microstructure is a very simple descriptor.
Of course, the volume fraction captures some but not all information needed to describe the microstructure.
Several other quantities matter, for example the size and shape of inclusions and the degree to which distinct phases are spatially clustered.
Besides these classical descriptors, in the light of increasing computational resources, recent research has been focused on more universal high-dimensional descriptors that are less dense in information, but have higher descriptive capabilities in total.
As an early example for high-dimensional descriptors, spatial correlations~\cite{yeong_reconstructing_1998} have proven to be a versatile tool that is still used today~\cite{bostanabad_computational_2018}.
A good introduction can be found in~\cite{jiao_modeling_2007}.
A differentiable generalization of spatial correlation is presented in~\cite{seibert_reconstructing_2021-1} and used in this work.
Spatial correlations have inspired a range of conceptually similar descriptors like the lineal path function~\cite{lu1992}, cluster correlation function~\cite{jiao_superior_2009} and polytope function~\cite{chen_novel_2019}.
The reader is referred to~\cite{bostanabad_computational_2018} for a comprehensive overview.
Finally, the Gram matrices of the feature maps of pre-trained convolutional neural networks have been shown to contain relevant microstructural information~\cite{lubbers_inferring_2017}.
Remarkable results in microstructure reconstruction have been achieved using such Gram matrices alone~\cite{li_transfer_2018,bostanabad_reconstruction_2020} and in combination with other descriptors~\cite{bhaduri_efficient_2021,seibert_descriptor-based_2022}.

Finding a microstructure description that is both dense and contains all relevant information is an active field of research~\cite{bostanabad_computational_2018}.
Examples are the recently developed entropic descriptors~\cite{piasecki_entropic_2010} or polytope functions~\cite{chen_novel_2019}.
Thus, besides the currently available descriptors listed in Table~\ref{tab:descriptors}, any researcher can add a descriptor plugin to \emph{MCRpy}.
If the descriptor plugin is defined with an indicator function as input, it is applied to the indicator function of each phase separately.
Furthermore, a 2D descriptor is automatically applied on and averaged over 2D slices of a 3D structure.
The only requirement posed on new descriptors is that they must be computable on a pixel or voxel geometry. 
More details on extensibility can be found in Section~\ref{sec:extensibility}.
Any of the available and added descriptors can be used from microstructure reconstruction, which is discussed in the following section.
\begin{table*}[h]
	\centering
	\caption{Microstructure descriptors that are implemented in~\emph{MCRpy}.}
	\label{tab:descriptors}
	\begin{tabular}{l | c  c  c  }
		\toprule
		 & Descriptor & Differentiable & Comment \\
		\midrule
		$v_\text{f}$ & \texttt{VolumeFractions} & \true & - \\
		$\tilde{S}$ & \texttt{Correlations} & \true & $\tilde{S}_2$ and $\tilde{S}_3$; see~\cite{seibert_reconstructing_2021-1} \\
		$S$ & \texttt{FFTCorrelations} & \false & only~$S_2$; FFT-based; from \emph{pyMKS}~\cite{brough_materials_2017} \\
		$G$ & \texttt{GramMatrices} & \true & using VGG19~\cite{simonyan_very_2015}; see~\cite{li_transfer_2018} \\
		$\mathcal{V}$ & \texttt{Variation} & \true & normalized total variation; see~\cite{seibert_descriptor-based_2022} \\
		$\tilde{L}$ & \texttt{LinealPath} & \true & see Appendix~\ref{sec:linealpath} \\
	\end{tabular}
\end{table*}

\subsection{Reconstruction}
\label{sec:reconstruction}
In \emph{MCRpy}, microstructure reconstruction is fundamentally regarded as an optimization problem
\begin{equation}
    M^\text{rec} = \underset{M}{\text{argmin}} \; \mathcal{L}\left( \{ (D_i(M), \; D_i^\text{des}) \}_{i=1}^{n_D} \right) \quad ,
    \label{eqn:optgeneral}
\end{equation}
where the reconstructed microstructure~$M^\text{rec}$ minimizes a loss function~$\mathcal{L}$.
The loss function depends on $n_D$ different descriptors~$\{D_i\}_{i=1}^{n_D}$ and quantifies the distance between their current and desired values.
Herein,~$D_i(M)$ denotes the value of the $i$-\emph{th} descriptor associated with the current microstructure and its desired value~$D_i^\text{des}$.
Naturally, as in the characterization step, any descriptors can be used, for example the volume fractions~$v_\text{f}$, the spatial correlations~$S$ or the Gram matrices~$G$.
For the loss function~$\mathcal{L}$, a simple choice is a weighted sum over the mean squared error norm.
Different loss functions are available in \emph{MCRpy} and the user can implement additional ones.
Finally, given a set of descriptors and a loss function, an optimization problem emerges as a special case of Equation~\ref{eqn:optgeneral}.
This optimization problem can be solved using an optimizer, which is provided as a plugin module.
If all descriptors are differentiable, then a gradient-based optimizer like L-BFGS-B~\cite{byrd_stochastic_2015} can be used, leading to the very efficient \emph{differentiable MCR}~\cite{seibert_reconstructing_2021-1,seibert_descriptor-based_2022}.
Otherwise, the choice is limited to gradient-free optimizers like simulated annealing.

As a simple example, if only the spatial two-point correlation~$S_2$ is used as a descriptor and the loss function is formulated as a mean squared error norm of the descriptor difference, the following optimization problem emerges:
\begin{equation}
    M^\text{rec} = \underset{M}{\text{argmin}} \; || S_2(M) - S_2^\text{des} ||_{\text{MSE}} \quad .
    \label{eqn:optyt}
\end{equation}
If simulated annealing is chosen as an optimizer, \emph{MCRpy} effectively performs the well-known Yeong-Torquato algorithm as used in~\cite{cule_generating_1999}.

As a more recent example, if the Gram matrices~$G$ of the feature maps of the VGG-19 CNN are chosen as a descriptor~\cite{lubbers_inferring_2017} for the same loss function, the emerging optimization problem
\begin{equation}
    M^\text{rec} = \underset{M}{\text{argmin}} \; || G(M) - G^\text{des} ||_{\text{MSE}} \quad 
    \label{eqn:optli}
\end{equation}
allows for a gradient-based optimizer. If L-BFGS-B~\cite{byrd_stochastic_2015} is chosen for this purpose, \emph{MCRpy} effectively performs the approach of Li~et~al.~\cite{li_transfer_2018}, which is a special case of differentiable MCR~\cite{seibert_reconstructing_2021-1}.

As a final example, the differentiable three-point correlations~$S_3$, the above-mentioned Gram matrices~$G$ and the normalized total variation~$\mathcal{V}$ are combined. The loss function accumulates the weighted mean squared error norm, where~$\lambda_{D_i}$ denotes the weight of the $i$-\emph{th} descriptor. If the resulting optimization problem
\begin{align}
    \nonumber M^\text{rec} = \underset{M}{\text{argmin}} \; & \lambda_S || S_3(M) - S_3^\text{des} ||_{\text{MSE}} + \\
    \nonumber + & \lambda_G || G(M) - G^\text{des} ||_{\text{MSE}} +  \\
    + &\lambda_\mathcal{V} || \mathcal{V}(M) - \mathcal{V}^\text{des} ||_{\text{MSE}}  
    \label{eqn:optdmcr}
\end{align}
is solved using the gradient-based L-BFGS-B optimizer, \emph{MCRpy} effectively performs the differentiable MCR algorithm as used in~\cite{seibert_descriptor-based_2022}.

As can be seen, different parameter settings allow to re-create well-known reconstruction algorithms as well as to try out new ones by simply changing the arguments.
As an overview, all descriptors, optimizers and loss functions are listed in Table~\ref{tab:plugins}.
\begin{table*}[h]
	\centering
	\caption{Microstructure descriptors, optimizers and loss functions that are implemented in~\emph{MCRpy}. Simulated annealing is the only optimizer in the list that is not gradient-based. More details on the descriptors is given in Table~\ref{tab:descriptors}.}
	\label{tab:plugins}
	\begin{tabular}{l | l | l }
		\toprule
		 Descriptors & Optimizers & Loss functions \\
		\midrule
		Volume fractions & L-BFGS-B~\cite{byrd_stochastic_2015} & 2D/3D weighted MSE \\
		Correlations & TNC~\cite{nash_newton-type_1984} & 2D/3D weighted RMS error \\
		Lineal path & Adam~\cite{kingma_adam_2017}, Adagrad~\cite{duchi_adaptive_2011}, Adadelta~\cite{zeiler_adadelta_2012}, ... & 2D/3D weighted L1 distance \\
		Gram matrices & RMSprop~\cite{abadi_tensorflow_2016} & 2D/3D weighted L2 distance \\
		Variation & SGD~\cite{abadi_tensorflow_2016} &  \\
		 & Simulated Annealing~\cite{yeong_reconstructing_1998} &  \\
	\end{tabular}
\end{table*}

\subsection{Extensibility}
\label{sec:extensibility}
\begin{figure*}[t]
    \centering
    \includegraphics[width=0.8\textwidth]{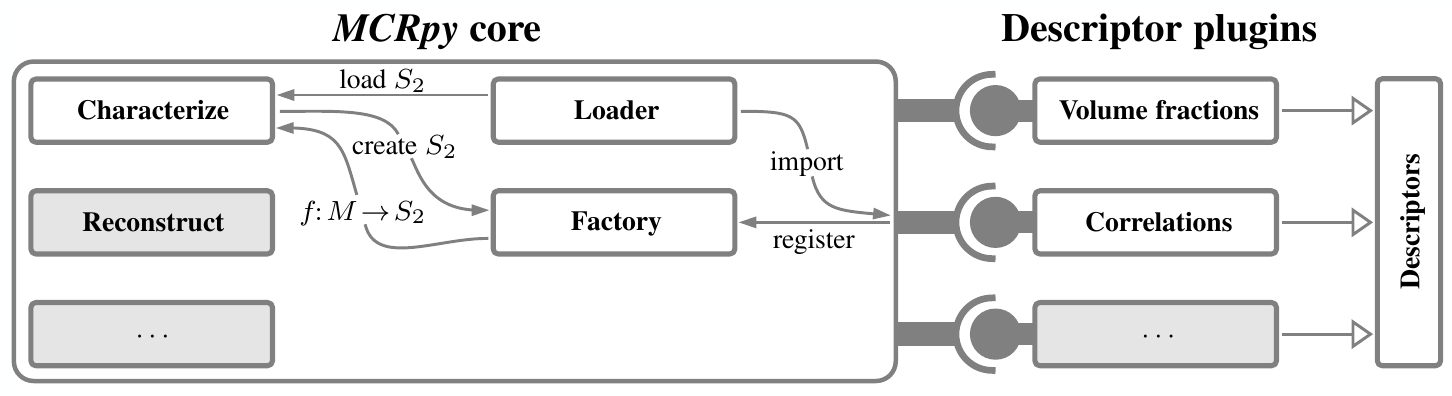}
    \caption{Schematic overview of the plugin architecture in \emph{MCRpy}.\label{fig:plugin_architecture}}
\end{figure*}

The central advantage of \emph{MCRpy} is its extensibility in that descriptors, loss functions and optimizers can be easily provided by anyone.
For example, new optimization-based reconstruction algorithms like the work of Cecen at al.~\cite{cecen_generalized_2021} can be implemented as an optimizer plugin to combine them with all the available microstructure descriptors.
This is achieved by a plugin architecture, which is sketched in Figure~\ref{fig:plugin_architecture}.
In this section, we explain the underlying software pattern, whereas exact instructions and an example on how to write a plugin are given in Section~\ref{sec:wf4}.
In the following, the plugin architecture is explained for the case of descriptors.
The same idea is employed for loss functions and optimizers.

A descriptor plugin can be written by simply inheriting from the abstract \texttt{Descriptor} class.
Consequently, the available descriptor plugins are not known at the time of writing the \emph{MCRpy} core code, so they must be loaded dynamically as soon as the characterization or reconstruction module demands the plugin.
This is done by means of a loader module based on \texttt{importlib}.
Upon import, a descriptor plugin registers itself at a descriptor factory.
After that, the descriptor factory can be queried to create descriptor instances from the plugin.
The descriptor factory then returns a callable which computes the descriptor value given a microstructure.
This callable can now be used to characterize microstructures, compose loss functions, compute gradients using automatic differentiation and to reconstruct microstructures.

Thus, adding a descriptor plugin to \emph{MCRpy} merely consists of adding a file with the plugin definition to the right directory, while the rest of the code does not need to be changed.
The descriptor immediately becomes available for characterization and for reconstruction in combination with any other descriptors, any loss function and any optimizer.

\section{Typical MCRpy workflows}
\label{sec:application}
Typical use-cases and workflows of \emph{MCRpy} are illustrated in this section by means of three representative examples.
First, in Section~\ref{sec:wf1}, a plausible 3D volume element is reconstructed from a 2D microstructure slice. 
This very relevant, since 3D information can be very time- and cost-intensive to obtain experimentally.
Secondly, in Section~\ref{sec:wf2}, a statistically similar set of small volume elements is generated from a single example.
This greatly reduces the computational effort for numerical homogenization.
Thirdly, in Section~\ref{sec:wf3}, descriptor values are directly manipulated and used for reconstructing novel structures.
Techniques like this may be explored in the future to augment data sets and explore PSP linkages.
These three examples are demonstrated in the three modes of operating \emph{MCRpy}, namely via a GUI, as a command line tool and as a Python library, respectively.
Note that this order is chosen for demonstration purposes only and it is possible to execute all three workflows with all three modes of operation.
Finally, in Section~\ref{sec:wf4}, it is demonstrated how to add a custom descriptor to \emph{MCRpy} and how to use it for characterization and reconstruction.
The original structures are taken from~\emph{pyMKS}~\cite{brough_materials_2017} for Sections~\ref{sec:wf1} to~\ref{sec:wf3} and from~\cite{li_transfer_2018} for Section~\ref{sec:wf4}.

\subsection{Obtaining a 3D domain from a 2D microstructure slice}
\label{sec:wf1}
As a first example, \emph{MCRpy} is used to reconstruct a plausible 3D volume element given a segmented 2D slice.
This is a common task since experimental observations are often available only in 2D.
The 3D volume element can be used for example for numerical simulations.
From an algorithmic perspective, this goal is achieved by computing the descriptor on the given slice and prescribing it on every slice of the microstructure, details cf.~\cite{seibert_descriptor-based_2022}.

\begin{figure*}[t]
    \centering
    \includegraphics[width=0.7\textwidth]{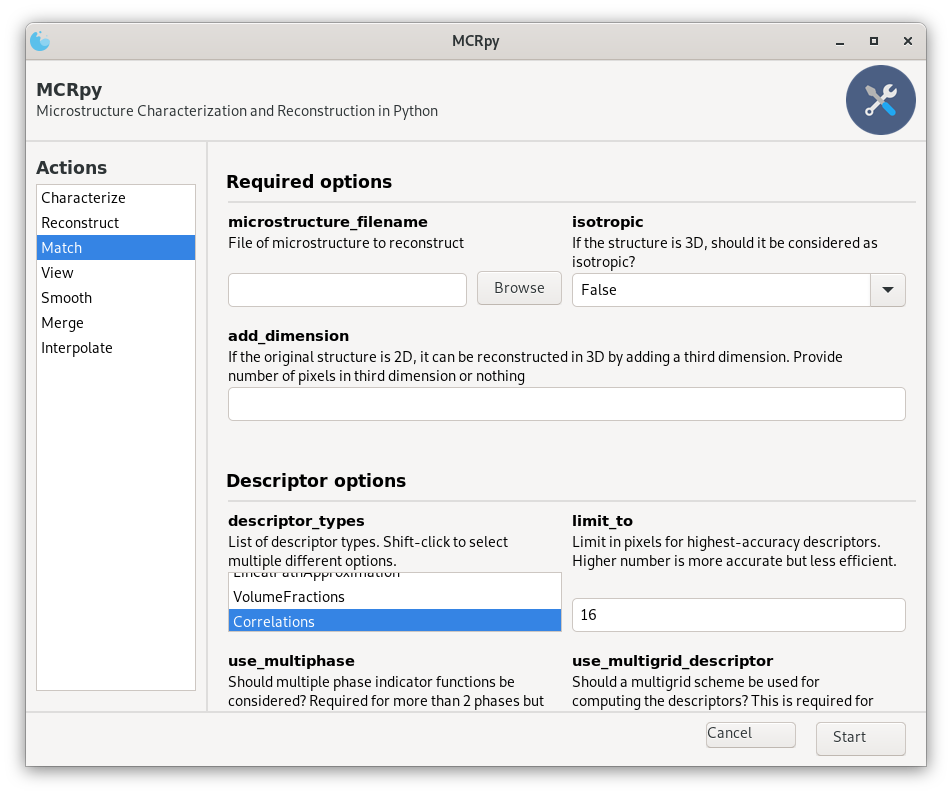}
    \caption{Screenshot of the \emph{MCRpy} graphical user interface. After selection an action on the left, all options can be set set in the center and performed upon clicking \emph{start}. The options are identical to the command line interface and the Python library.\label{fig:gui}}
\end{figure*}

This task is solved using the \emph{MCRpy} GUI as shown in Figure~\ref{fig:gui}.
A simple approach would be characterization and immediate reconstruction, but as mentioned in Table~\ref{tab:functions}, \emph{MCRpy} provides a shortcut for this in the \texttt{match} function.
After selecting the \emph{match}-action on the left, the relevant options can be set in the center.
The name of each option is identical to the command line and the Python library, allowing users to easily switch interfaces.
By default, a 2D structure is reconstructed in 2D.
However, by using the option \texttt{add\_dimension}, the extent of the reconstructed structure in $z$-direction is set to the desired value.
The differentiable three-point correlations~$\tilde{S}_3$ as proposed in~\cite{seibert_reconstructing_2021-1} are chosen as descriptor.
Furthermore, as discussed in~\cite{seibert_descriptor-based_2022}, the variation~$\mathcal{V}$ is employed as a descriptor in order to suppress noise in the 3D reconstruction.
The weights of~$\tilde{S}_3$ and~$\mathcal{V}$ are empirically set to~1 and~100, respectively\footnote{The role of the weights is discussed in~\cite{seibert_descriptor-based_2022}. If the weight of the variation is too small, the noise is not suppressed well enough. If it is too large, the optimization problem becomes harder to solve and more iterations are needed for convergence.}.
Finally, the role of the setting \texttt{limit\_to} needs to be discussed.
The parameter is introduced in~\cite{seibert_reconstructing_2021-1} as~$P$ and~$Q$ and quantifies the length in pixels up to which spatial correlations are computed with the highest-possible precision.
All longer-ranged correlations are computed on a lower-resolution version of the structure in order to save computational resources, cf.~\cite{seibert_reconstructing_2021-1}.
With a default of~16, it allows a flexible trade-off between accuracy and efficiency. 
In this example, it is lowered to~8 in order to accelerate the computations.

\begin{figure}[t]
    \centering
    \subfloat[Original slice]{\includegraphics[width=0.4\linewidth]{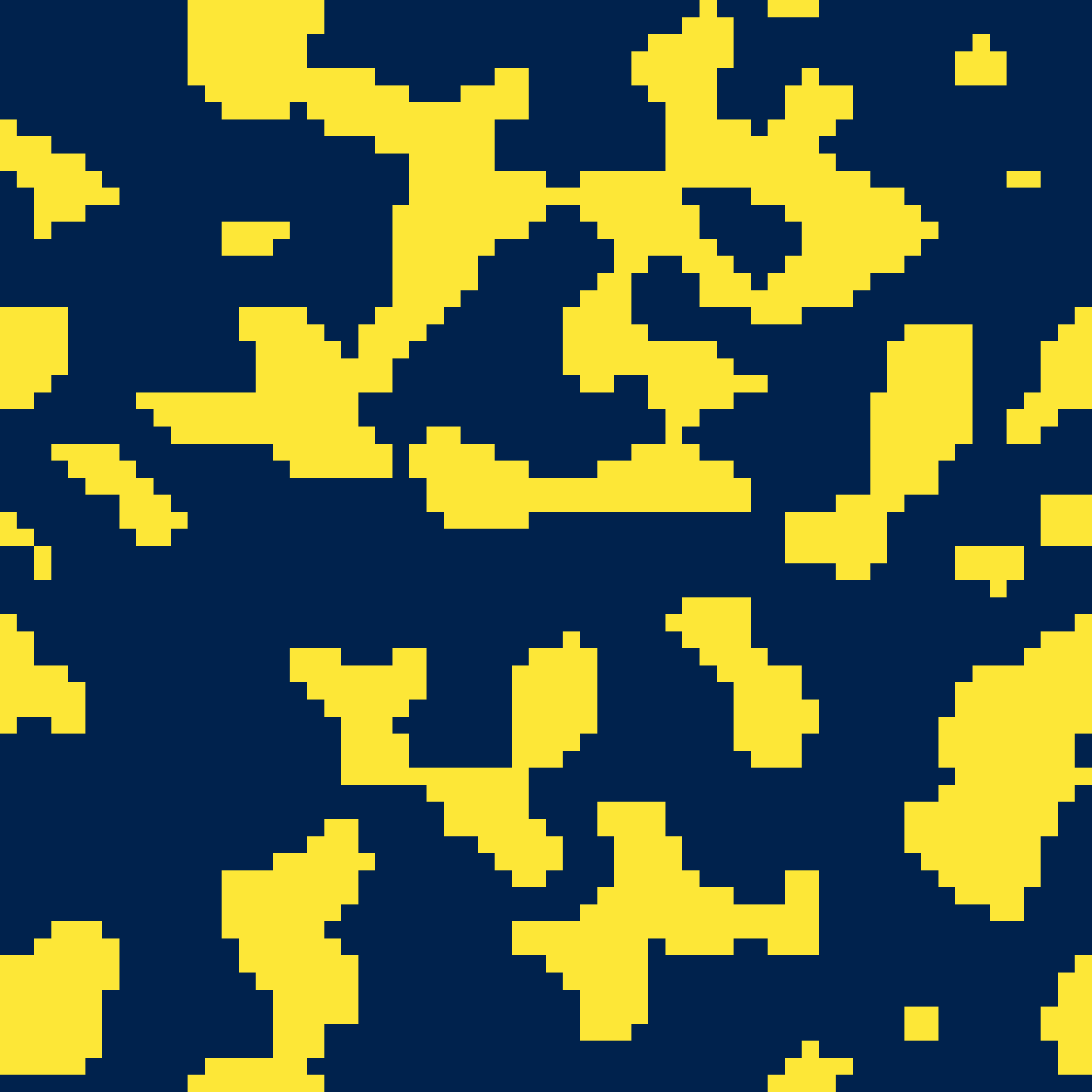}}
    \hspace{1cm}
    \subfloat[Reconstructed structure]{\includegraphics[width=0.4\linewidth]{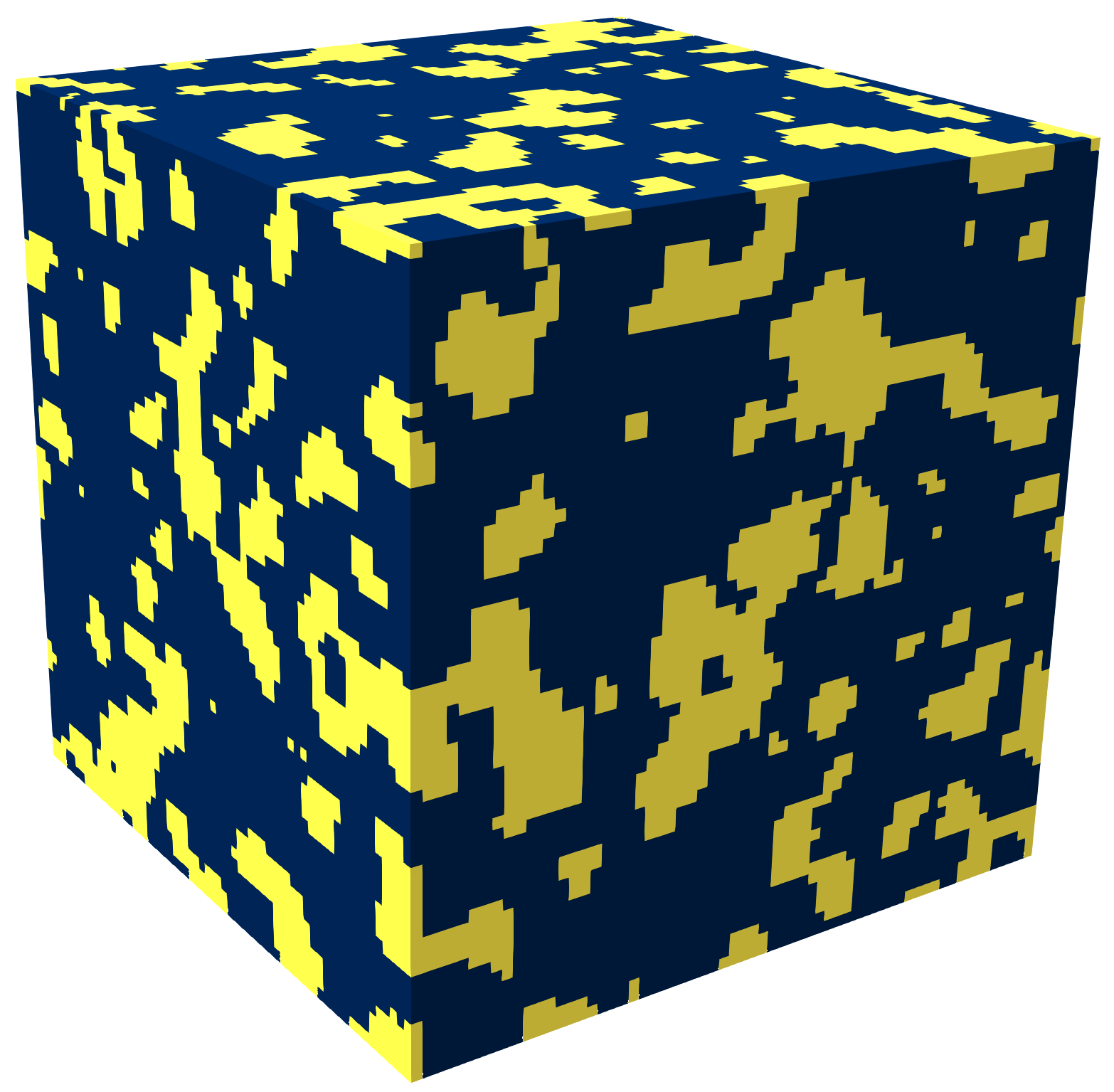}}
    \caption{Results for the example in Section~\ref{sec:wf1}.\label{fig:ms_gui}}
\end{figure}

After setting all options, the reconstruction can be started and the results can be viewed from the GUI by selecting the \emph{view}-action on the left.
2D microstructures are plotted directly, whereas 3D structures are exported to and opened in~\emph{ParaView}~\cite{ahrens2005paraview}.
The original 2D slice and the reconstructed 3D volume are shown in Figure~\ref{fig:ms_gui}.

\begin{figure*}[t]
    \centering
    \includegraphics[width=0.7\textwidth]{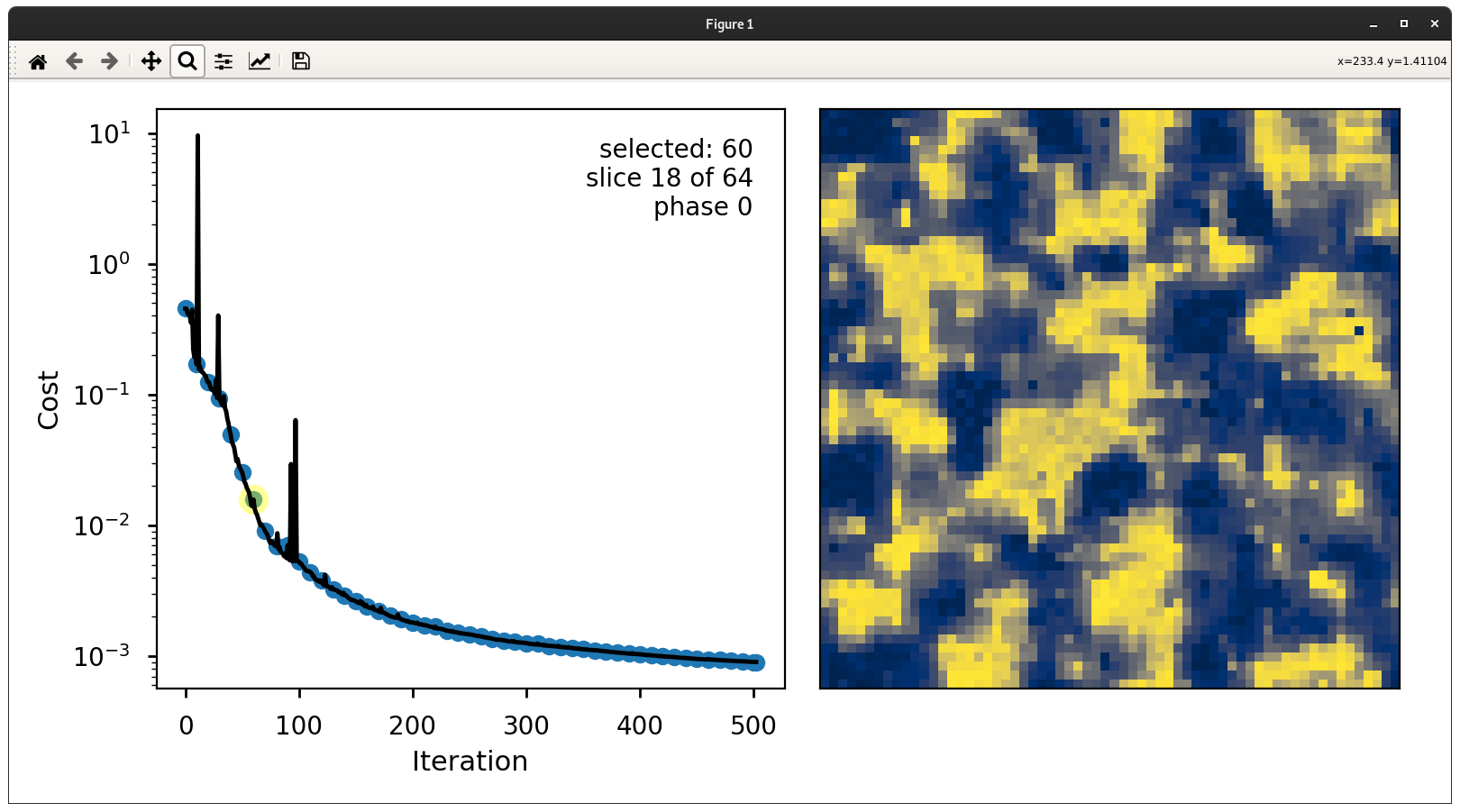}
    \caption{Interactive window for inspecting convergence data. By selecting the highlighted dot on the left at iteration~60, a slice of the intermediate result is displayed on the right. In this example, the indicator function of phase~1 is displayed for slice~18 of~64.\label{fig:convergence}}
\end{figure*}

In addition to the final microstructure, a convergence data file is written, which can be viewed interactively with~\emph{MCRpy} as shown in Figure~\ref{fig:convergence}. 
On the left, the loss is plotted over iterations along with blue dots indicating intermediate results.
The user can click on any of these dots to have the corresponding microstructure displayed on the right.
For 3D structures, only one slice is plotted and the user can scroll through the microstructure using the mouse wheel.
For displaying the raw phase indicator functions of multiphase structures and other functionalities, the user is referred to the documentation. 
In summary, the~\emph{MCRpy} GUI constitutes an easily accessible solution for microstructure reconstruction.

\subsection{Obtaining a set of similar volume elements}
\label{sec:wf2}
As a second example, a statistically similar set of volume elements is created from a single microstructure example.
In numerical homogenization, a volume element can only be called representative if it is large enough for the stochasticity of the microstructure to have no effect on the effective properties.
In practice, this requirement can imply unfeasible computational effort.
If smaller volume elements are used, it is still possible to quantify the effective behavior by using sufficiently many smaller volume elements and statistically aggregating the results.
An example for structure-property linkages based on this idea can be found in~\cite{rasloff_accessing_2021-1}.
From an MCR perspective, this requires characterizing the given structure and reconstruct different random realizations from it\footnote{In~\cite{seibert_reconstructing_2021-1}, the reconstruction was shown to converge to the exact same microstructure that was used for characterization. The same can be seen for some cases in Figure~\ref{fig:linealpathresults}. However, this only happens in 2D reconstruction (not in 3D) and only for certain descriptors and microstructures. In these cases, the desired descriptor prescribed for reconstruction needs to be varied statistically in order to create a diverse set of microstructure realizations. This aspect is not considered in the following because in application, it is most useful to reconstruct 3D structures.}.

\begin{figure*}
\subfloat[Original 2D structure]{
\begin{minipage}[t][8cm][c]{0.45\linewidth}
\includegraphics[width=\linewidth]{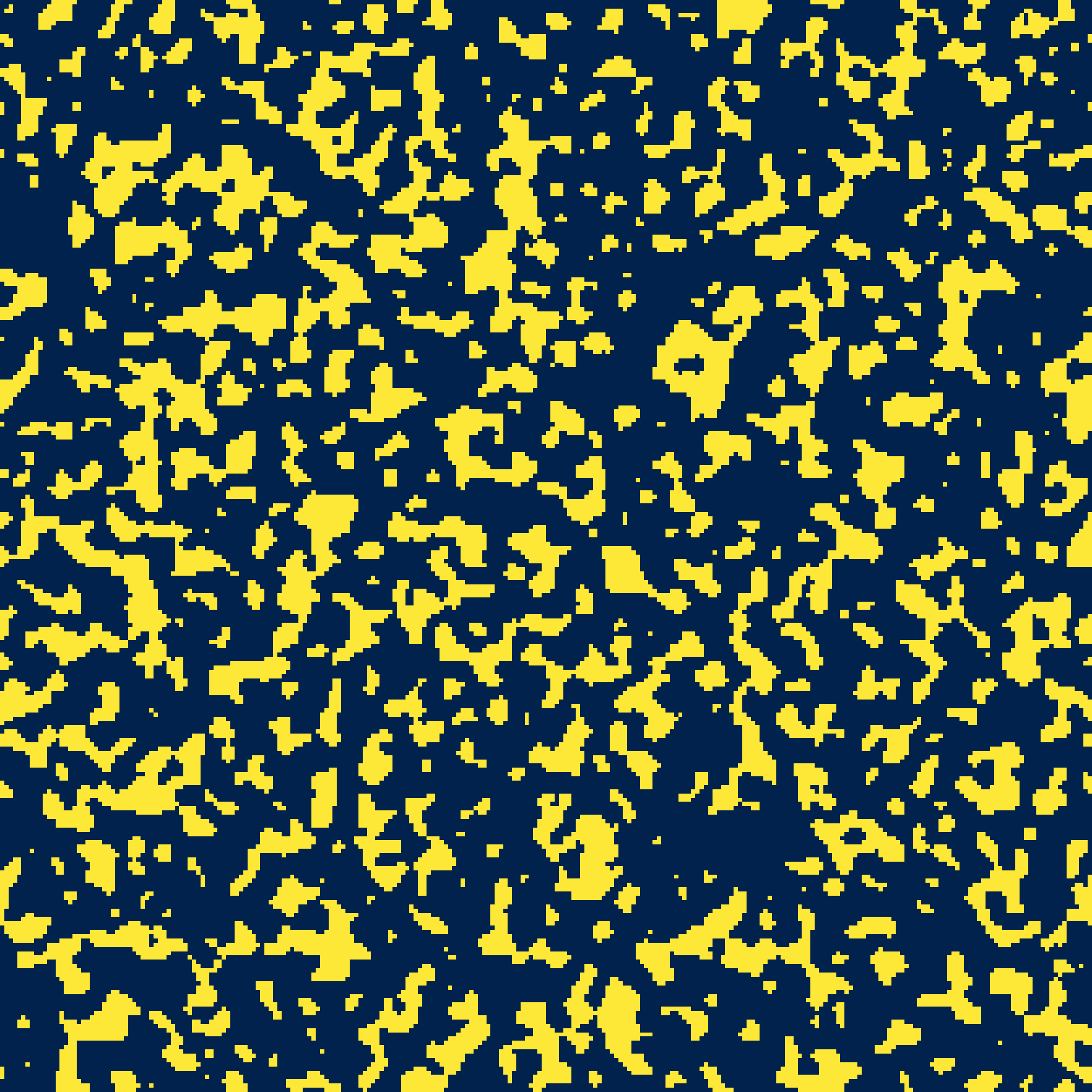}
\end{minipage}
}
\hfill
\subfloat[Small set of reconstructed 3D volumes]{
\begin{minipage}[t][8cm][c]{0.5\linewidth}
\includegraphics[width=0.3\linewidth]{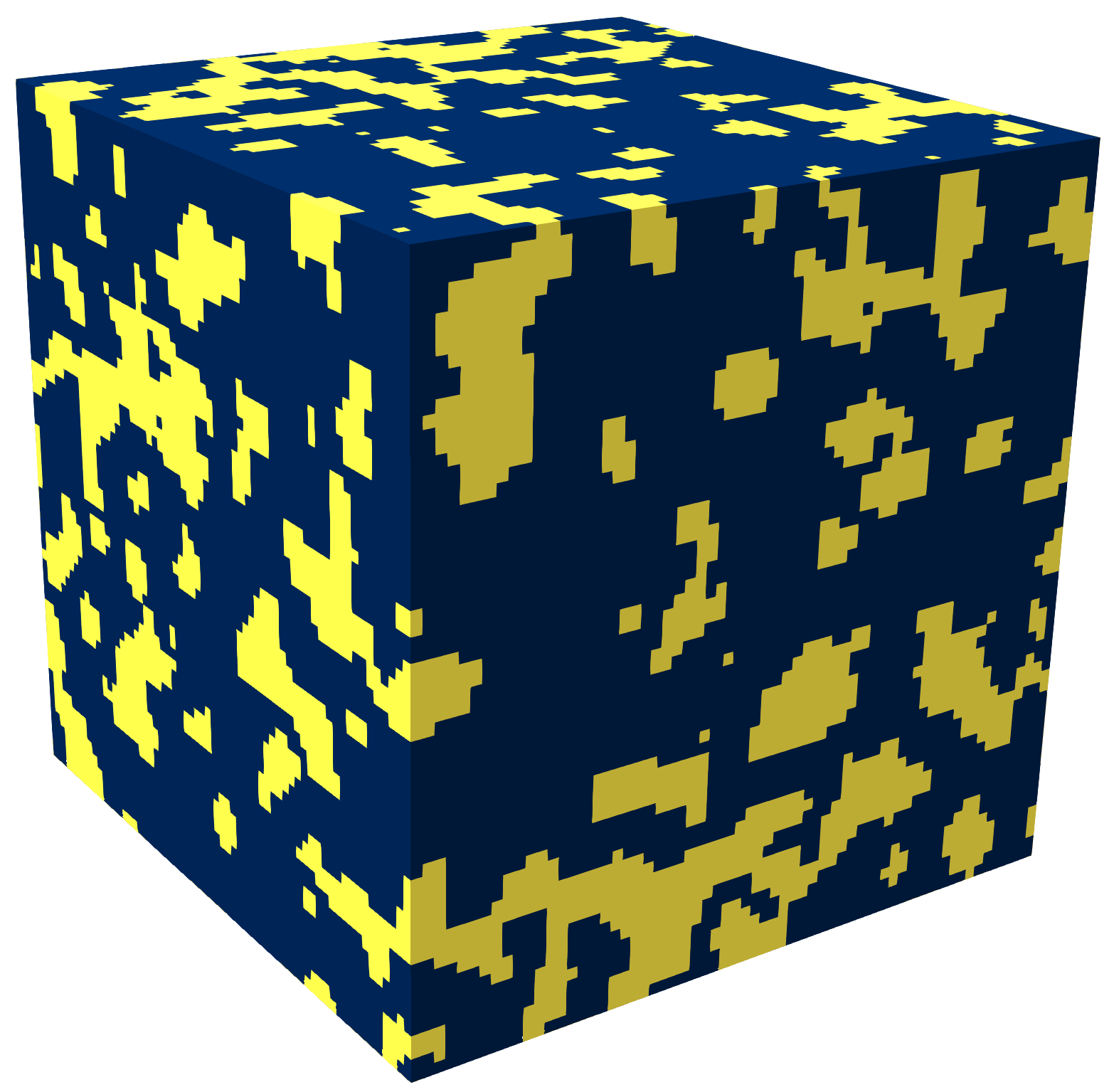}
\hfill
\includegraphics[width=0.3\linewidth]{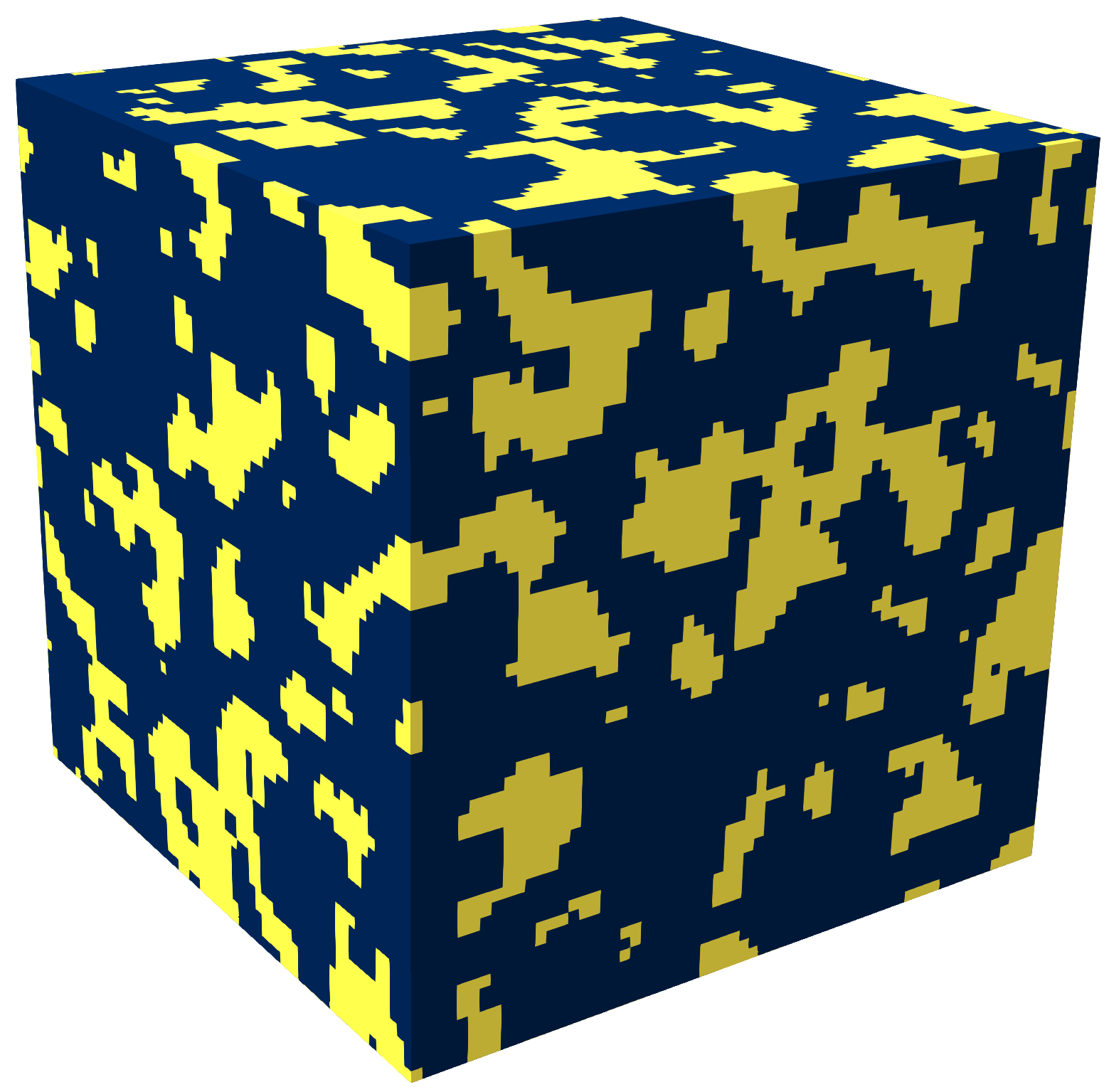}
\hfill
\includegraphics[width=0.3\linewidth]{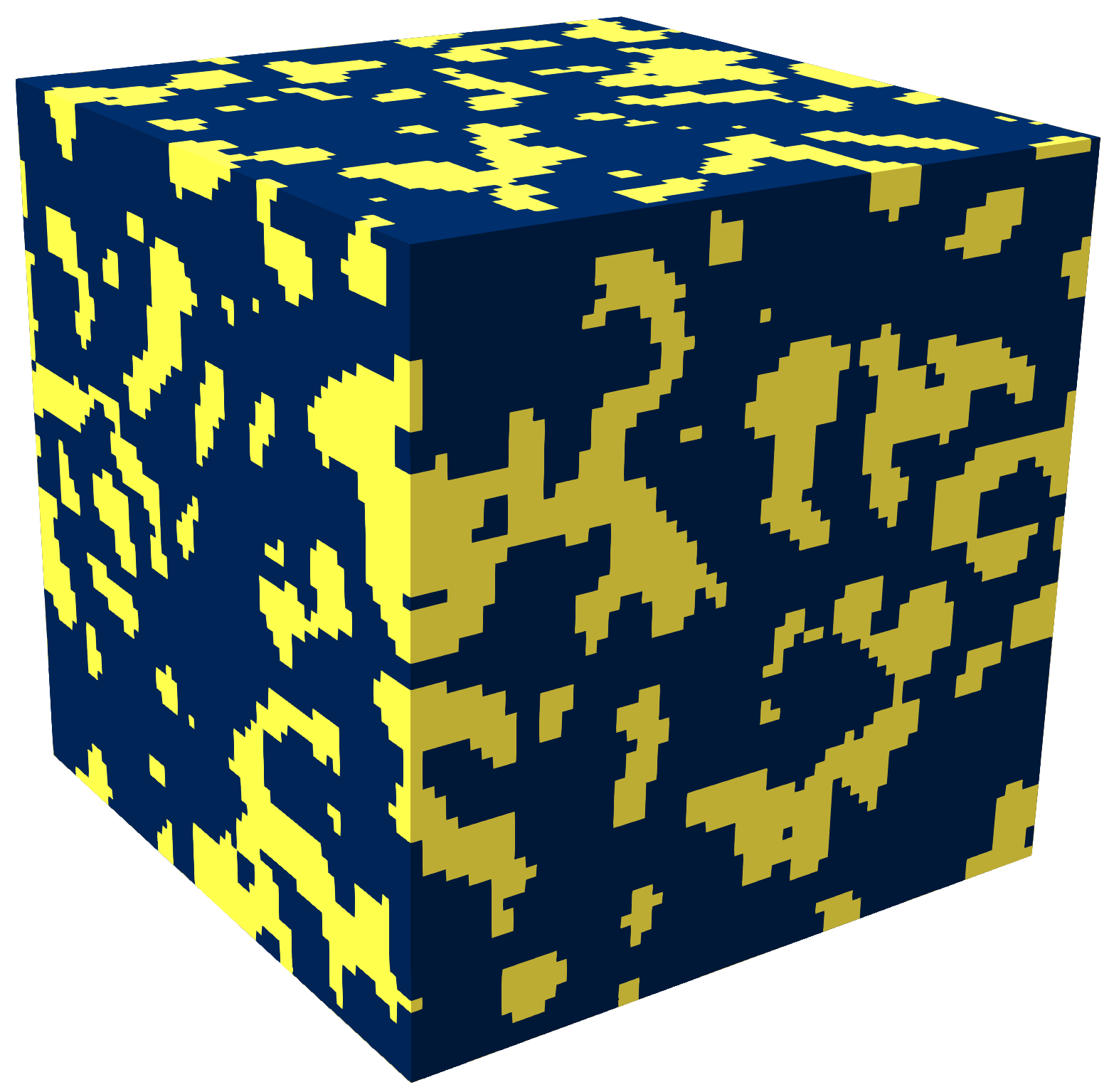}
\vfill
\includegraphics[width=0.3\linewidth]{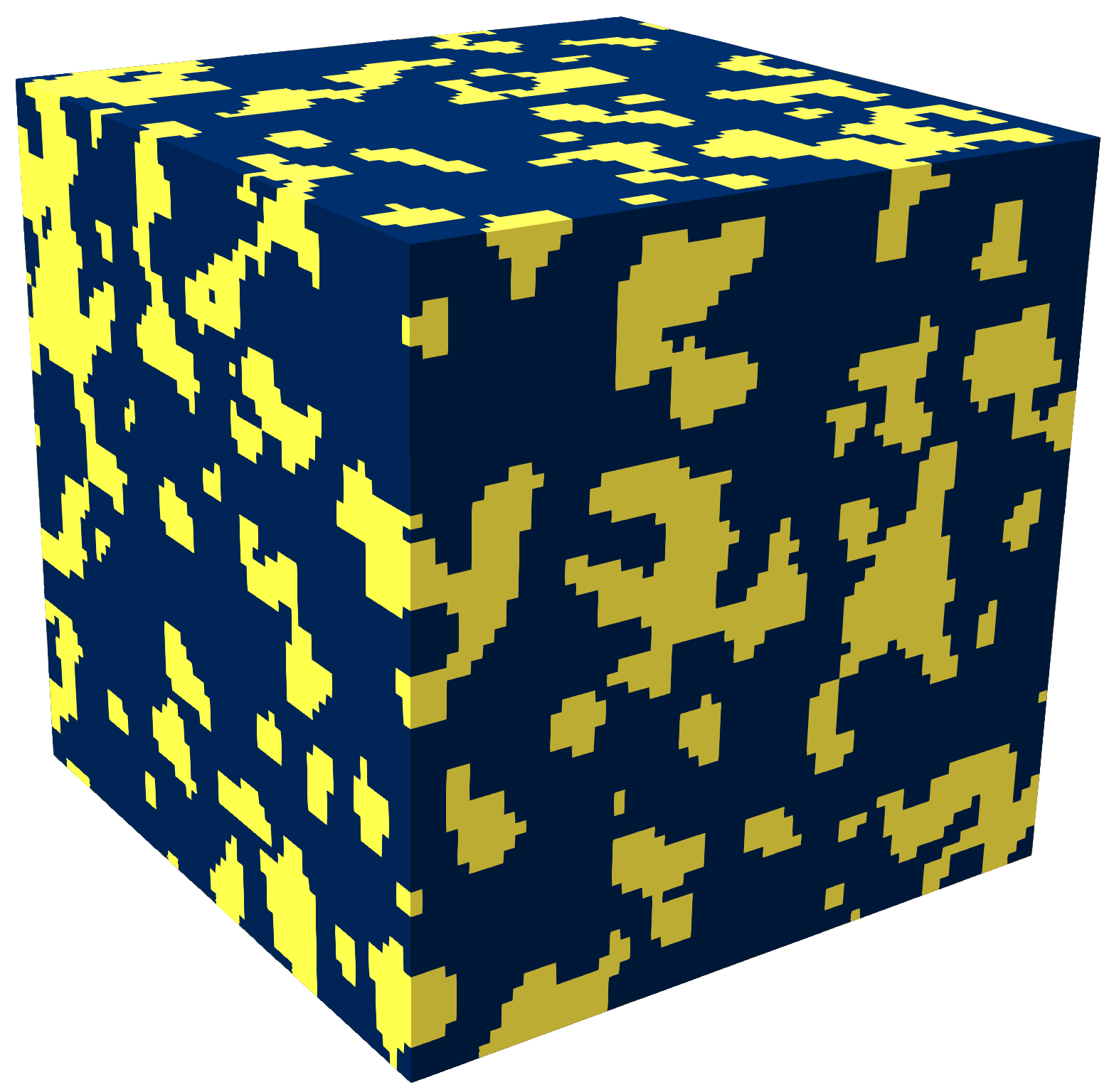}
\hfill
\includegraphics[width=0.3\linewidth]{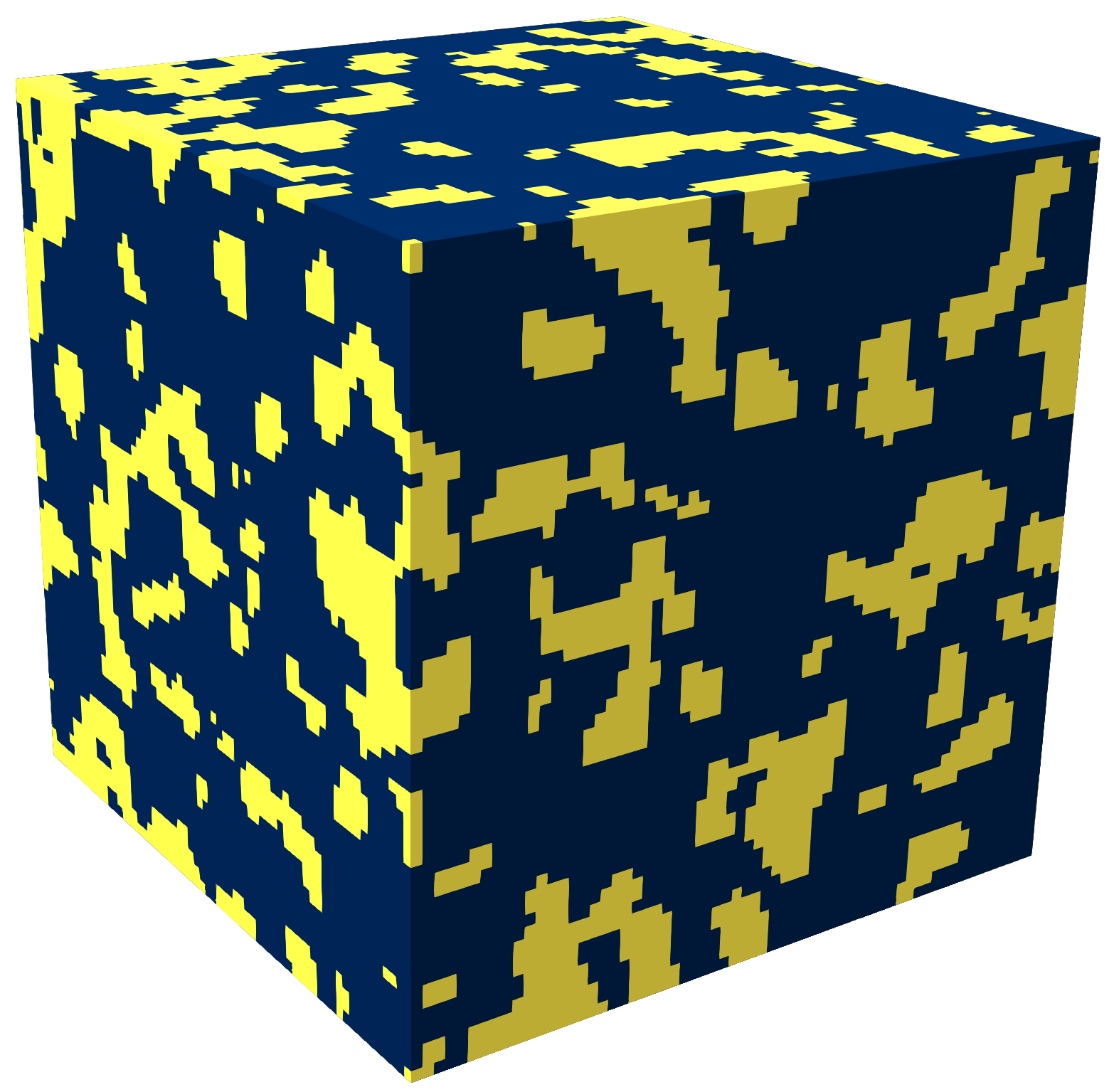}
\hfill
\includegraphics[width=0.3\linewidth]{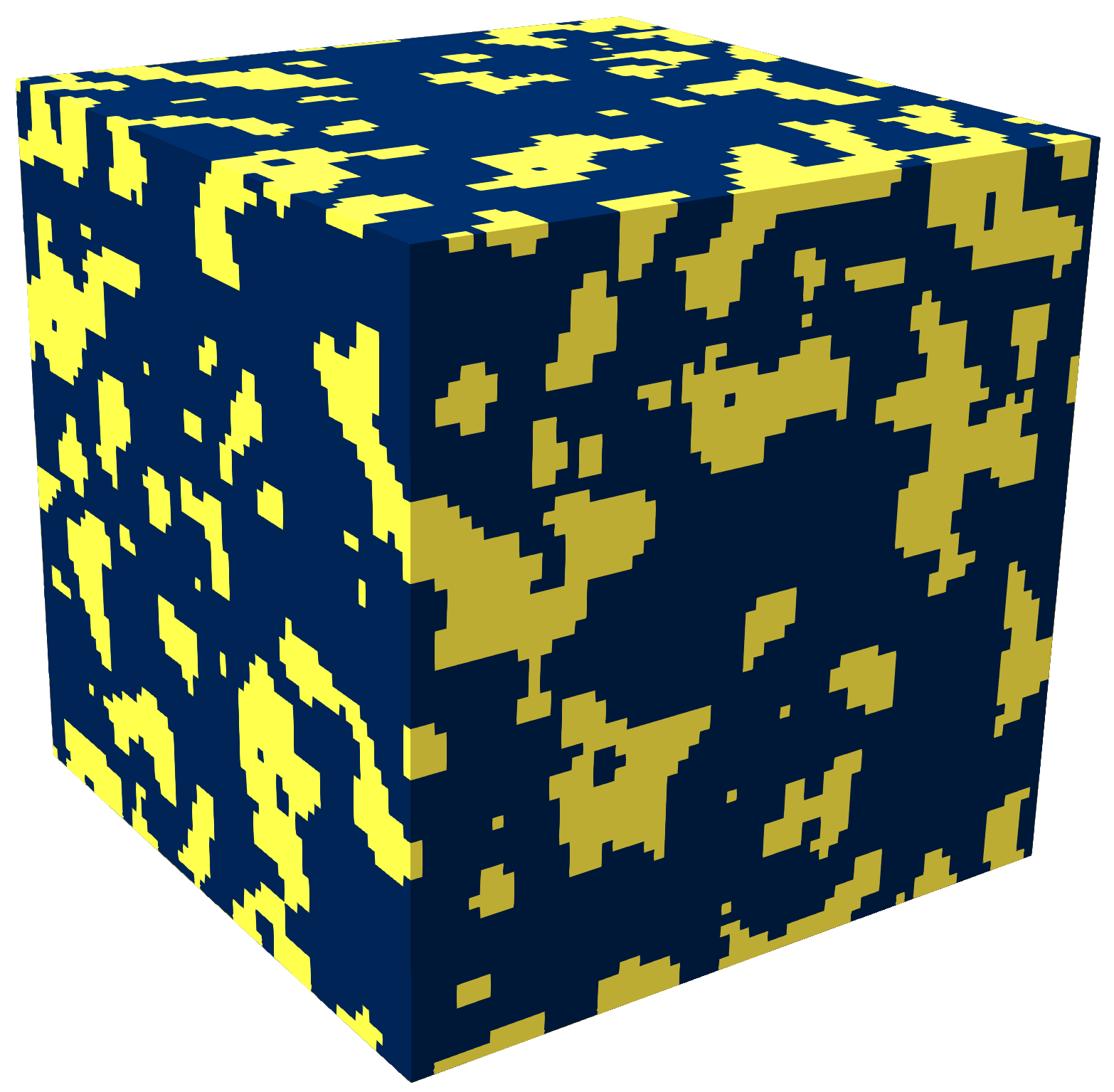}
\vfill
\includegraphics[width=0.3\linewidth]{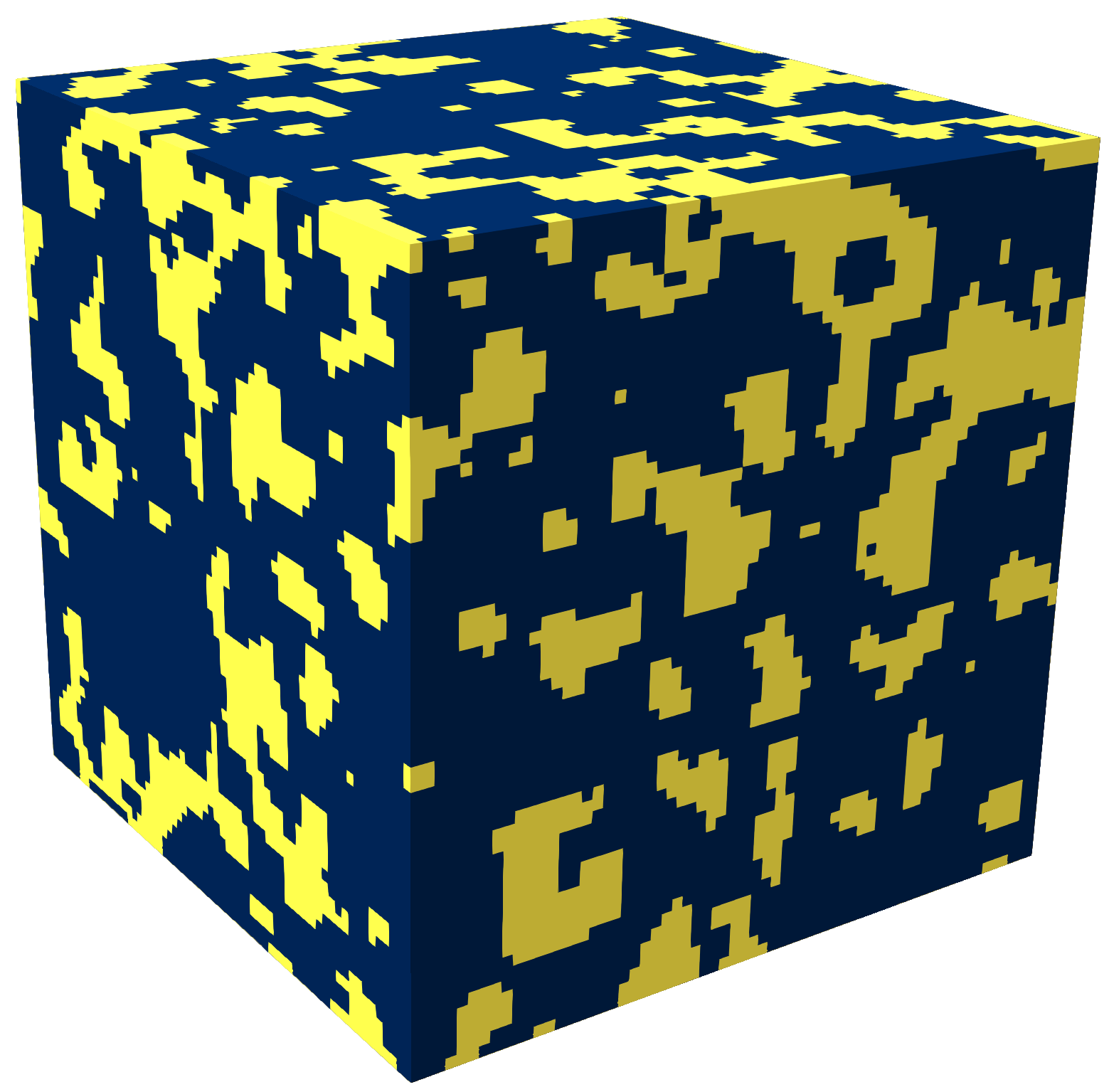}
\hfill
\includegraphics[width=0.3\linewidth]{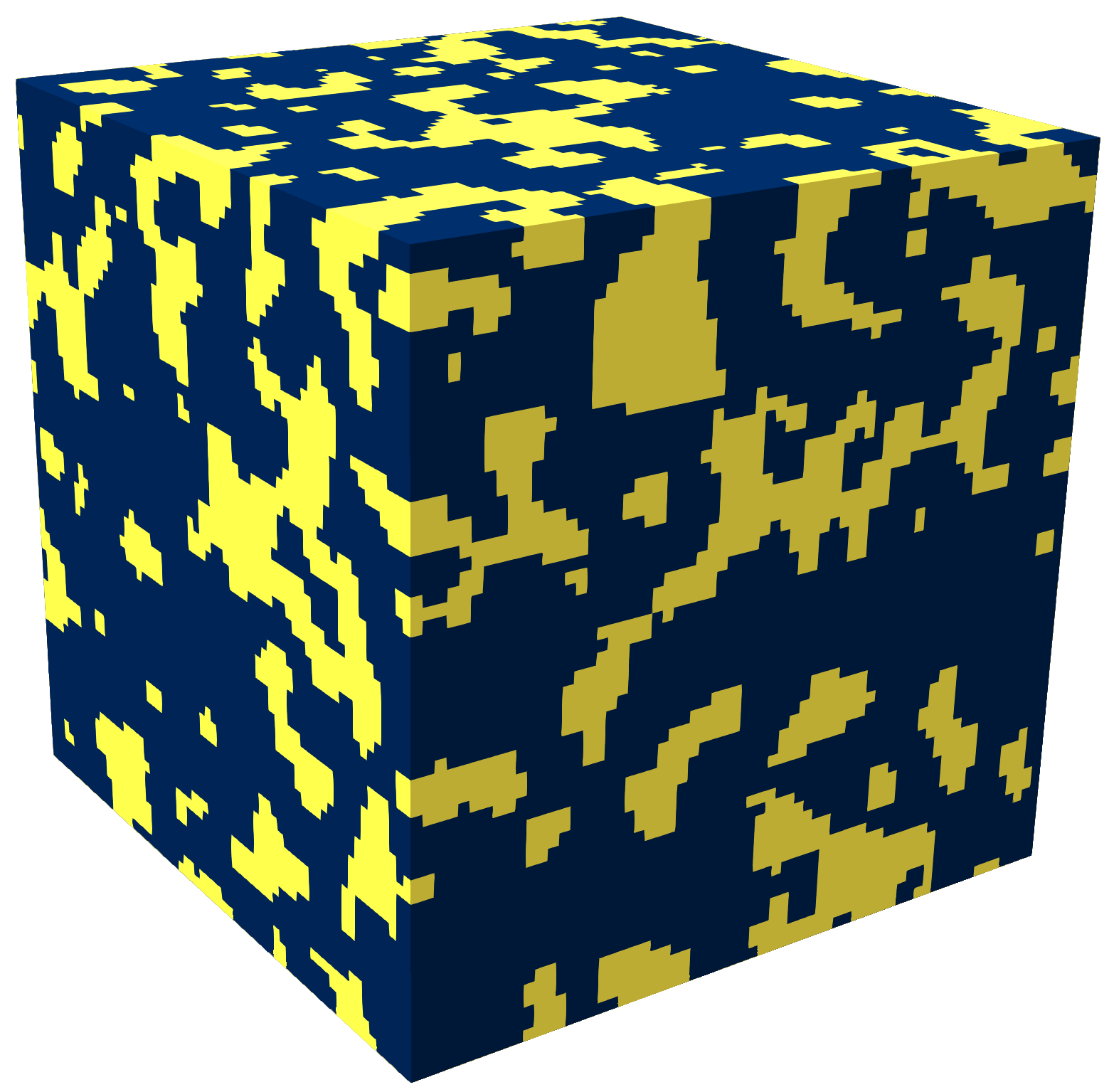}
\hfill
\includegraphics[width=0.3\linewidth]{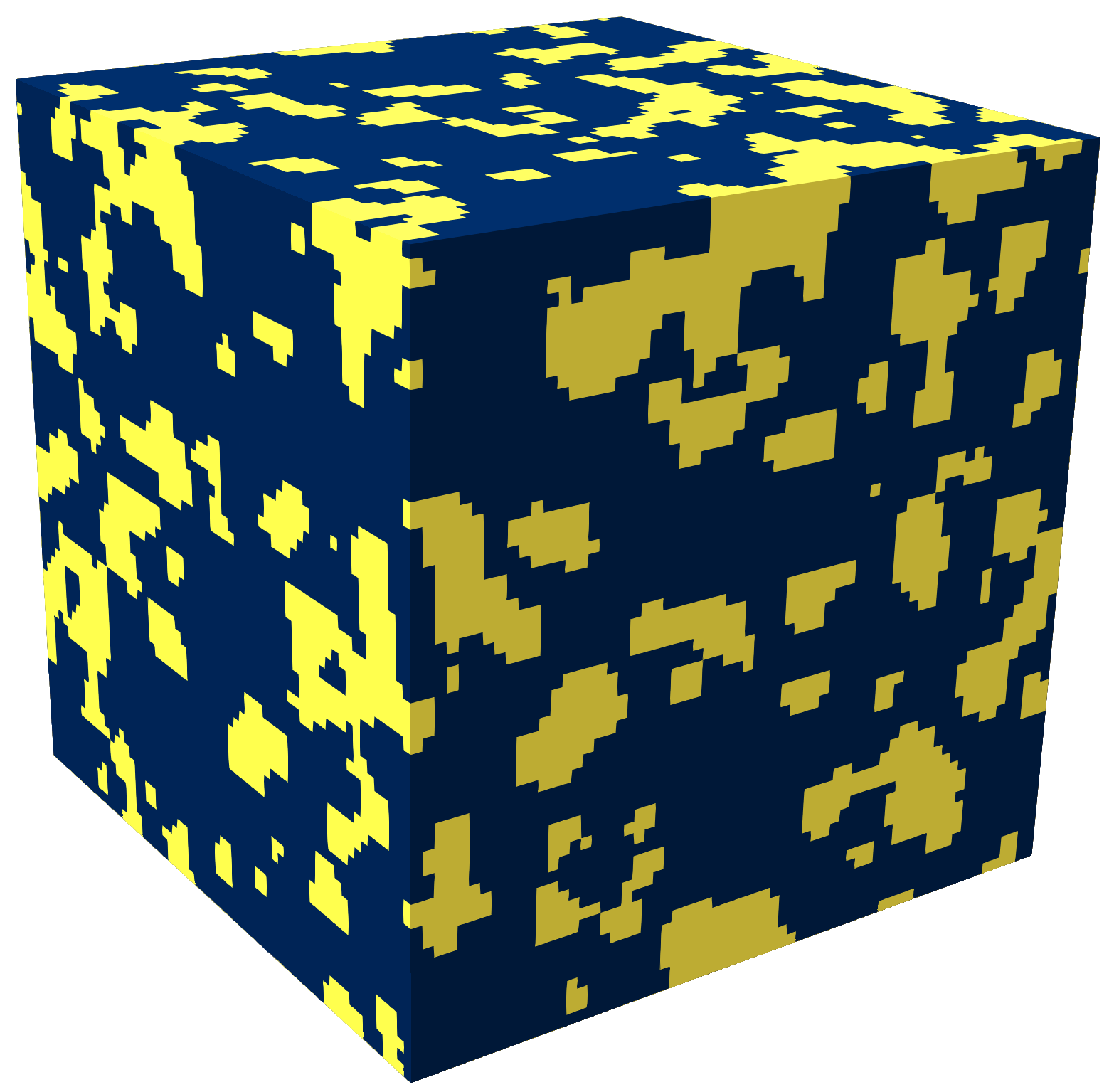}
\end{minipage}
}
\caption{Input and results generated from the code in Listing~\ref{lst:bash}.\label{fig:sve}}
\end{figure*}

This task is solved using \emph{MCRpy} as a command line tool as shown in Listing~\ref{lst:bash}.
First, the original 2D microstructure stored as \texttt{ms\_slice.npy} is characterized using the same parameters as in Section~\ref{sec:wf1}~(line 1).
Then,~9 different 3D structures are generated by a simple loop over the reconstruction script~(lines 2-7).
Note that the extent of the reconstructed structures in voxels is set independently of the original slice~(line 5).
Furthermore, the loop index is passed to the reconstruction script in order to have it added to all result filenames and prevent to overwrite previous results~(line 5).
Because the chosen descriptors are differentiable, the standard optimizer L-BFGS-B~\cite{byrd_stochastic_2015} can be used, allowing to harness the computational efficiency of DMCR~\cite{seibert_reconstructing_2021-1,seibert_descriptor-based_2022}.
On an \emph{Nvidia A100} GPU, the reconstructions take around 25~minutes per structure for 500~iterations.
The original structure and the results can be seen in Figure~\ref{fig:sve}.
In summary, the command line interface is analogous to the GUI and allows for easy automation and large-scale application.
\begin{figure*}
\begin{lstlisting}[language=bash,caption={Simple bash script for automating reconstruction using the \emph{MCRpy} command line interface. The results are shown in Figure~\ref{fig:sve}},label={lst:bash}]
python characterize.py ms_slice.npy --limit_to 8 --descriptor_types Correlations Variation
for i in {1..9} 
do
    python reconstruct.py --descriptor_filename results/ms_slice_characterization.pickle \
        --extent_x 64 --extent_y 64 --extent_z 64 --limit_to 8 --information ${i} \ 
        --descriptor_types Correlations Variation --descriptor_weights 1 100
done
\end{lstlisting}
\end{figure*}

\subsection{Manipulating the descriptor space}
\label{sec:wf3}
As a third example, the explicit availability of the descriptor is exploited by directly manipulating it.
Specifically, \emph{MCRpy} is used to interpolate between two given microstructures in a morphologically meaningful way.
Consider two microstructures that could stem from different sets of process parameters.
It can be interesting to create a morphology that is a mix between these two structures.
For example, if numerical simulations and homogenization of the interpolated structure predict favorable effective properties, it might be worth to fine-tune the process parameters or try to establish a PSP linkage to manufacture these structures.
Direct interpolation of the microstructures in terms of pixel values is meaningless.
As a simple alternative, we interpolate linearly in the descriptor space and reconstruct microstructures from the interpolated descriptors.

\begin{figure*}[t]
    \centering
	\subfloat[Isotropic~$\tilde{S}_3(\vec{r},\vec{r})$]{\includegraphics[width=0.19\textwidth]{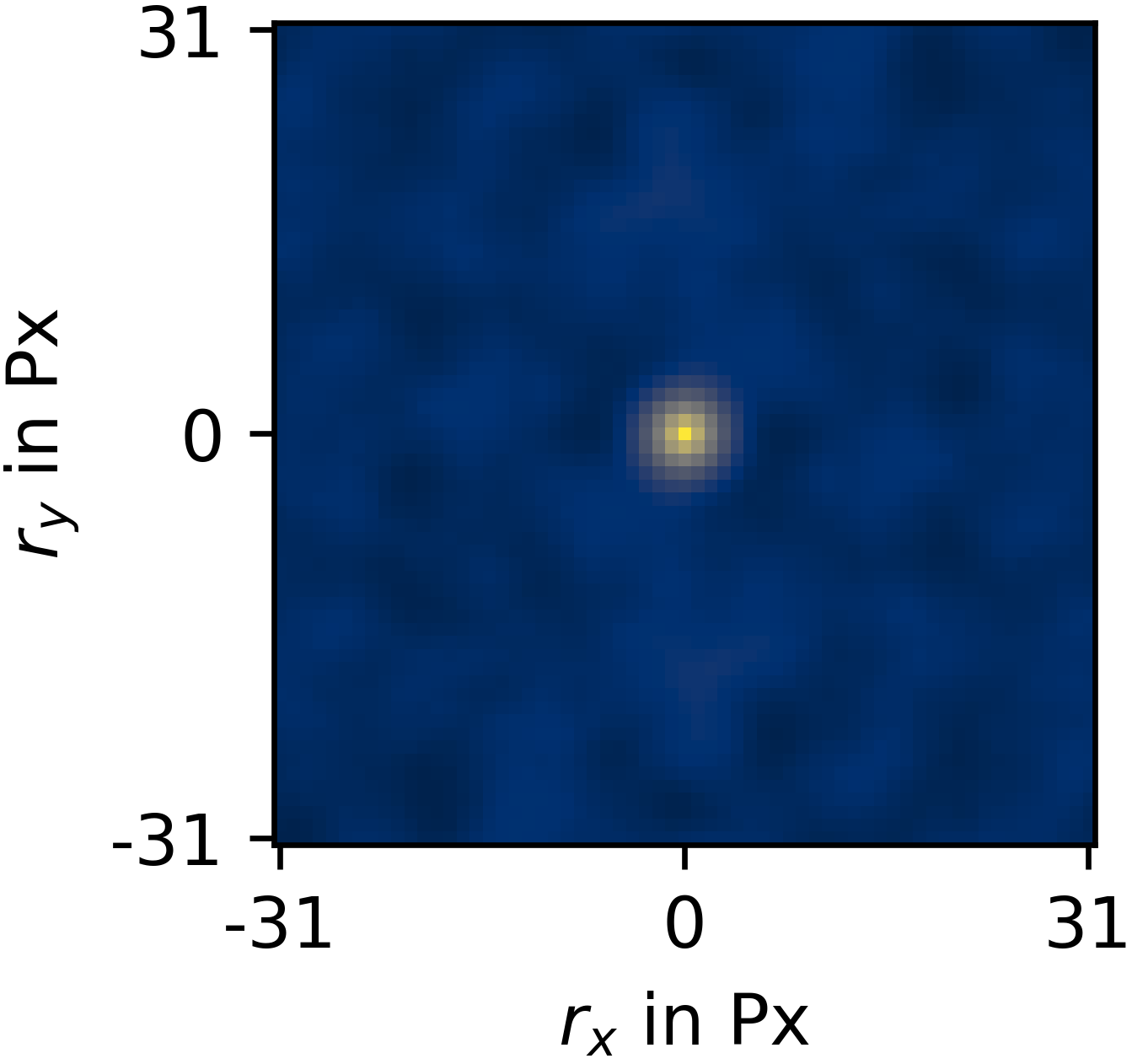}}
	\hfill
	\subfloat[$25\%$-interpolation]{\includegraphics[width=0.19\textwidth]{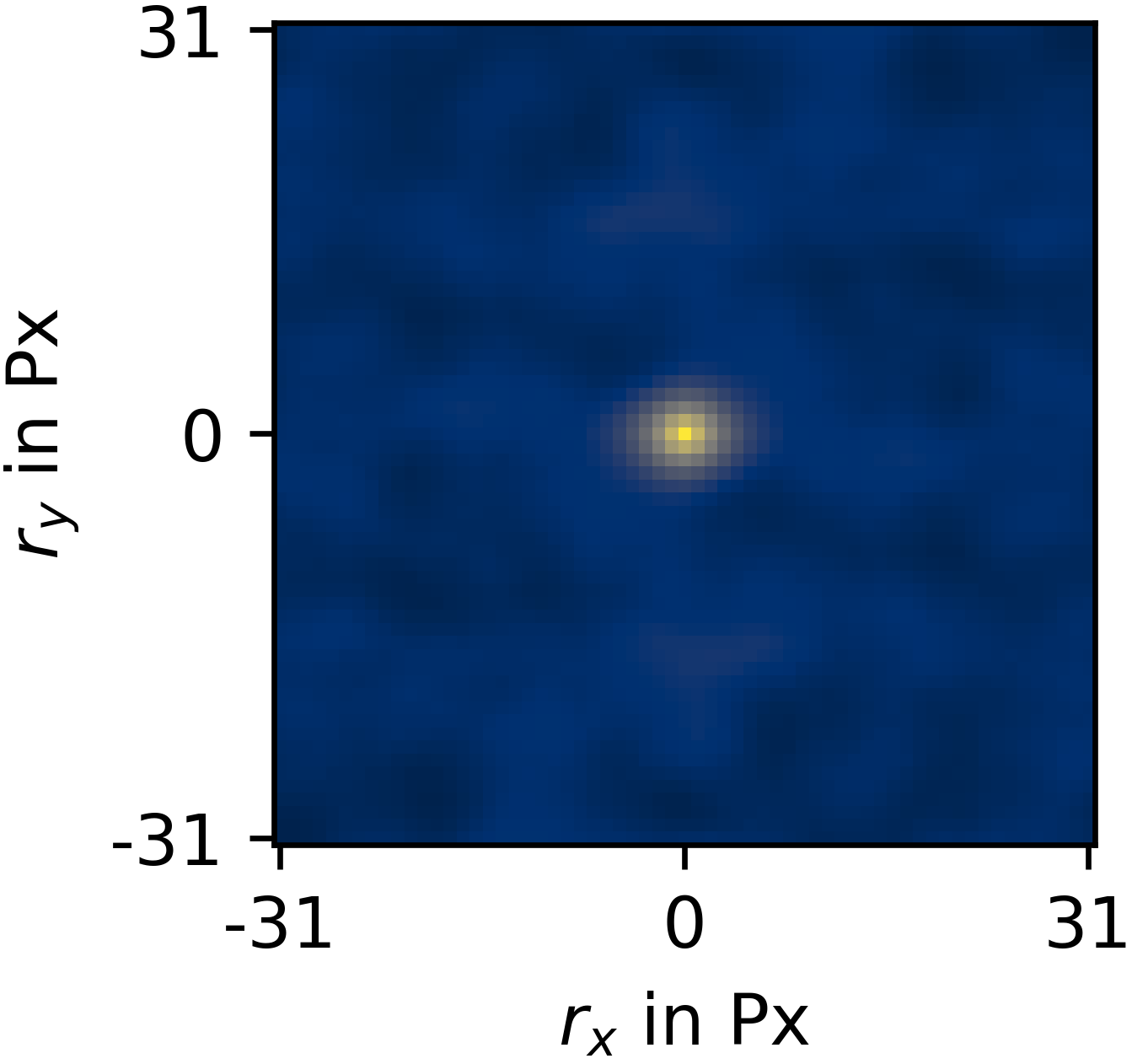}}
	\hfill
	\subfloat[$50\%$-interpolation]{\includegraphics[width=0.19\textwidth]{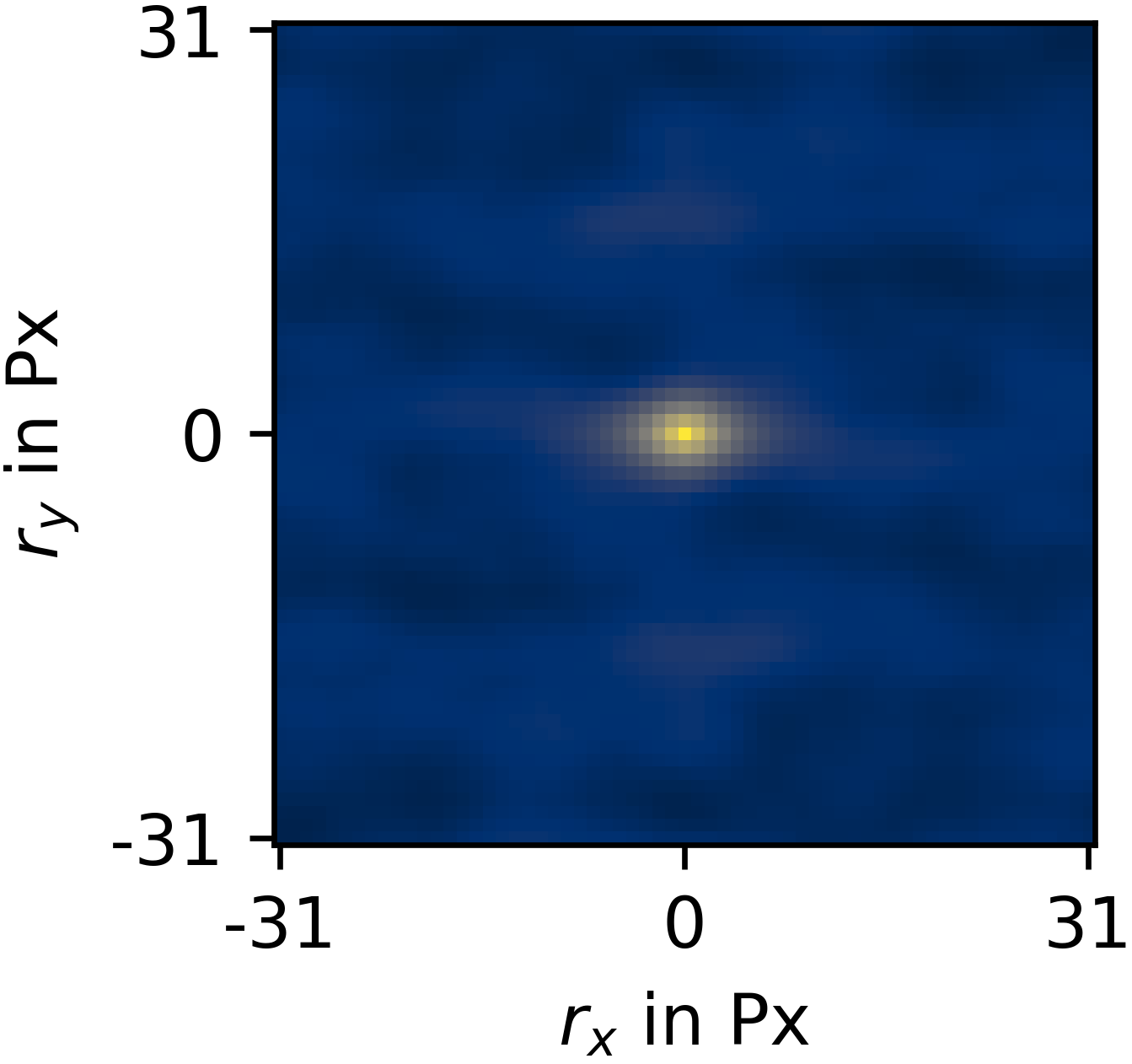}}
	\hfill
	\subfloat[$75\%$-interpolation]{\includegraphics[width=0.19\textwidth]{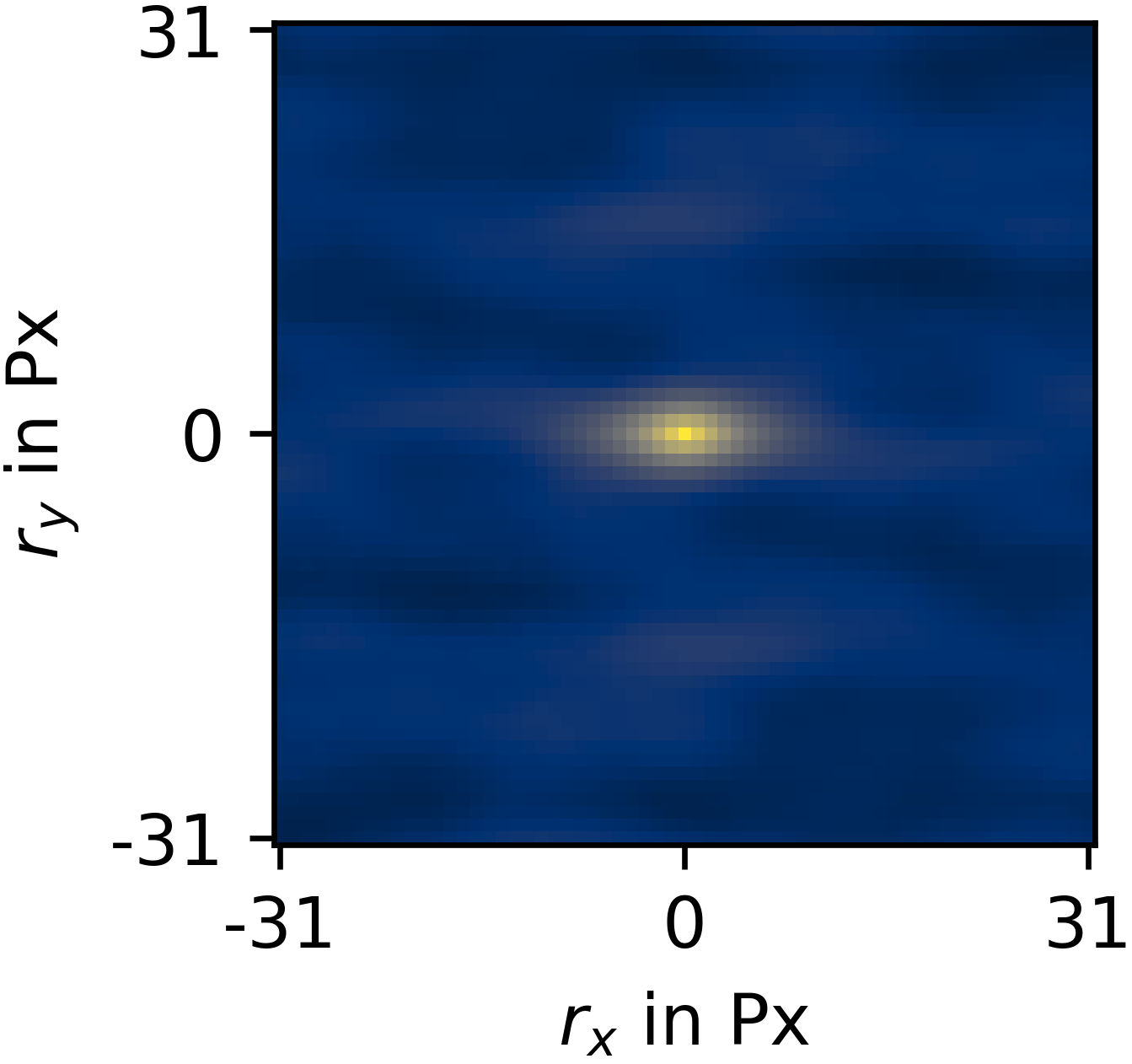}}
	\hfill
	\subfloat[Elongated~$\tilde{S}_3(\vec{r},\vec{r})$]{\includegraphics[width=0.19\textwidth]{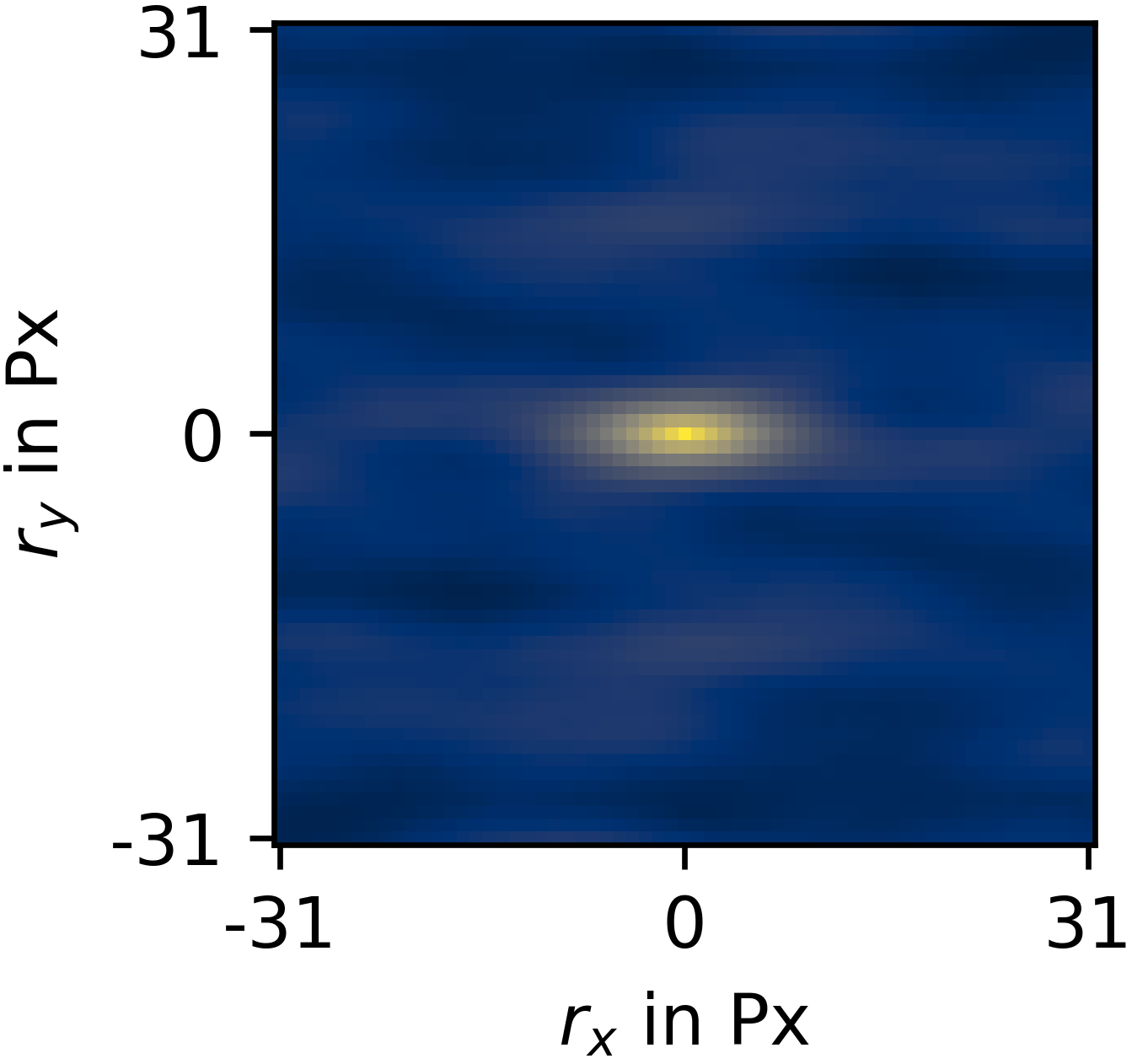}}
	\vfill
	\subfloat[Reconstruction from (a)]{\includegraphics[width=0.19\textwidth]{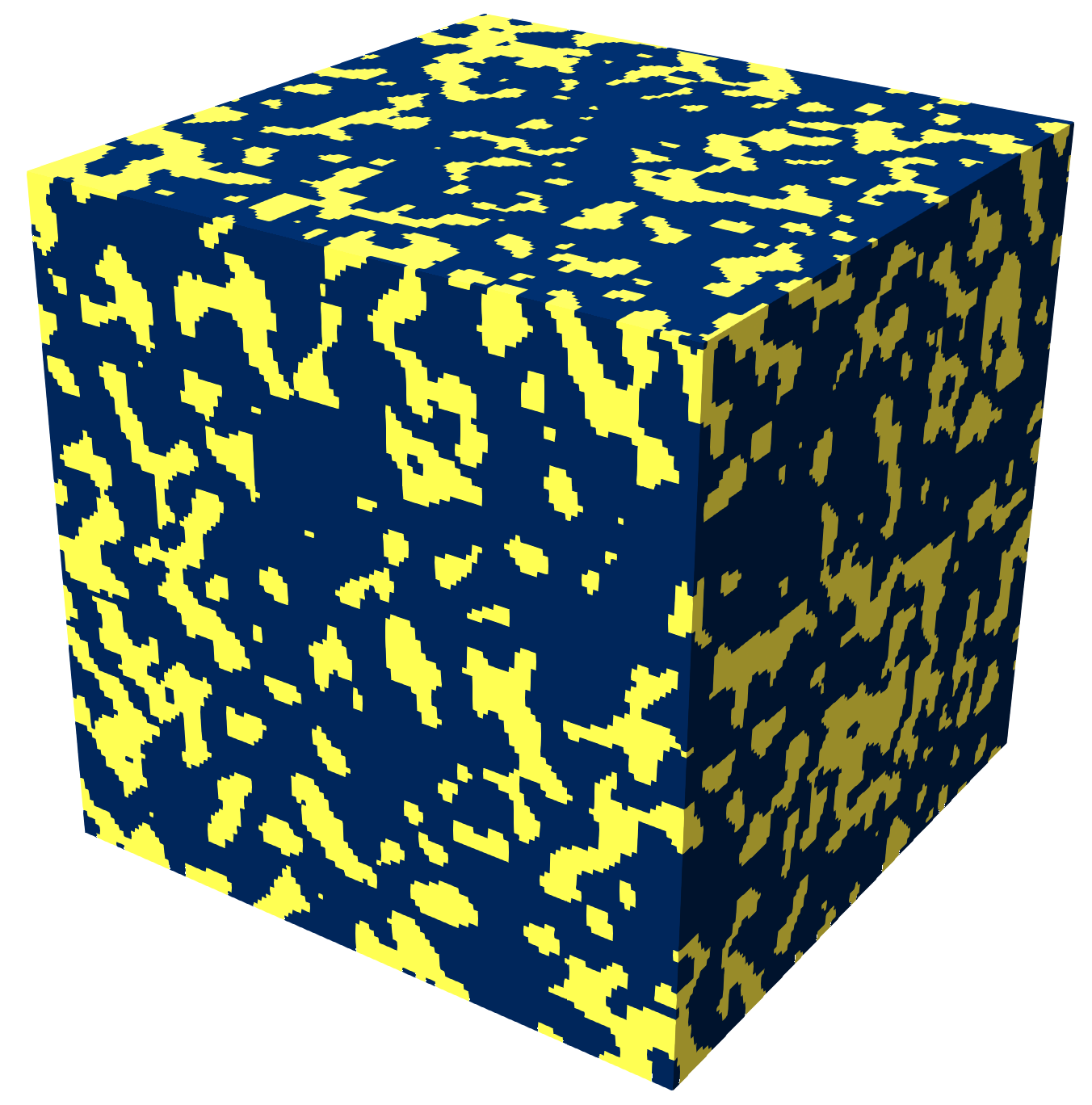}}
	\hfill
	\subfloat[Reconstruction from (b)]{\includegraphics[width=0.19\textwidth]{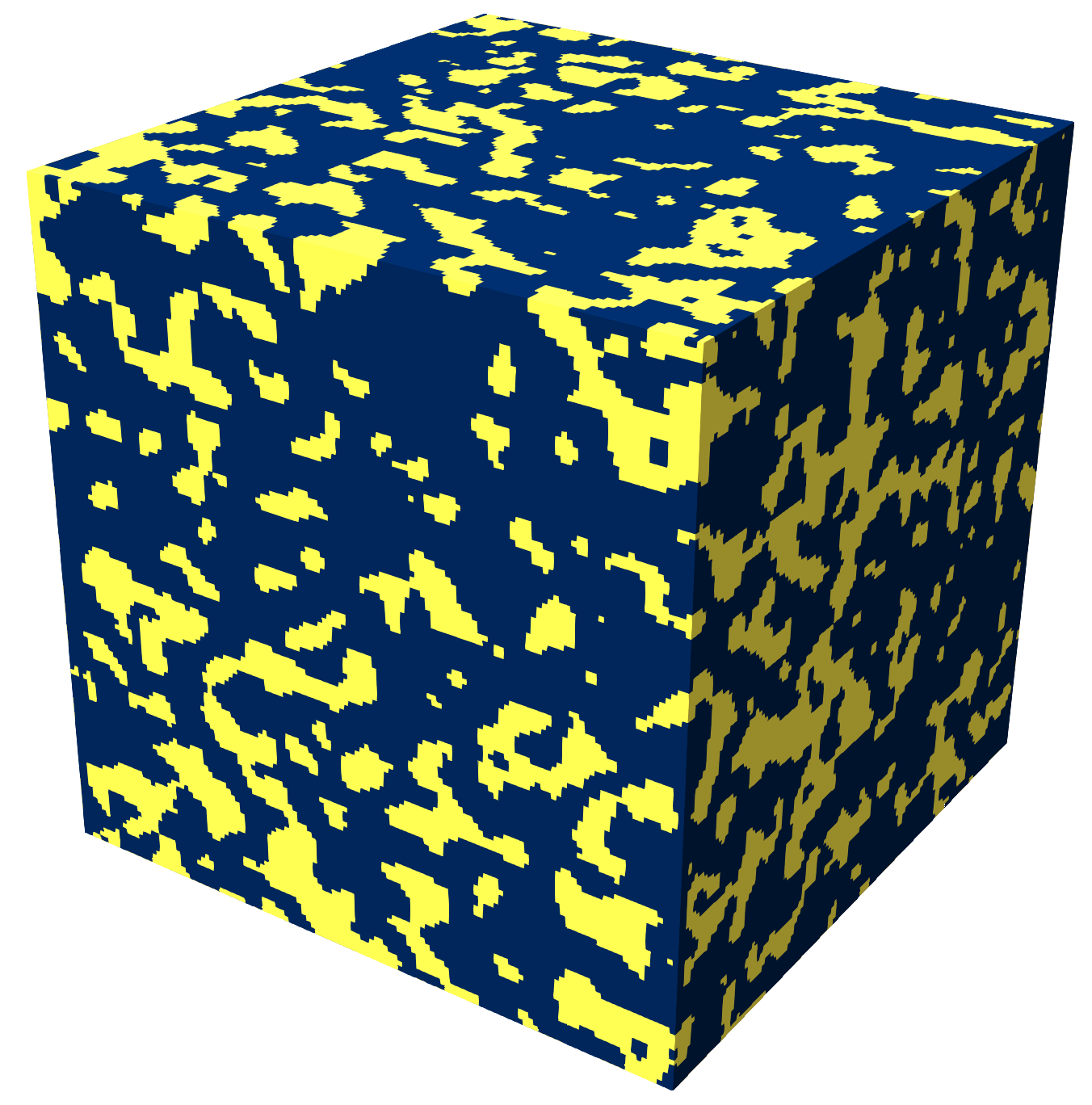}}
	\hfill
	\subfloat[Reconstruction from (c)]{\includegraphics[width=0.19\textwidth]{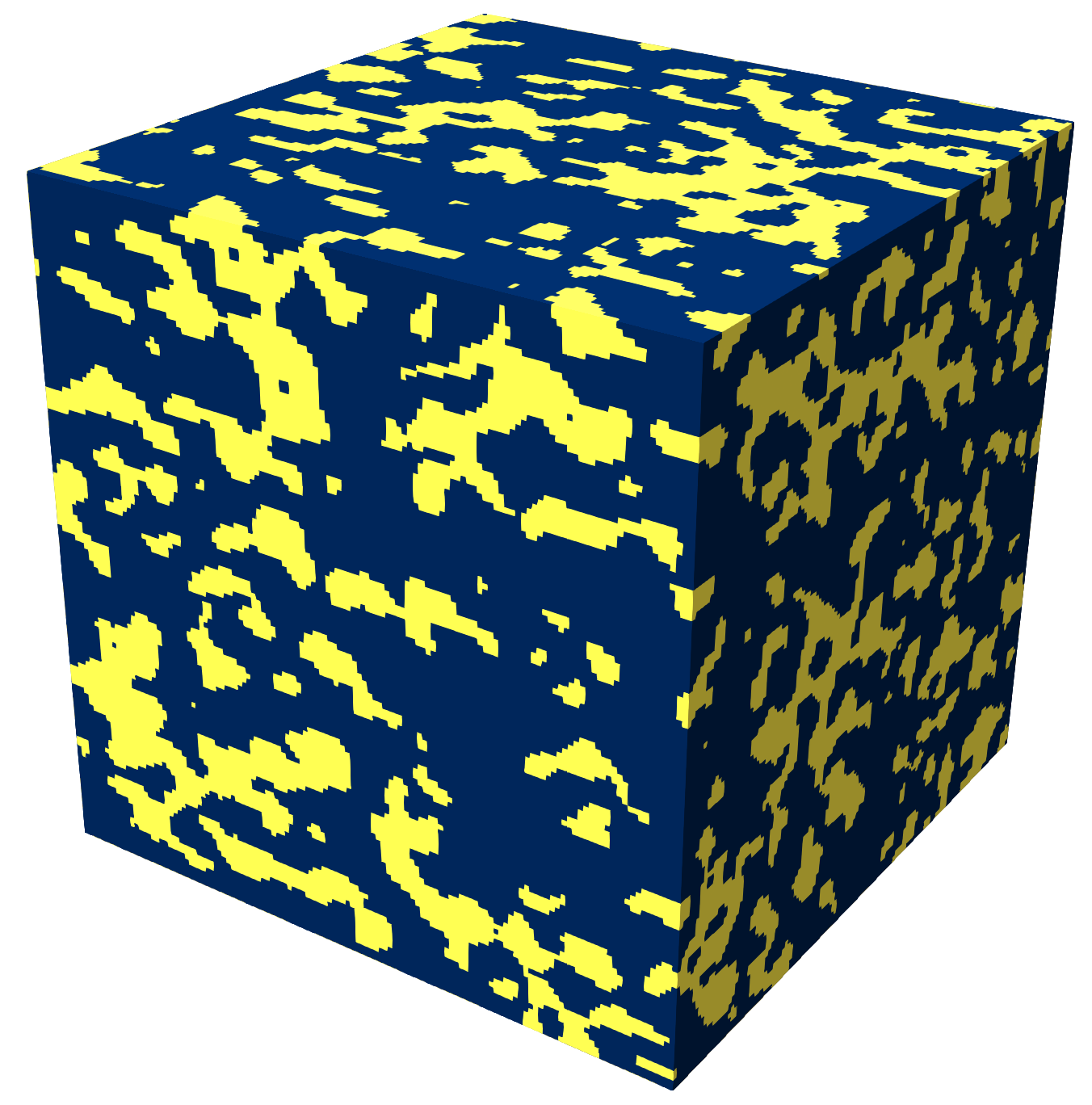}}
	\hfill
	\subfloat[Reconstruction from (d)]{\includegraphics[width=0.19\textwidth]{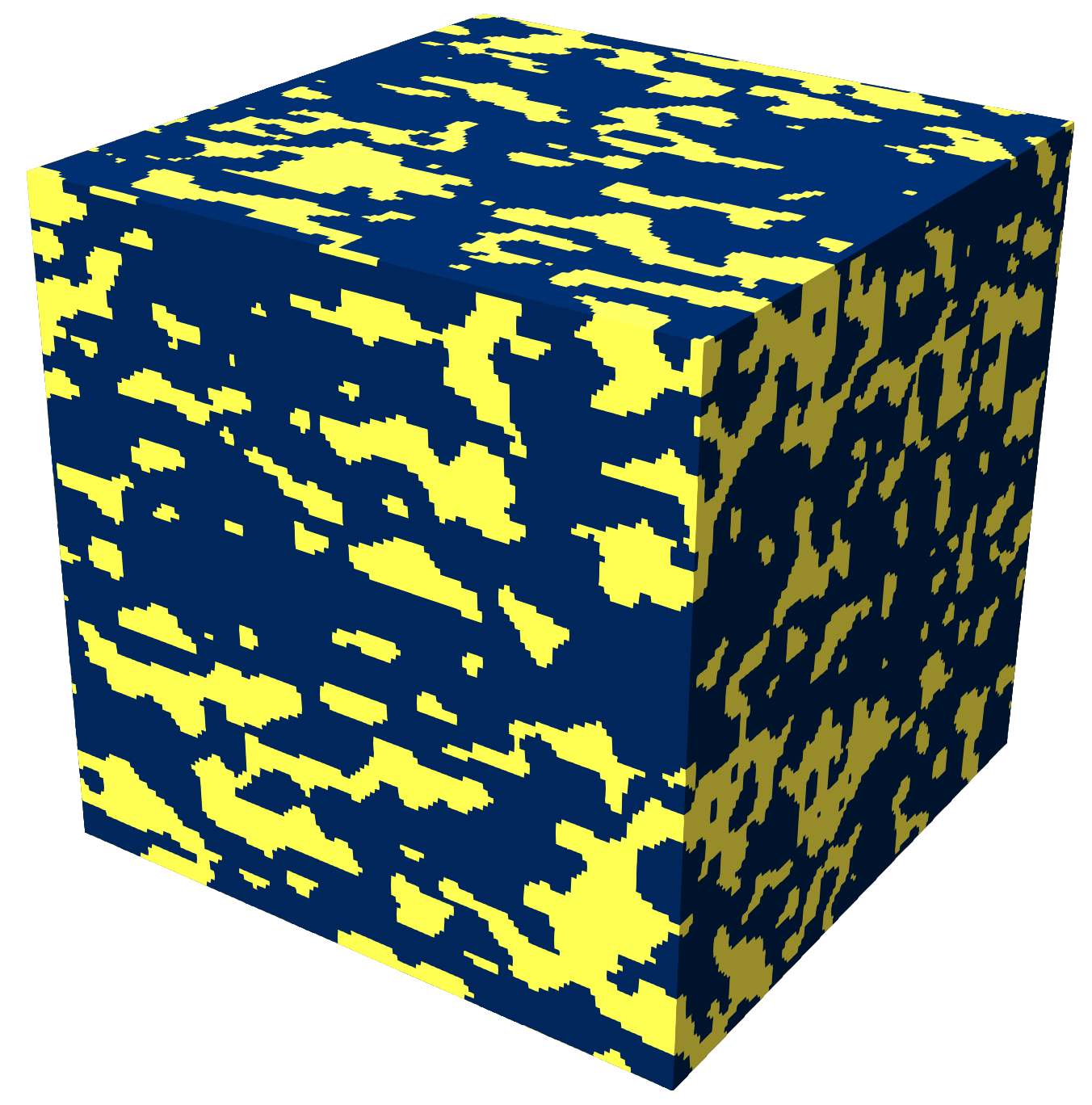}}
	\hfill
	\subfloat[Reconstruction from (e)]{\includegraphics[width=0.19\textwidth]{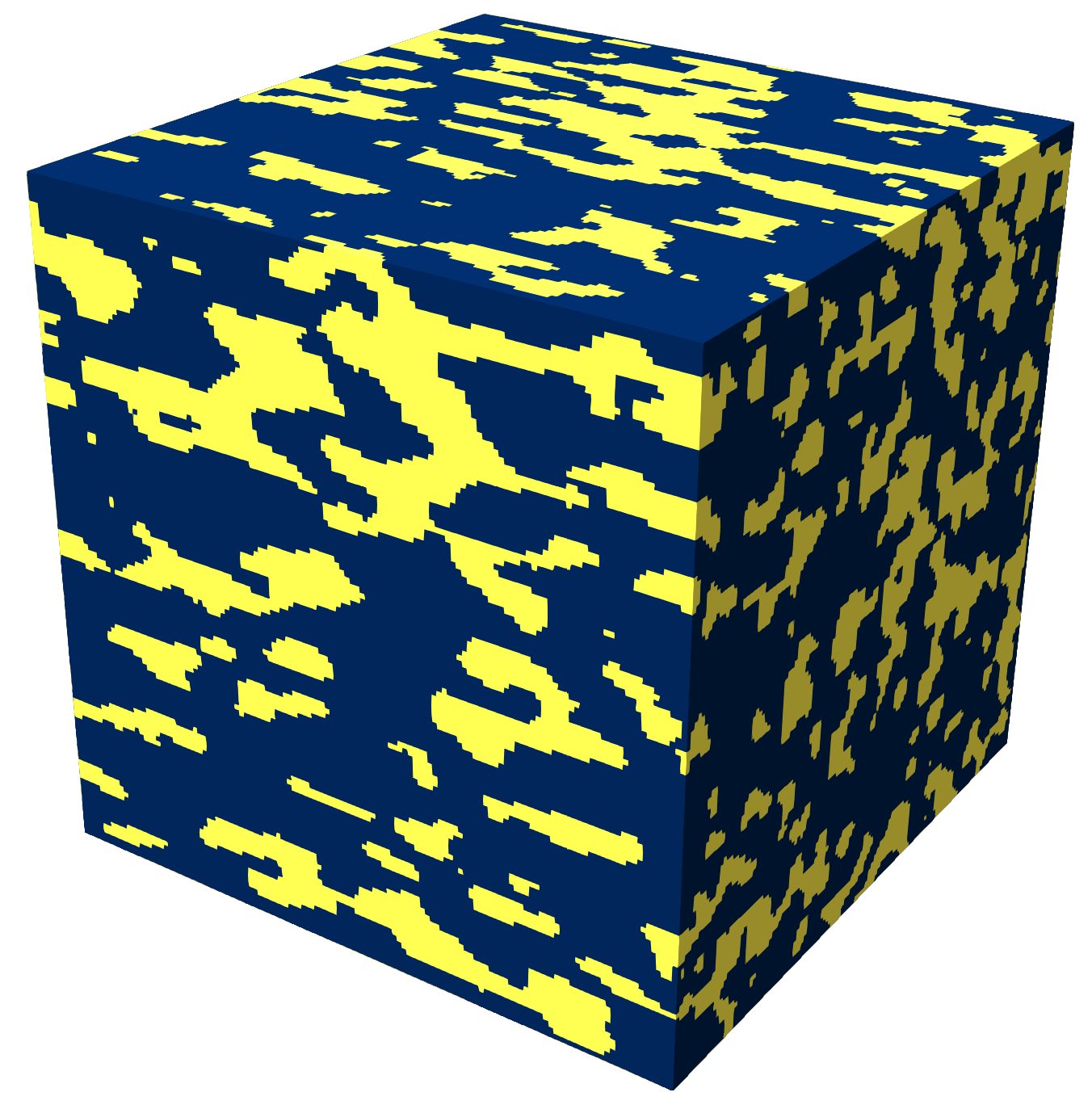}}
    \caption{The original descriptors (a, e) are linearly interpolated (b-d) and used for reconstruction to create a smooth transition between an isotropic and an elongated microstructure (a-f). For the descriptor~$\tilde{S}_3(\vec{r}_a,\vec{r}_b)$, only the case that~$\vec{r}_a=\vec{r}_b=\vec{r}$ is plotted for clarity.\label{fig:ms}}
\end{figure*}

This task is solved using \emph{MCRpy} as a Python library as shown in Listing~\ref{lst:python}.
After defining the settings~(lines 4-10), the original 2D microstructure slices are loaded~(lines 13-14) and characterized~(lines 17-18).
For reconstructing elongated 3D structures, the 2D descriptors need to be combined such that different descriptors are used in different directions.
The order thereby matters and mistakes can lead to geometrically unrealizable descriptors\footnote{For example, consider the three planes $x-y$, $x-z$ and $y-z$. Structures that are elongated in $x$-direction can be created by prescribing horizontally elongated descriptors in planes~1 and~2 and an isotropic descriptor in plane~3. However, if the same horizontally elongated descriptors are prescribed in planes~1 and~3, the structure cannot be realized: Plane~1 requires elongations in the~$y$-direction, whereas plane~3 requires the elongations to be in the~$z$-direction and not in~$y$.}.
In order to avoid confusion and mistakes, \emph{MCRpy} provides the function \texttt{merge} for this task.
The merged descriptors~(lines 21-22) are then interpolated in~5 steps including start and end~(line 25).
Each descriptor is used for a 3D reconstruction, which returns the convergence data and the final microstructure~(line 29). 
The convergence data is viewed in an interactive window as shown in Figure~\ref{fig:convergence}~(line 31).
Finally, the microstructures are smoothed by a Gaussian filter~(line 32) and saved to a file~(line 33).
The results are shown in Figure~\ref{fig:ms}.
It can be confirmed that linear interpolation in the descriptor space leads to a visually reasonable transition between the corresponding microstructures.
\begin{figure*}
\begin{lstlisting}[language=python,caption={Simple Python script that uses \emph{MCRpy} to characterize two microstructure slices, interpolate between them in the descriptor space and reconstruct the corresponding 3D structures.},label={lst:python}]
import mcrpy

# define settings
limit_to = 8
descriptor_types = ['Correlations', 'Variation']
descriptor_weights = [1.0, 10.0]
characterization_settings = mcrpy.CharacterizationSettings(descriptor_types=descriptor_types, 
    limit_to=limit_to)
reconstruction_settings = mcrpy.ReconstructionSettings(descriptor_types=descriptor_types, 
    descriptor_weights=descriptor_weights, limit_to=limit_to, use_multigrid_reconstruction=True)

# load microstructures
ms_from = mcrpy.load('microstructures/ms_slice_isotropic.npy')
ms_to = mcrpy.load('microstructures/ms_slice_elongated.npy')

# characterize microstructures
descriptor_isotropic = mcrpy.characterize(ms_from, characterization_settings)
descriptor_elongated = mcrpy.characterize(ms_to, characterization_settings)

# merge descriptors
descriptor_from = mcrpy.merge([descriptor_isotropic])
descriptor_to = mcrpy.merge([descriptor_elongated, descriptor_isotropic])

# interpolate in descriptor space
d_inter = mcrpy.interpolate(descriptor_from, descriptor_to, 5)

# reconstruct from interpolated descriptors and save results
for i, interpolated_descriptor in enumerate(d_inter):
    convergence_data, ms = mcrpy.reconstruct(interpolated_descriptor, (128, 128, 128), 
        settings=reconstruction_settings)
    mcrpy.view(convergence_data)
    smoothed_ms = mcrpy.smooth(ms)
    mcrpy.save_microstructure(f'ms_interpolated_{i}.npy', smoothed_ms)
\end{lstlisting}
\end{figure*}

\subsection{Defining a custom descriptor}
\label{sec:wf4}
\emph{MCRpy} can be easily extended by adding custom plugin modules.
In this section, the procedure is demonstrated by means of a descriptor plugin.
Similar concepts apply to loss functions and optimizers.
First, the implementation of a descriptor plugin is discussed for the volume fraction~$v_\text{f}$.
Secondly, a differentiable approximation to the lineal path function is developed and tested.

Listing~\ref{lst:descriptor} shows the plugin source code for the volume fraction~$v_\text{f}$.
Like all descriptors in \emph{MCRpy}, the volume fraction must inherit from the abstract \texttt{Descriptor} class~(line 5).
This base class provides 
\begin{itemize}
    \item[\textit{(i)}] a wrapper that applies descriptors defined for single phases to the indicator function of each phase,
    \item[\textit{(ii)}] a wrapper to compute multigrid descriptors as discussed in~\cite{seibert_reconstructing_2021-1} and
    \item[\textit{(iii)}] default functions for visualizing descriptors\footnote{By default, low-dimensional descriptors are visualized via bar plots and high-dimensional descriptors are reshaped to an approximately quadratic array and plotted as a heatmap. This behavior can be overwritten for each descriptor subclass separately.}.
\end{itemize}
In this case, it is reasonable to define the descriptor for a single phase and let the superclass handle the generalization to multiple phases.
For this purpose, the subclass function \texttt{make\_singlephase\_descriptor} is defined~(line 9).
This function receives information about the microstructure, like the resolution, which is not needed in this case and therefore summarized via~\texttt{**kwargs}.
It returns a~\texttt{callable} which computes the descriptor given the indicator function of a phase~(line 12).
In order to allow for automatic differentiation of the descriptor with respect to the microstructure, this \texttt{callable} needs to be implemented in \emph{TensorFlow}.
In contrast, a non-differentiable descriptor would be implemented in \emph{Numpy} and integrated into the computation graph by \emph{MCRpy} using the \emph{TensorFlow} function \texttt{tf.py\_function}.
Finally, the plugin is required to register itself at the descriptor factory using its class name~(lines 16-17).
\begin{figure*}[h]
\begin{lstlisting}[language=python,caption={Simple definition of a descriptor plugin for computing the volume fraction~$v_\text{f}$. This descriptor is differentiable with respect to each pixel in the microstructure, so it is implemented in \emph{TensorFlow} to allow for automatic differentiation. Non-differentiable descriptors can be implemented in~\emph{Numpy}.},label={lst:descriptor}]
import tensorflow as tf
from mcrpy.src import descriptor_factory
from mcrpy.descriptors.Descriptor import Descriptor

class VolumeFractions(Descriptor):
    is_differentiable = True

    @staticmethod
    def make_singlephase_descriptor(**kwargs) -> callable:

        @tf.function
        def compute_descriptor(indicator_function: tf.Tensor) -> tf.Tensor:
            return tf.math.reduce_mean(indicator_function)
        return compute_descriptor

def register() -> None:
    descriptor_factory.register("VolumeFractions", VolumeFractions)
\end{lstlisting}
\end{figure*}

In the following, the same procedure is applied to a differentiable approximation~$\tilde{L}$ to the lineal path function~$L$, which is developed in Appendix~\ref{sec:linealpath}.
Naturally, the code for defining~$\tilde{L}$ is much longer than Listing~\ref{lst:descriptor} and is not given in this paper.
Instead, the reader is referred to the \emph{GitHub} repository for viewing the code. 
After adding the descriptor definition to the \texttt{mcrpy/descriptors} directory, it is accessible for characterization and reconstruction via the~\emph{MCRpy} GUI, the command line interface and the Python library.

\begin{figure*}
    \centering
	\subfloat[Original alloy]{\includegraphics[width=0.21\textwidth]{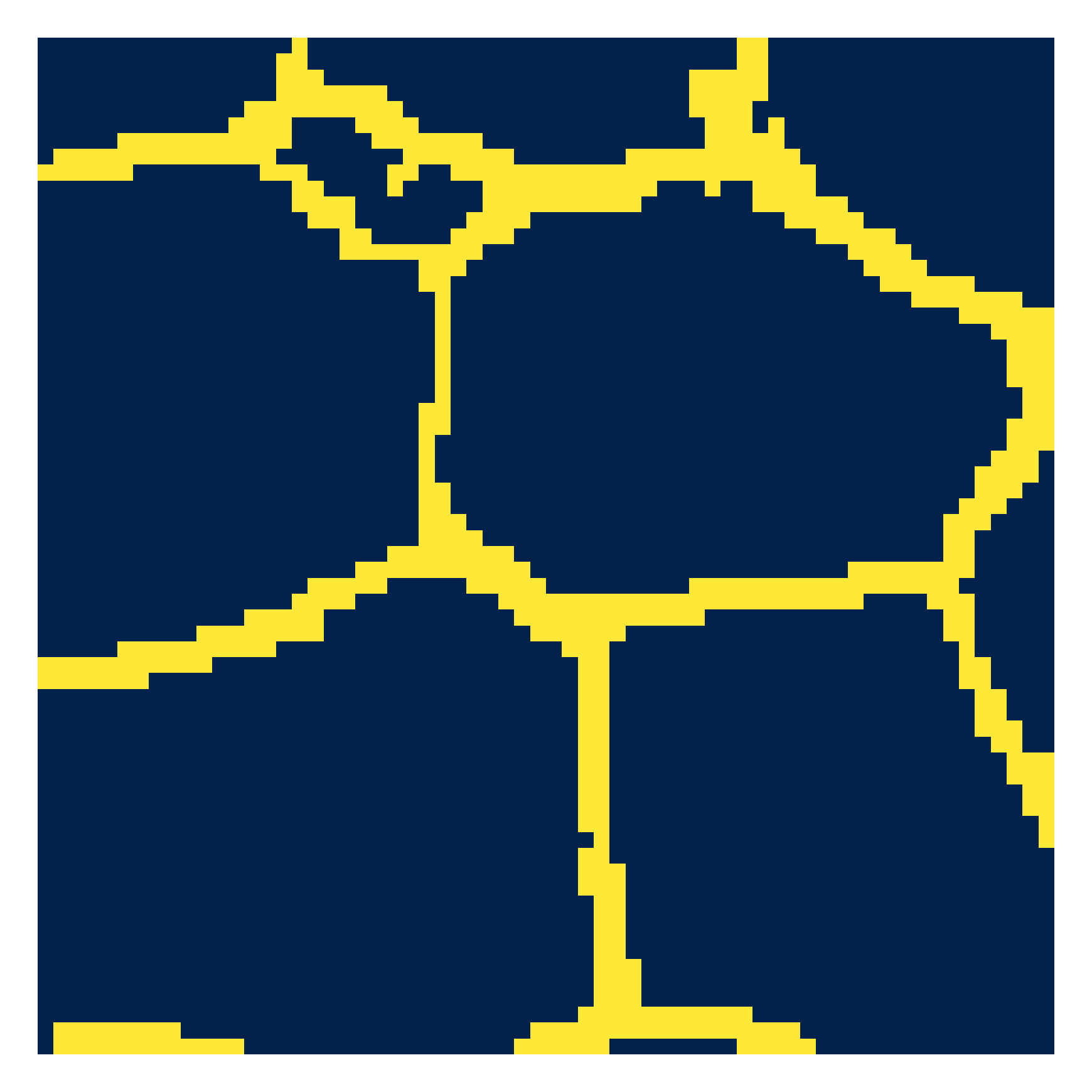}}
	\hfill
	\subfloat[Reconstruction from $\tilde{S}_2,\tilde{L}$]{\includegraphics[width=0.21\textwidth]{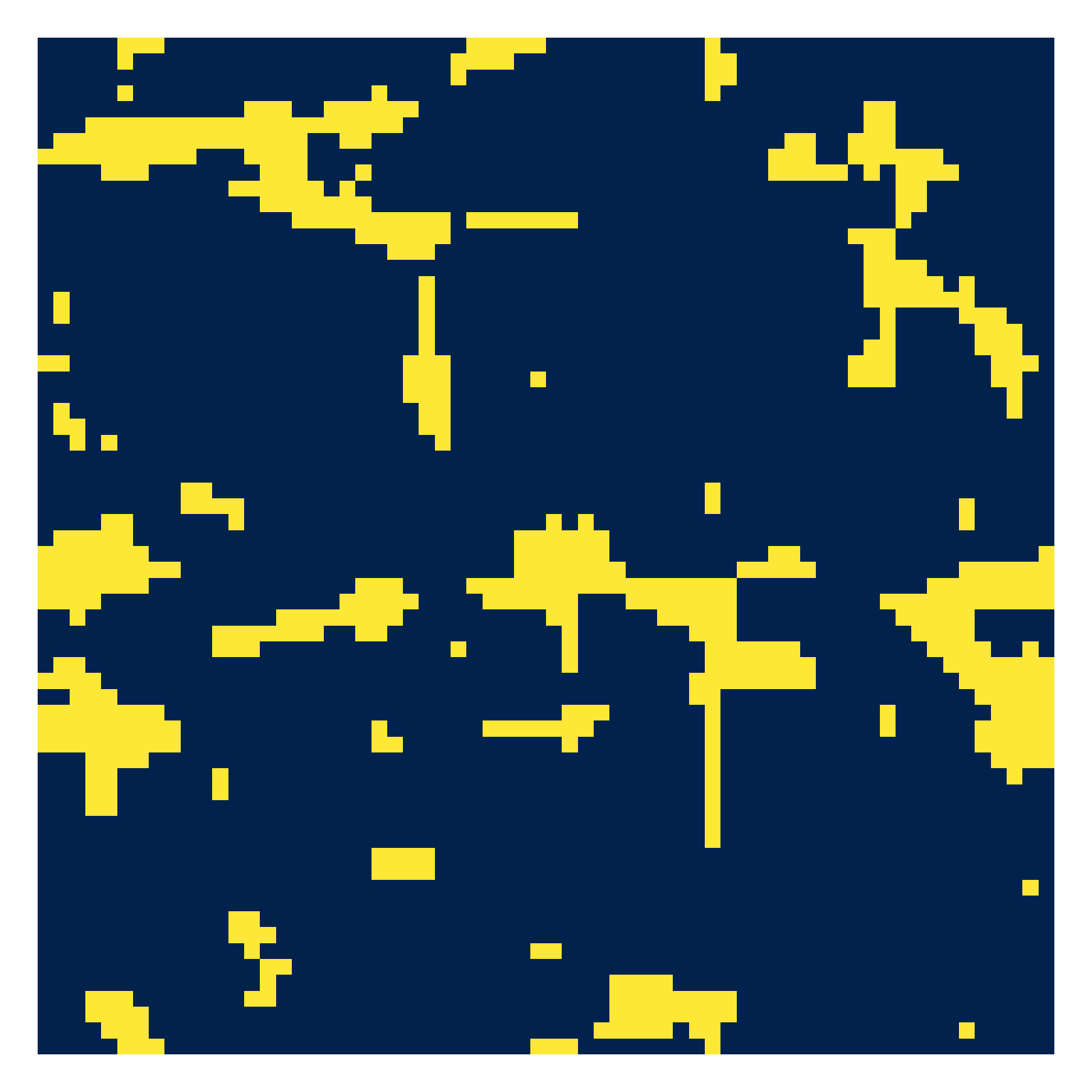}}
	\hfill
	\subfloat[Reconstruction from $\tilde{S}_3$]{\includegraphics[width=0.21\textwidth]{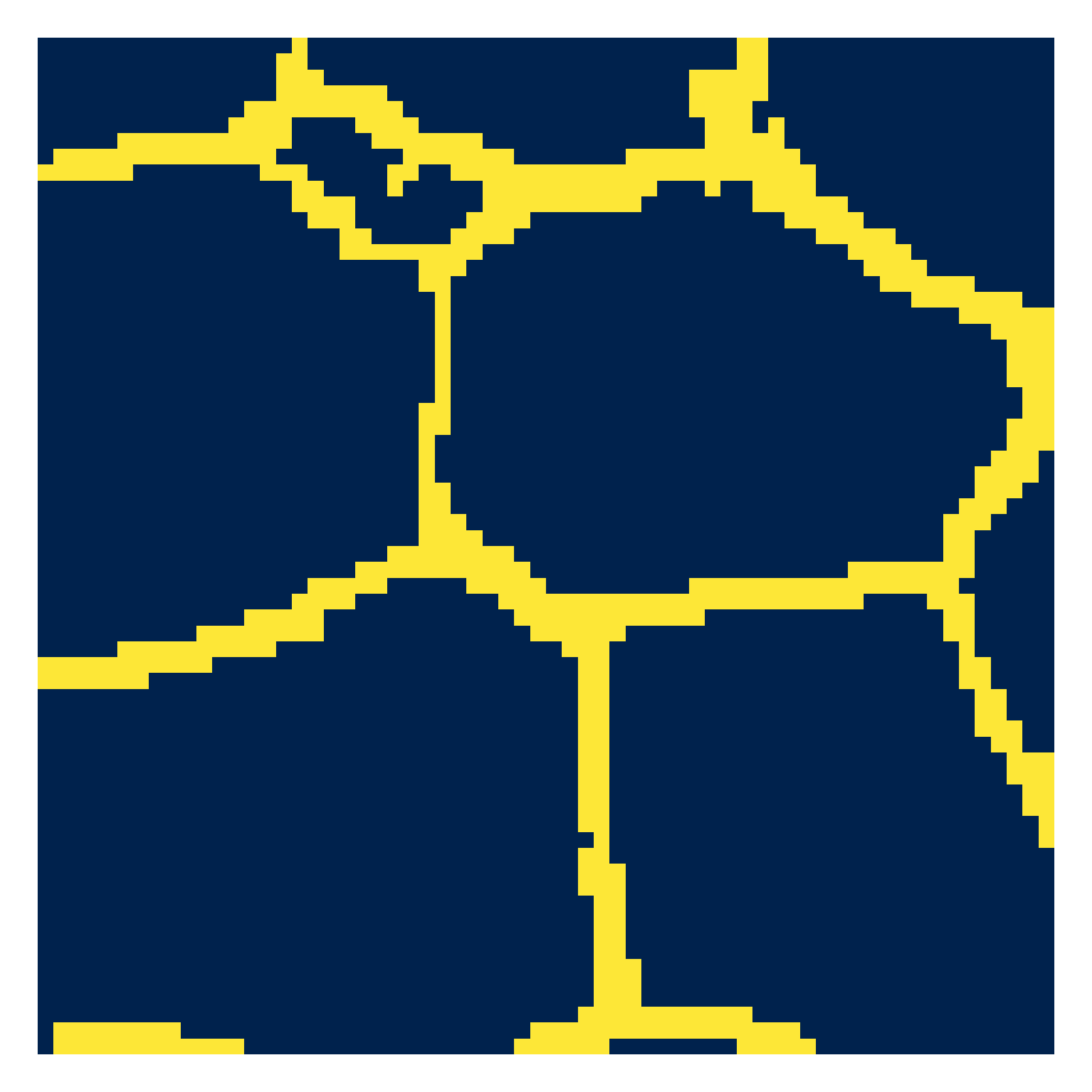}}
	\hfill
	\subfloat[Reconstruction from $G$]{\includegraphics[width=0.21\textwidth]{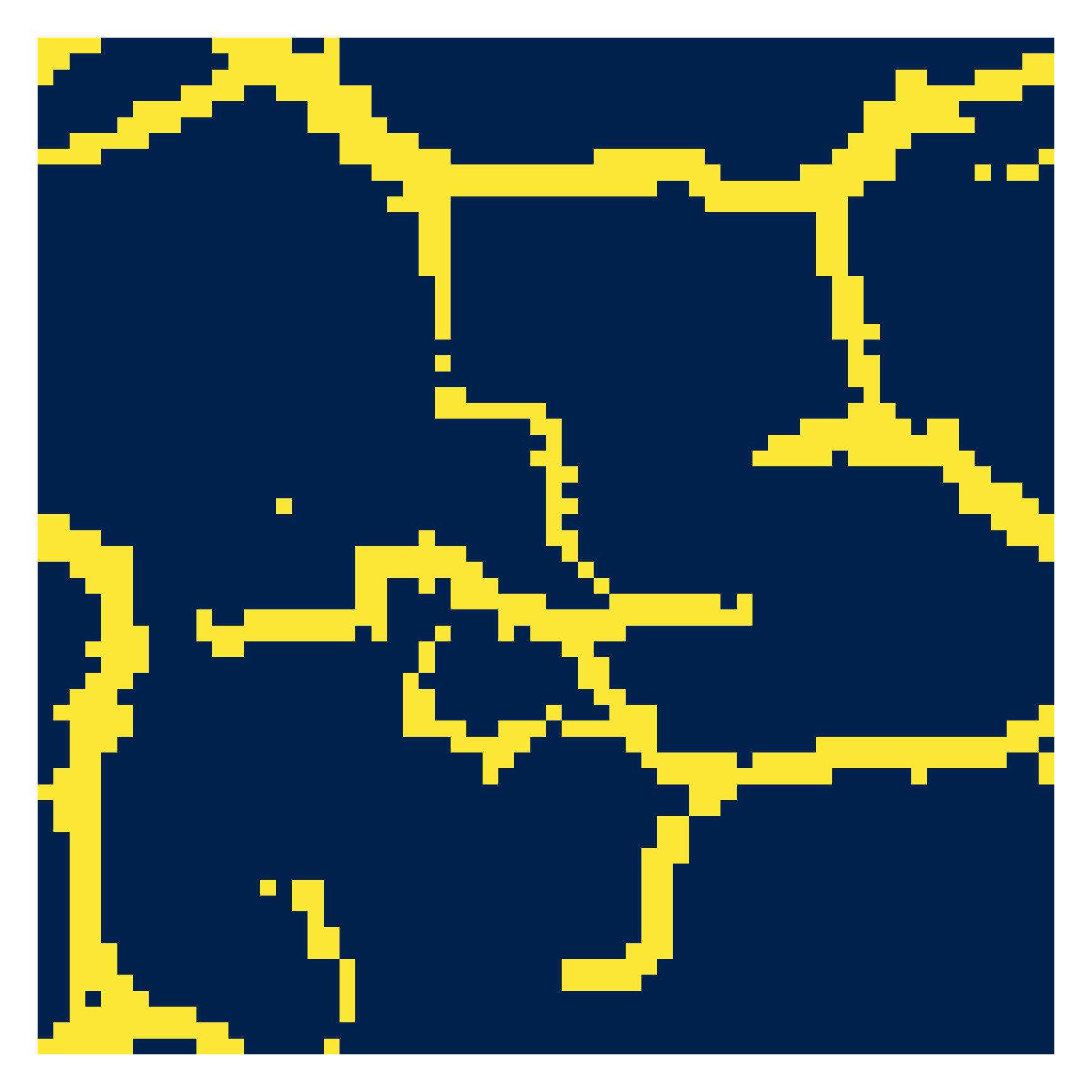}}
	\vfill
	\subfloat[Original carbonate]{\includegraphics[width=0.21\textwidth]{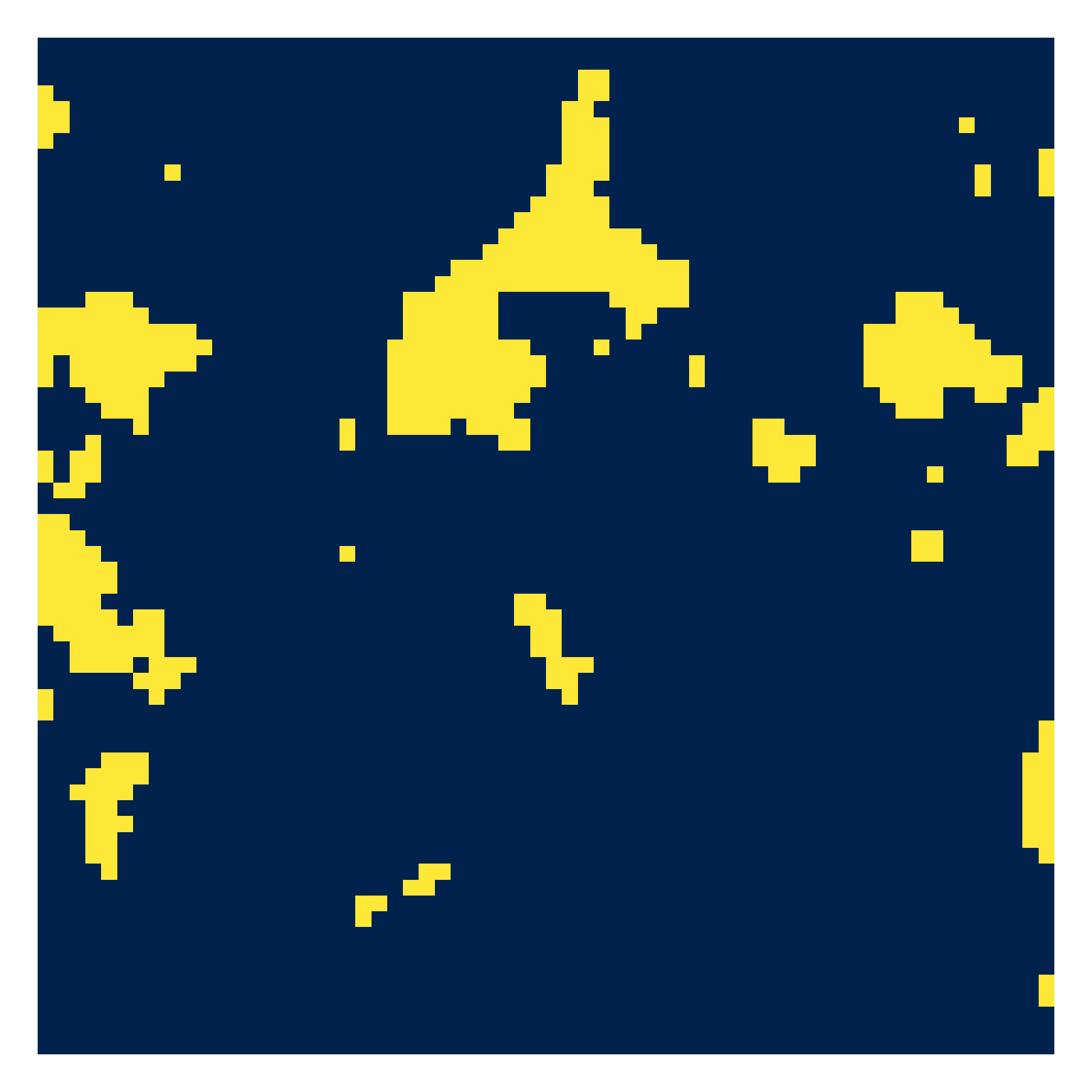}}
	\hfill
	\subfloat[Reconstruction from $\tilde{S}_2,\tilde{L}$]{\includegraphics[width=0.21\textwidth]{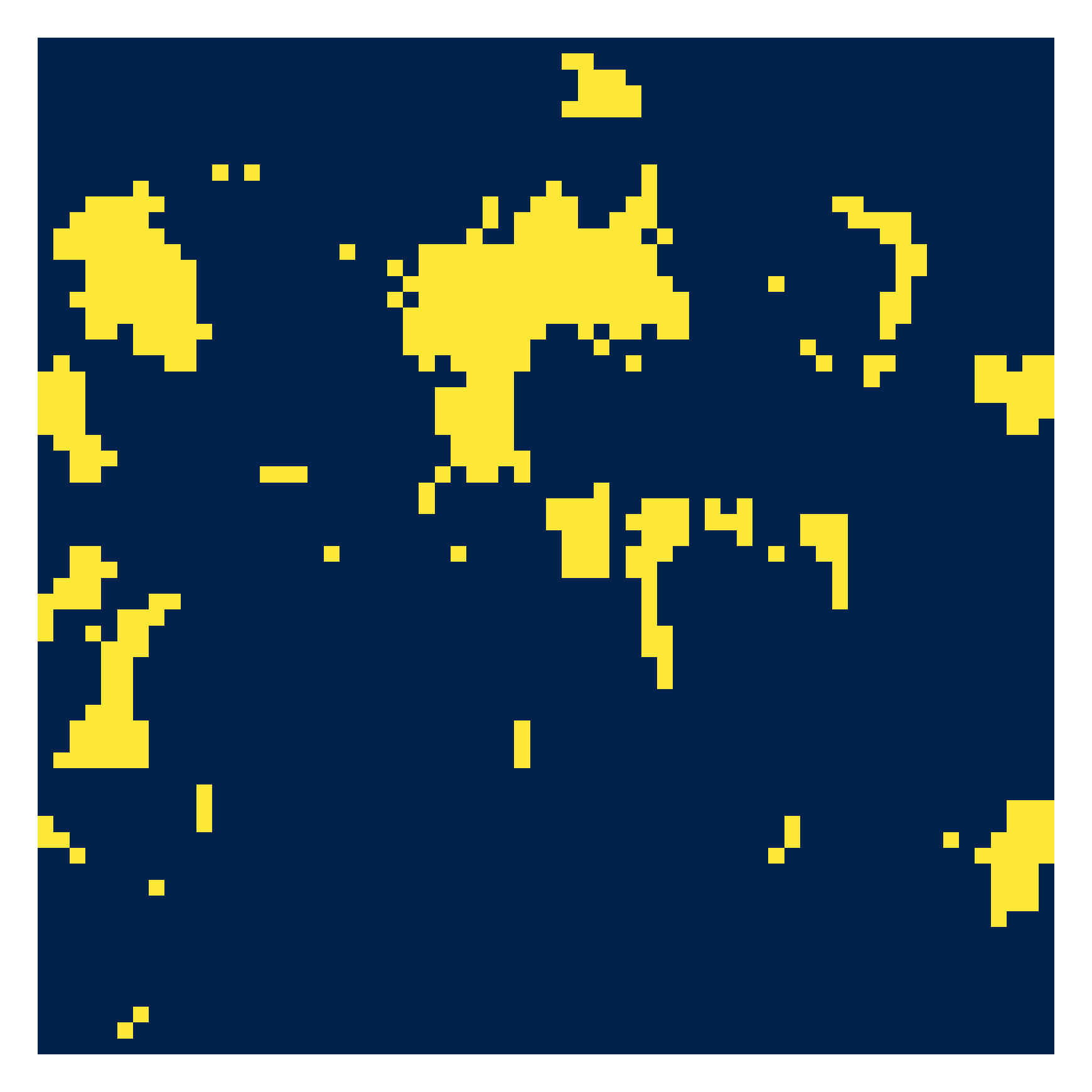}}
	\hfill
	\subfloat[Reconstruction from $\tilde{S}_3$]{\includegraphics[width=0.21\textwidth]{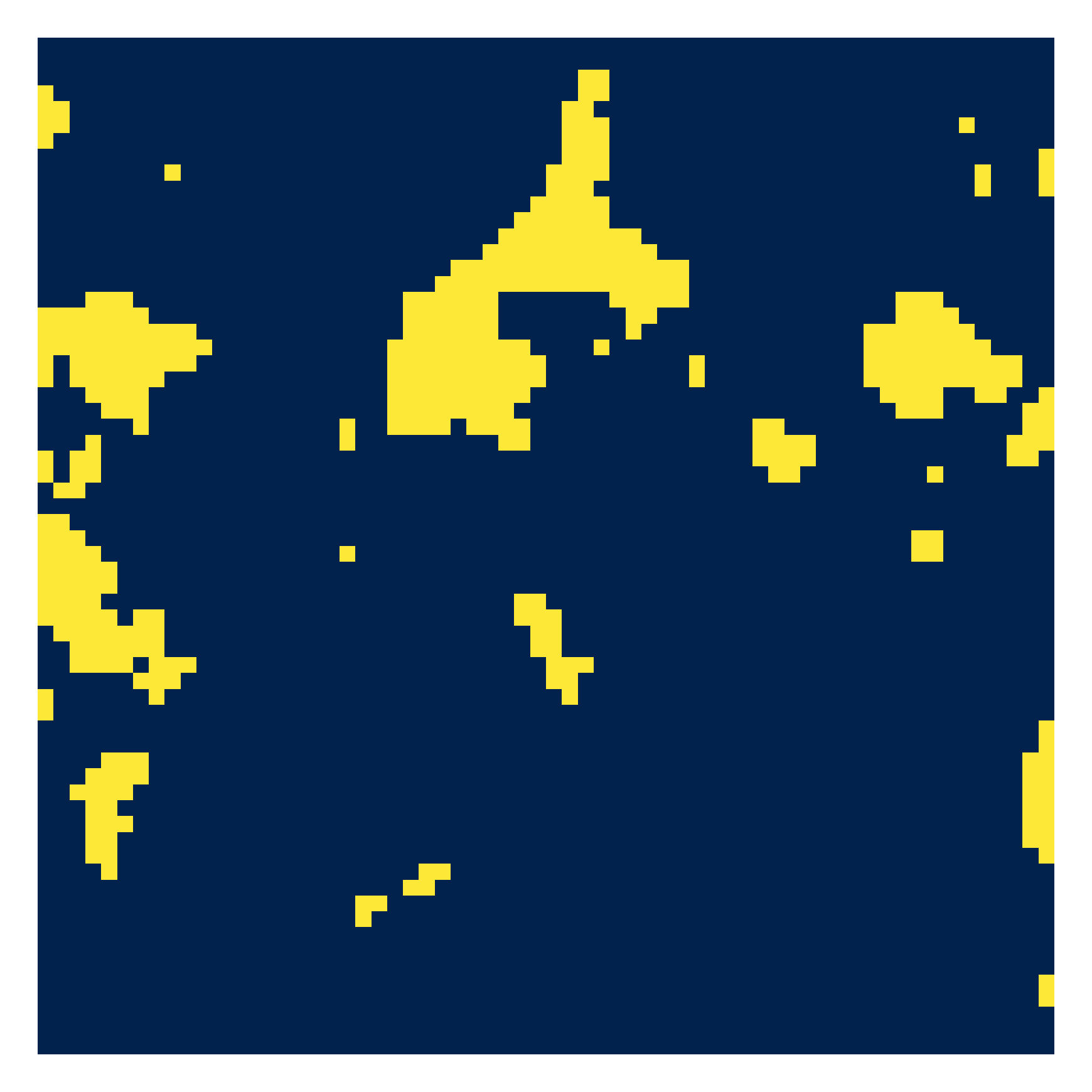}}
	\hfill
	\subfloat[Reconstruction from $G$]{\includegraphics[width=0.21\textwidth]{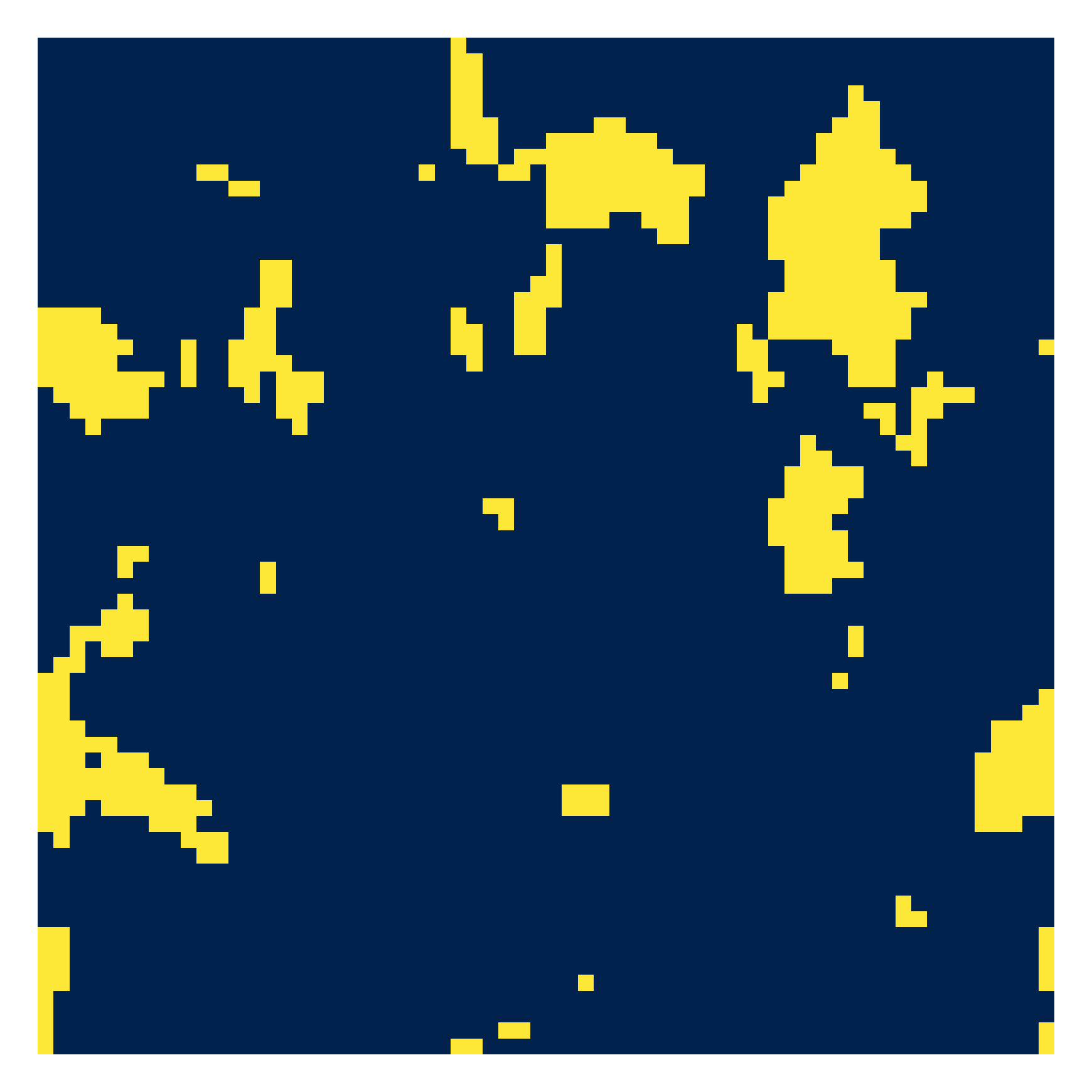}}
	\vfill
	\subfloat[Original ceramics]{\includegraphics[width=0.21\textwidth]{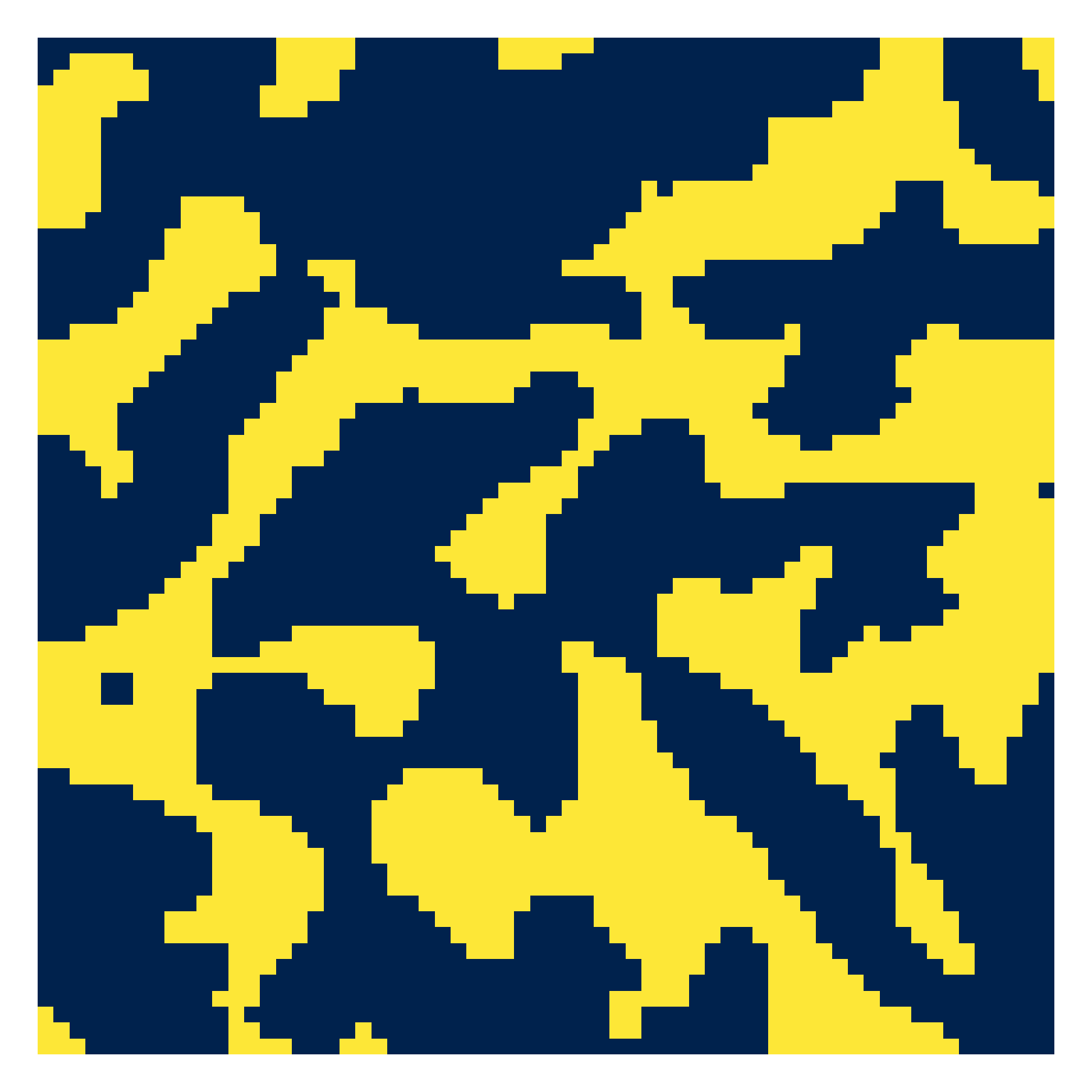}}
	\hfill
	\subfloat[Reconstruction from $\tilde{S}_2,\tilde{L}$]{\includegraphics[width=0.21\textwidth]{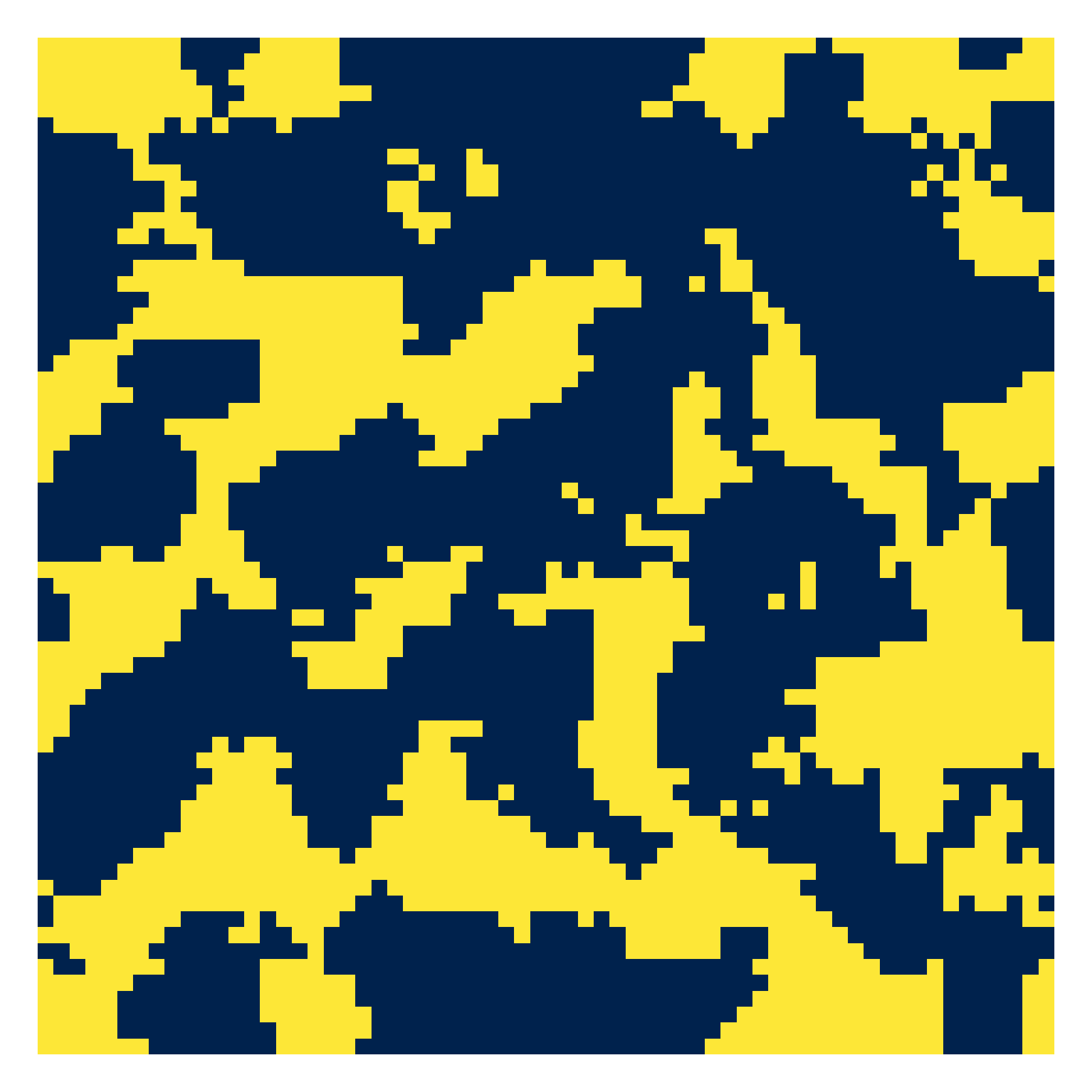}}
	\hfill
	\subfloat[Reconstruction from $\tilde{S}_3$]{\includegraphics[width=0.21\textwidth]{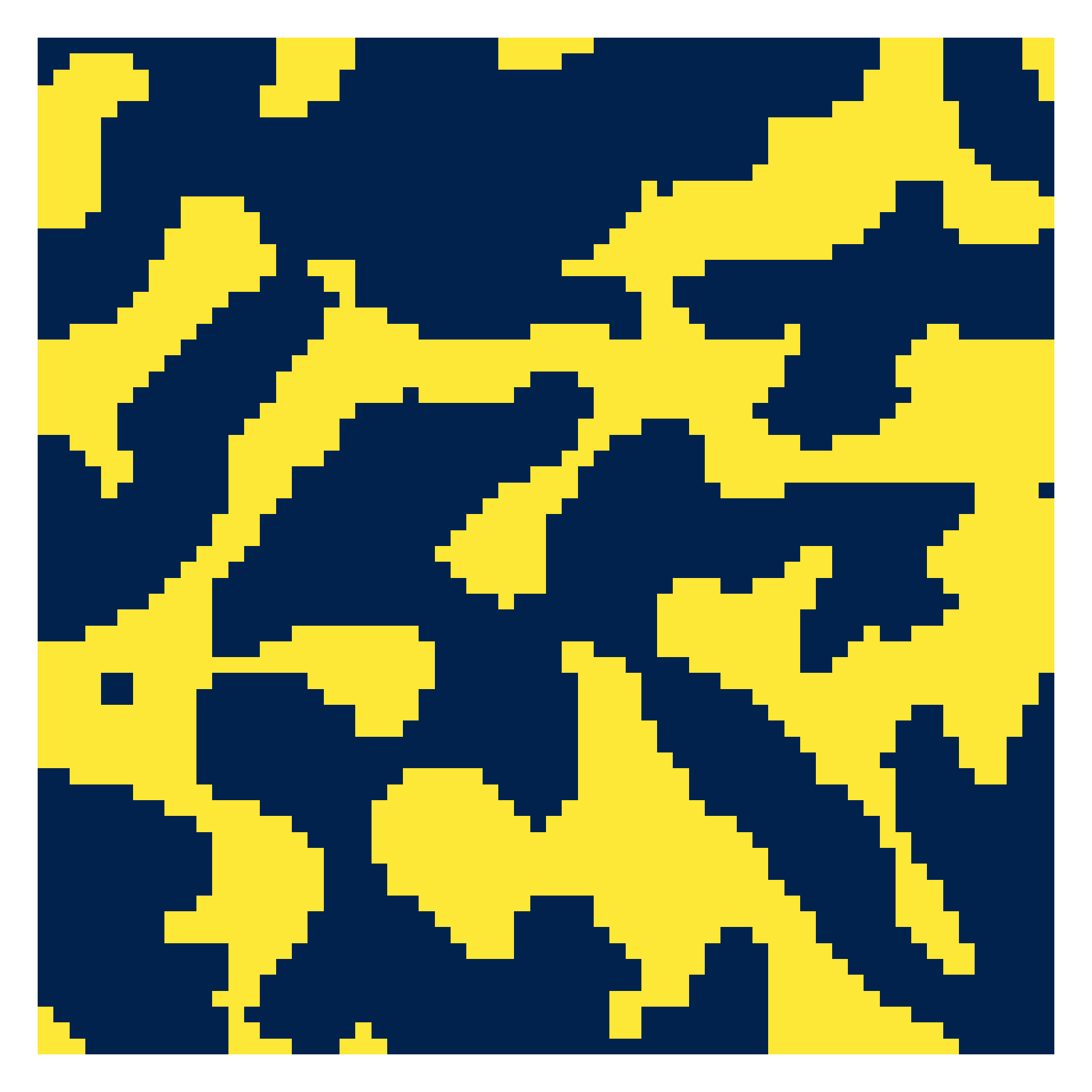}}
	\hfill
	\subfloat[Reconstruction from $G$]{\includegraphics[width=0.21\textwidth]{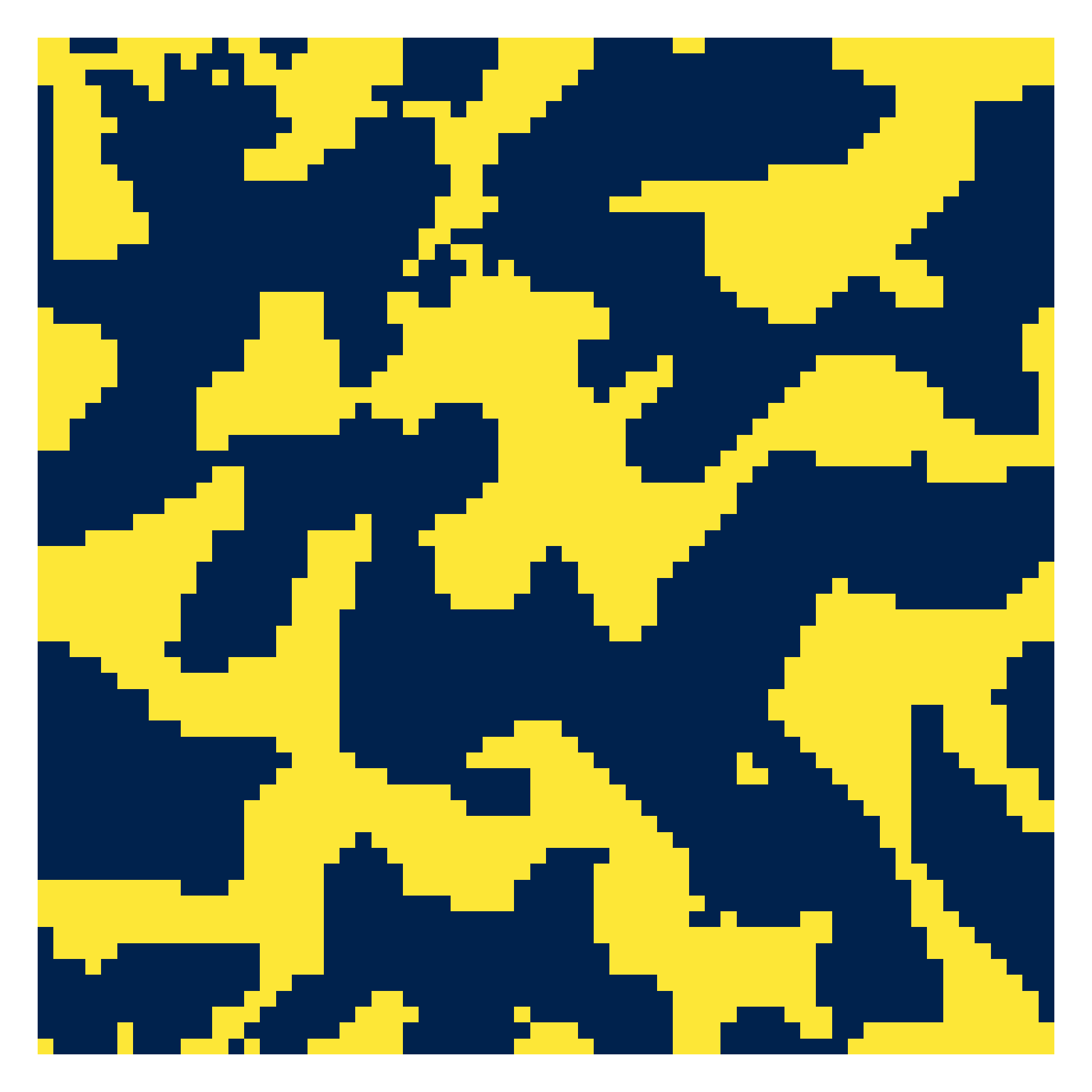}}
	\vfill
	\subfloat[Original copolymer]{\includegraphics[width=0.21\textwidth]{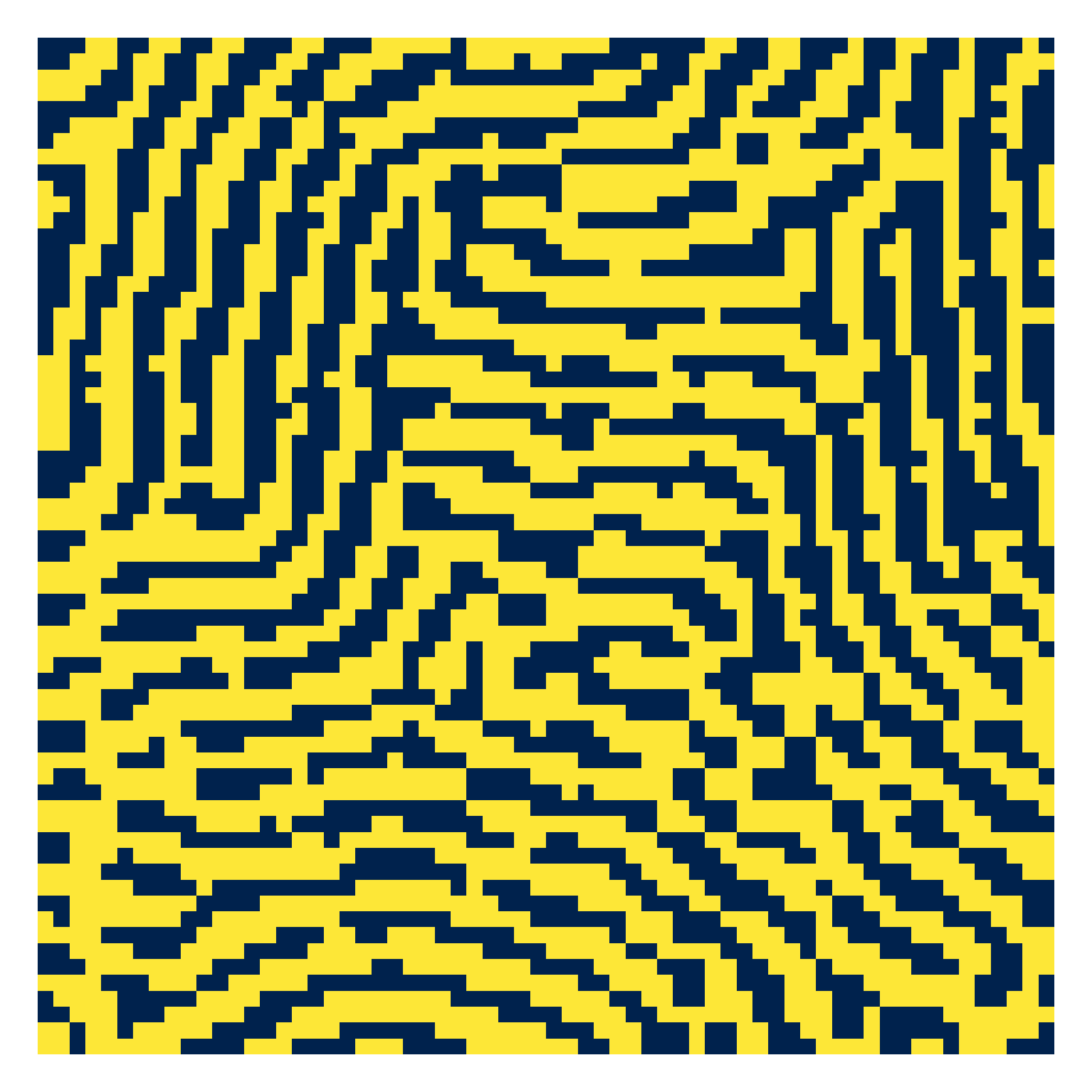}}
	\hfill
	\subfloat[Reconstruction from $\tilde{S}_2,\tilde{L}$]{\includegraphics[width=0.21\textwidth]{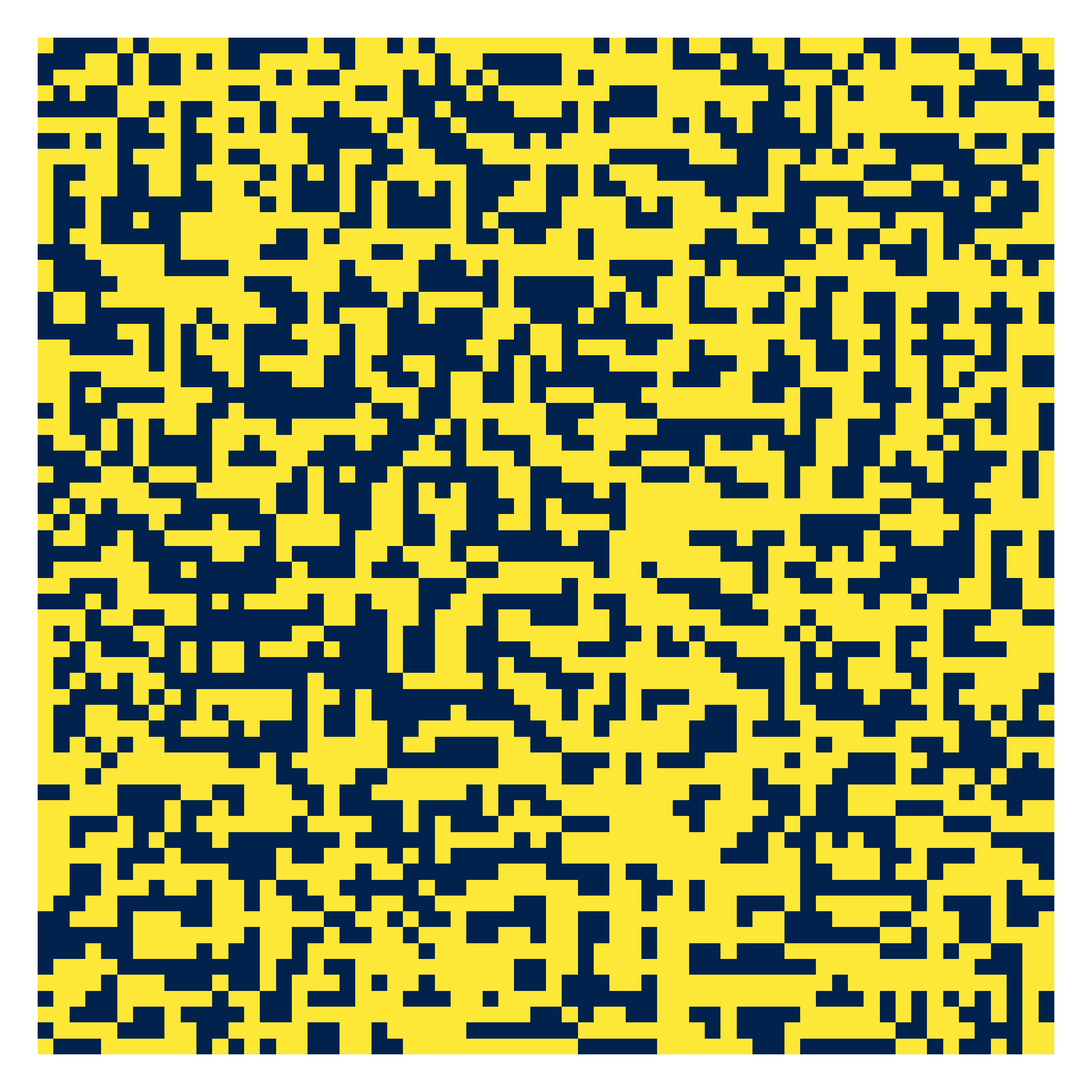}}
	\hfill
	\subfloat[Reconstruction from $\tilde{S}_3$]{\includegraphics[width=0.21\textwidth]{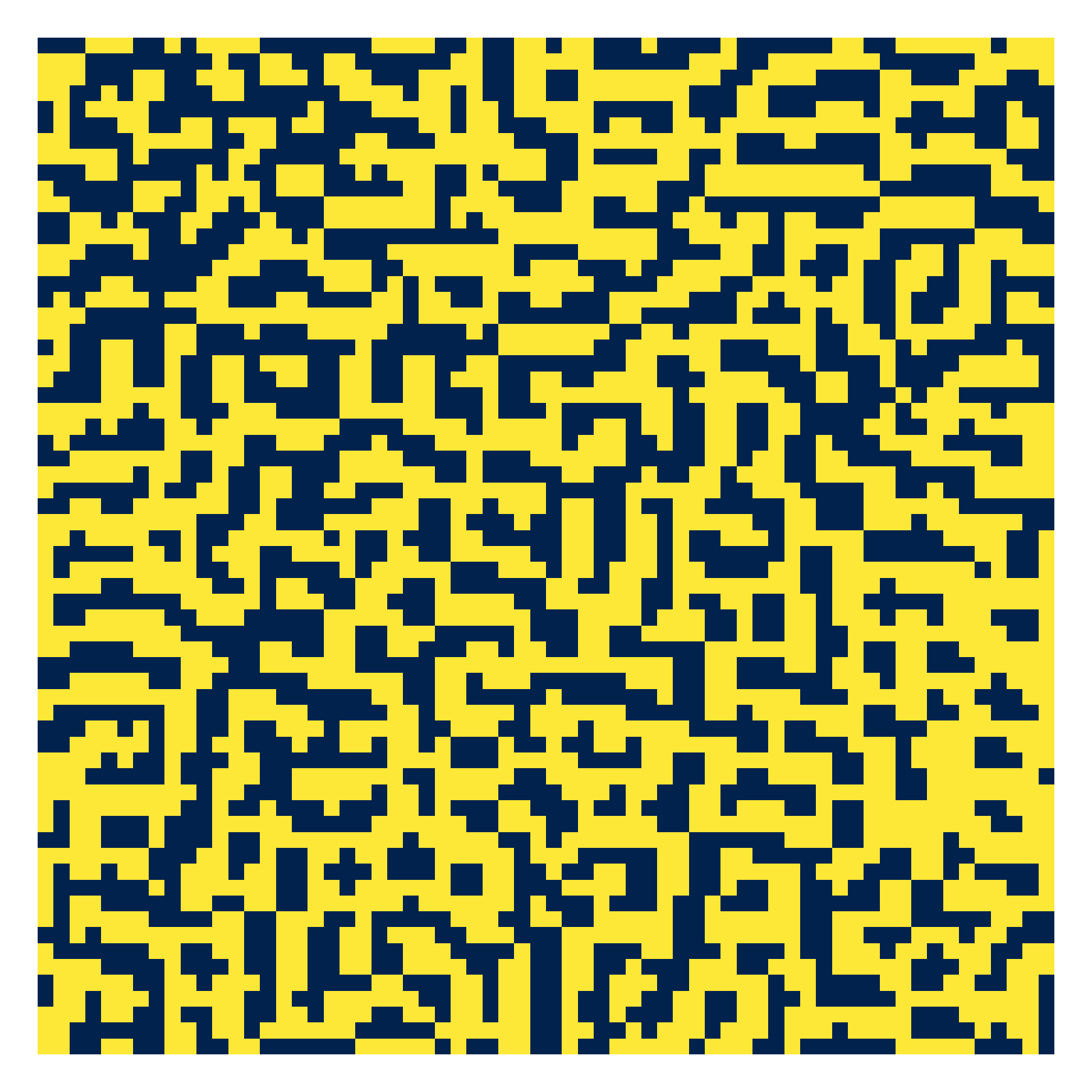}}
	\hfill
	\subfloat[Reconstruction from $G$]{\includegraphics[width=0.21\textwidth]{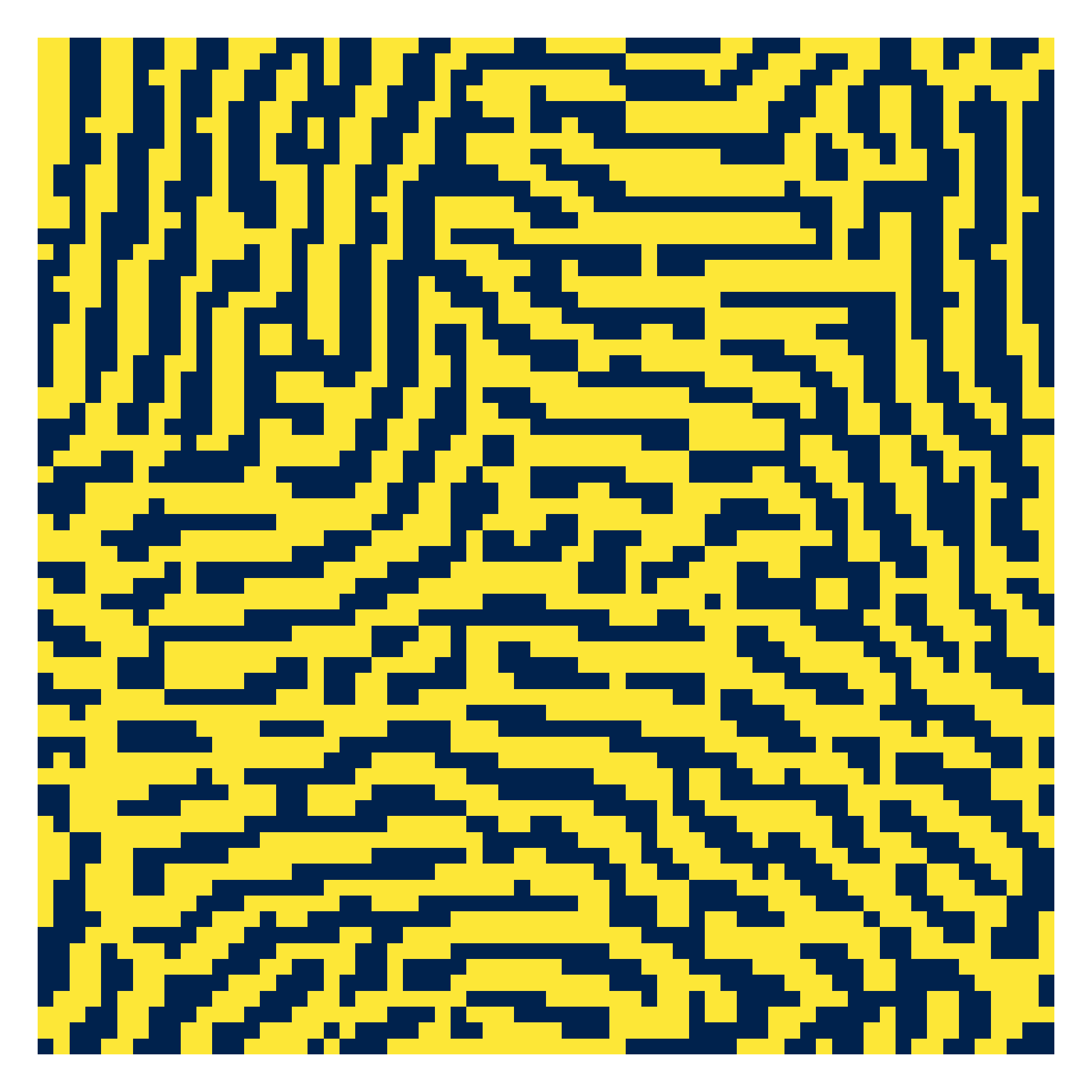}}
	\vfill
	\subfloat[Original polymer comp.]{\includegraphics[width=0.21\textwidth]{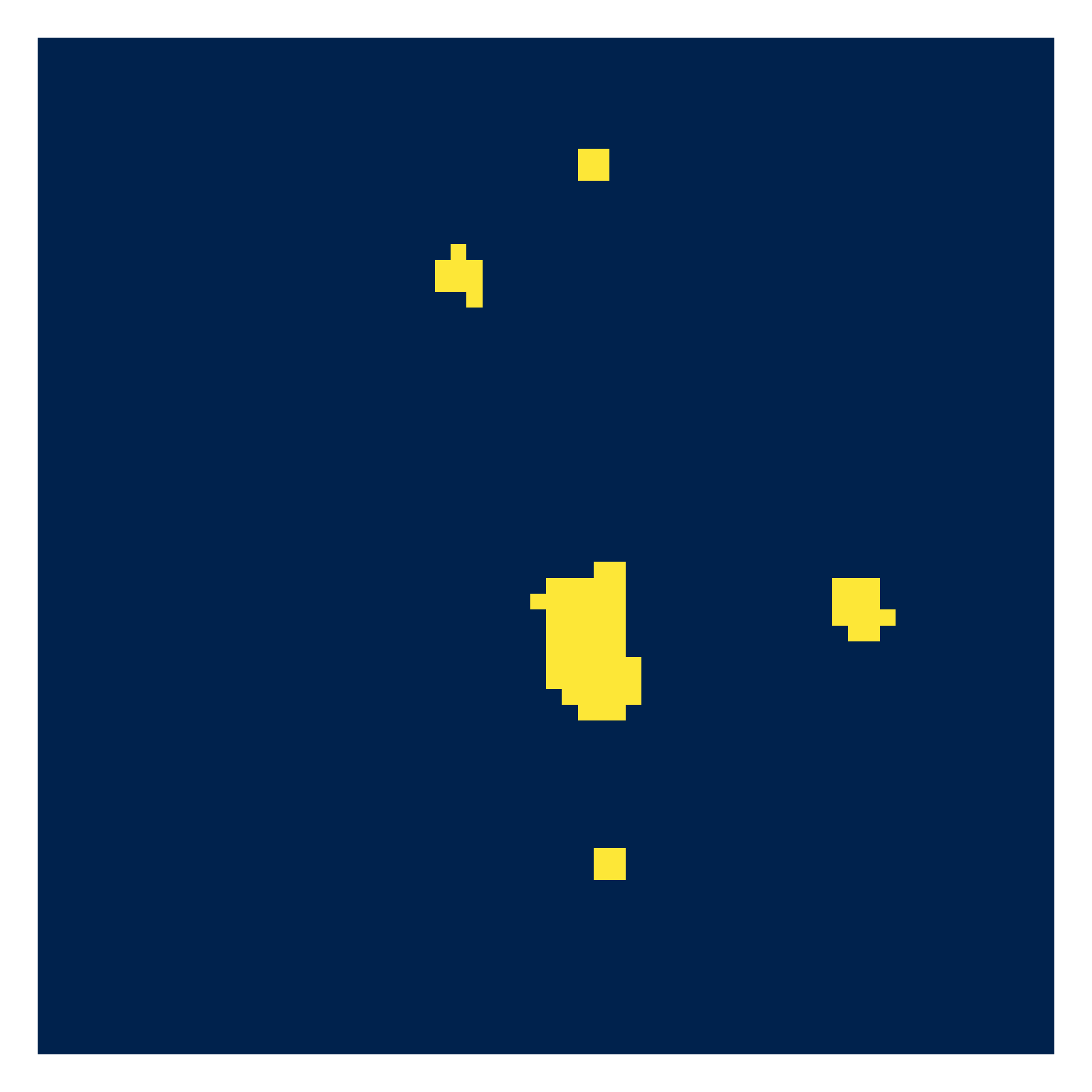}}
	\hfill
	\subfloat[Reconstruction from $\tilde{S}_2,\tilde{L}$]{\includegraphics[width=0.21\textwidth]{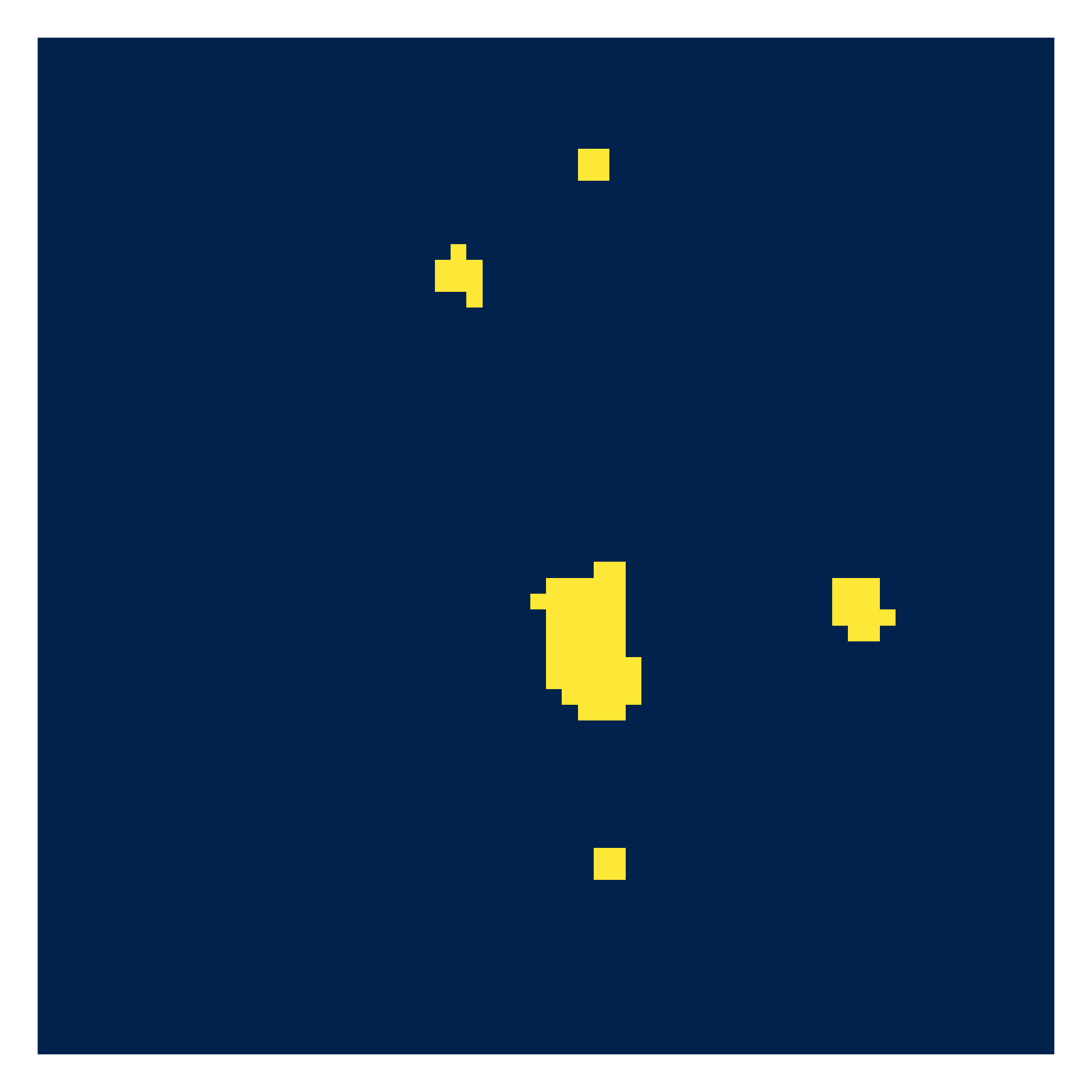}}
	\hfill
	\subfloat[Reconstruction from $\tilde{S}_3$]{\includegraphics[width=0.21\textwidth]{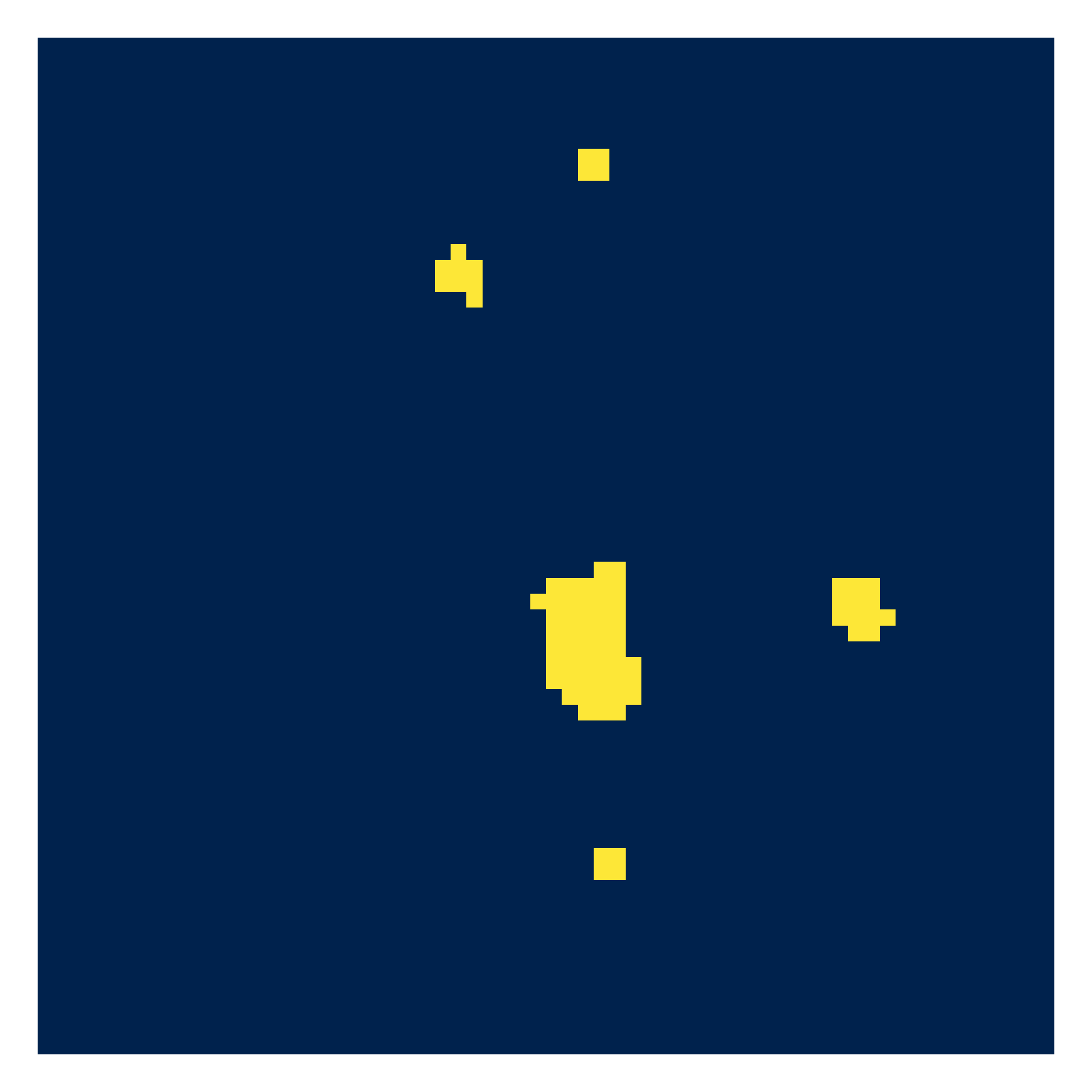}}
	\hfill
	\subfloat[Reconstruction from $G$]{\includegraphics[width=0.21\textwidth]{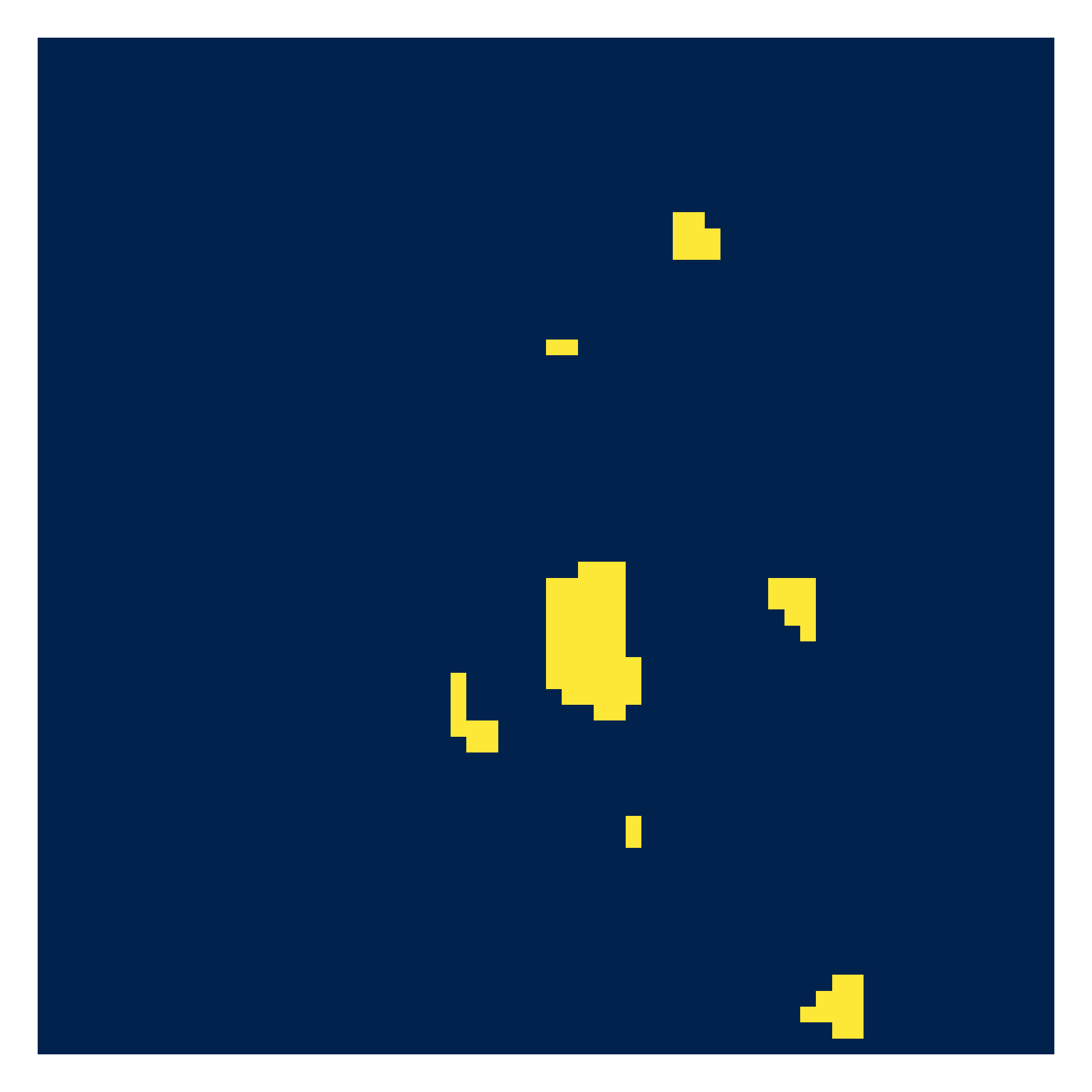}}
	\vfill
    \caption{Comparison between original microstructures and reconstruction results from different descriptors. For a clearer visualization, the reconstructed microstructures are shifted periodically to match the original structure as closely as possible.\label{fig:linealpathresults}}
\end{figure*}

In the Yeong-Torquato algorithm, the lineal path function is often employed to compensate for the shortcomings of the two-point correlation~$S_2$ alone~\cite{yeong_reconstructing_1998, bostanabad_computational_2018}.
As an alternative approach to enriching~$S_2$, the differentiable three-point correlations~$\tilde{S}_3$ are used in~\cite{seibert_reconstructing_2021-1}.
Furthermore, Gram matrices~$G$ have become a common descriptor recently~\cite{li_transfer_2018,bostanabad_reconstruction_2020,seibert_descriptor-based_2022,bhaduri_efficient_2021}.
In order to determine a best-practice for gradient-based reconstruction, $\tilde{S}_3$ is compared to~$G$ and a combination of~$\tilde{S}_2$ and~$\tilde{L}$ in Figure~\ref{fig:linealpathresults}.
It can be seen that~$\tilde{S}_3$ yields perfect reconstructions except for the copolymer, which can only be reconstructed well from~$G$.
In contrast,~$G$ yields acceptable results for all structures.
The combination of $\tilde{S}_2$ and $\tilde{L}$ performs very poorly for the alloy and the copolymer and is relatively noisy for the carbonate and ceramics.
However, it outperforms~$G$ for the polymer composite.
In summary, the results in Figure~\ref{fig:linealpathresults} indicate that including higher-order information to~$\tilde{S}_2$ via~$\tilde{S}_3$ is more promising for gradient-based reconstruction than via the newly proposed differentiable approximation to the lineal path function~$\tilde{L}$.

The more relevant aspect, however, is how easily new descriptors can be assessed using~\emph{MCRpy}.
After defining a plugin as shown in Listing~\ref{lst:descriptor}, it can be used for characterization and reconstruction and evaluated seamlessly.
This extensibility facilitates quick and easy experimentation, allowing researchers to assess new MCR techniques easily and provide them to their colleagues.

\section{Conclusions and Outlook}
\label{sec:summary}
\emph{MCRpy} is a powerful and extensible open-source Python library and toolkit for microstructure characterization and reconstruction. 
Besides these core features, \emph{MCRpy} provides a plethora of convenient tools for inspecting and comparing descriptors, analyzing reconstruction results and controlling the descriptor space.
It be easily applied via a GUI and brought to automated large-scale application on high-performance computers through a command line interface.
For advanced and custom operations in the descriptor space,~\emph{MCRpy} can be imported and used as a Python module with direct access to the structures and descriptors.
Typical workflows for these interfaces are presented in this work by means of different MCR tasks.

A central design aspect in~\emph{MCRpy} is its extensibility in that descriptors, loss functions and optimizers can be provided by the community as simple plugin modules.
An example for a simple plugin is given in this work.
We hope that the open source nature of the code and the plugin architecture will make~\emph{MCRpy} an international collaborative project with contributions from numerous researchers.
This growth can leverage the potential of the presented tool to facilitate faster and easier MCR research and ultimately help accelerating materials development.

\section*{Acknowledgements}
The group of M. Kästner thanks the German Research Foundation DFG which supported this work under Grant number KA 3309/18-1.
Furthermore, this work is partially funded by the European Regional Development Fund (ERDF) and co-financed by tax funds based on the budget approved by the members of the Saxon State Parliament under Grants 100373334.
All presented computations were performed on a PC-Cluster at the Center for Information Services and High Performance Computing (ZIH) at TU Dresden.
The authors thus thank the ZIH for generous allocations of computer	time.



\section*{Code availability}
\emph{MCRpy} is available on the \emph{GitHub} repository \url{https://github.com/NEFM-TUDresden/MCRpy}.

\section*{Competing interests}
The authors declare no competing interests.

\section*{Author contributions}
\textbf{P. Seibert:} Conceptualization, Formal analysis, Investigation, Methodology, Software, Validation, Visualization, Writing - original draft.
\textbf{A. Raßloff:} Conceptualization, Software, Writing - review and editing.
\textbf{K. Kalina:} Supervision, Visualization, Writing - review and editing.
\textbf{M. Ambati:} Supervision, Writing - review and editing.
\textbf{M. Kästner:} Funding acquisition, Resources, Supervision, Writing - review and editing.


\appendix

\section{Differentiable approximation to lineal path function}
\label{sec:linealpath}
The lineal path function~$L$ is a well-established microstructure descriptor~\cite{lu1992} that is often used in the Yeong-Torquato algorithm to compensate for the shortcomings of the two-point correlation~$S_2$ alone~\cite{yeong_reconstructing_1998, bostanabad_computational_2018}.
Given a vector~$\vec{r} = (r_x, r_y)$, it yields the probability that~$\vec{r}$ lies entirely within a single phase if it is placed randomly in the structure.
In contrast to~$S_2$, which considers only the start and end point of the vector,~$L$ incorporates information about the connectedness of the phases.
In this section,~$\tilde{L}$ is presented as a differentiable approximation to~$L$.

\begin{figure}[h]
    \centering
    \includegraphics[width=0.6\linewidth]{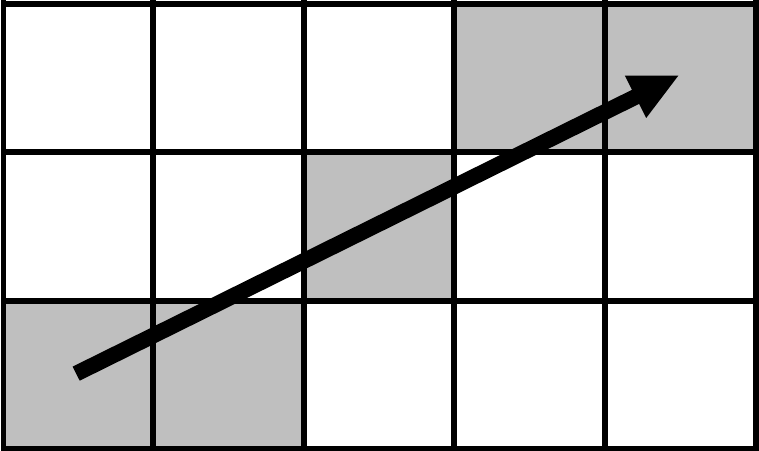}
    \caption{A discretization of the vector~$\vec{r}$ to a discrete pixel grid using Bresenham's method~\cite{bresenham_algorithm_1965}.\label{fig:bresenham}}
\end{figure}

The lineal path function is approximated using a \emph{convolve} - \emph{threshold} - \emph{reduce} pipeline similarly to~\cite{seibert_reconstructing_2021-1}.
For this purpose, the vector or line~$\vec{r}$ is discretized to a pixel grid as shown in Figure~\ref{fig:bresenham} and divided by the length of the line. 
In this work, the Bresenham line algorithm~\cite{bresenham_algorithm_1965} is used, but alternative approaches like the Xiaolin Wu line algorithm~\cite{wu_efficient_1991} might be equally viable options that can be investigated in future works.
In the \emph{convolve} step, the discretized line from Figure~\ref{fig:bresenham} is used as a mask for a convolution with periodic boundary conditions.
The output of the convolution is an image where each pixel corresponds to the discretized line being placed at the pixel's location.
The pixel value is~0 if no part of the line lies in phase~1 and~1 if the line lies completely in phase~1.
If only parts of the line are in phase~1, the value is between~0 and~1.
These pixels can be set to zero by thresholding the image with a value~$t$, where~$1 / (1-||\vec{r}||_{\infty}) < t < 1$.
The \emph{threshold} step thus yields an image which can be interpreted as an ensemble of realizations, where each pixel takes the value~0 or~1.
In the \emph{reduce} step, this ensemble is averaged to obtain the probability of a randomly placed line being entirely in phase~1.

\begin{figure}[h]
    \centering
    \includegraphics[width=0.7\linewidth]{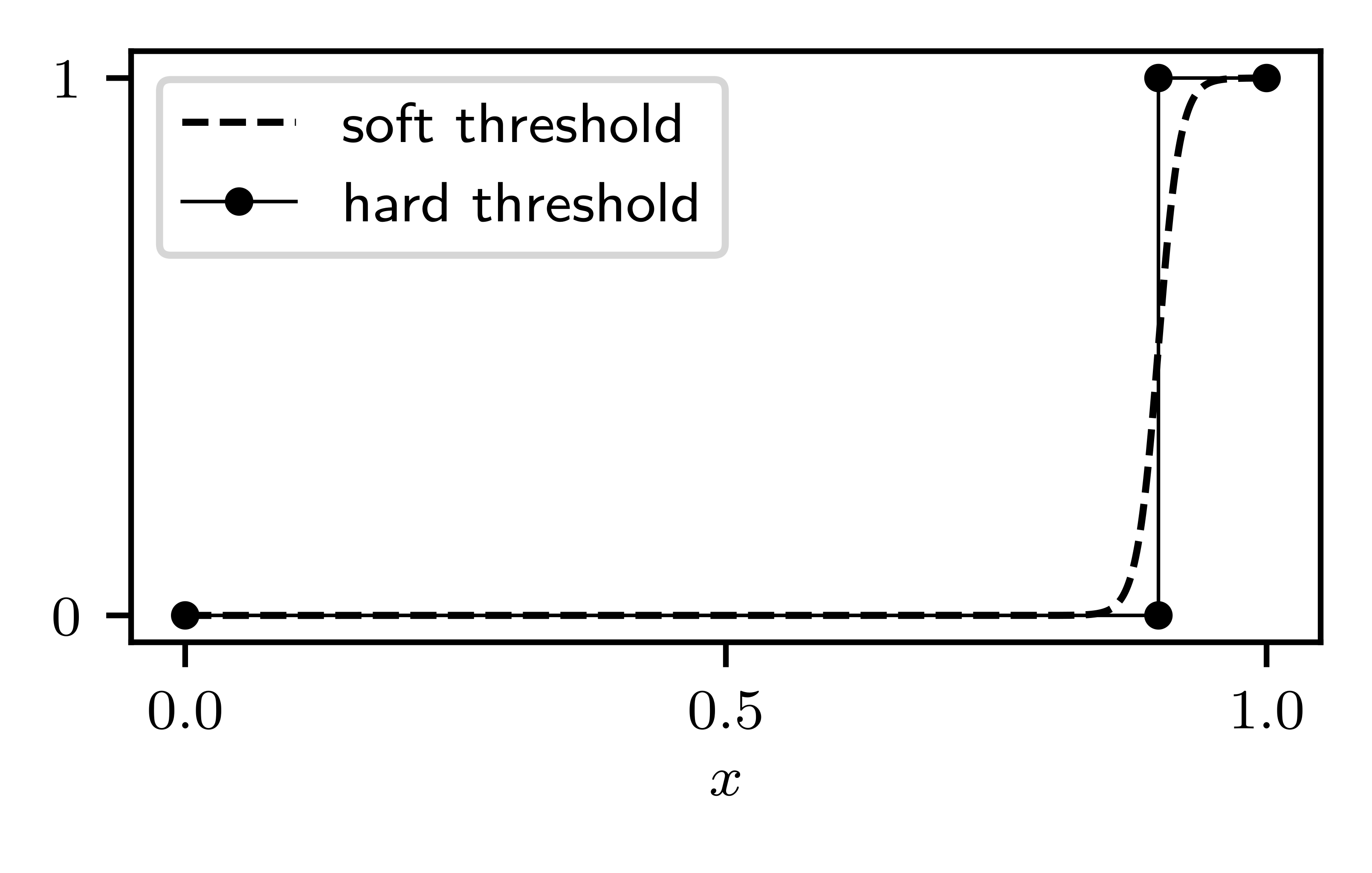}
    \caption{Approximation of a hard threshold by a scaled and shifted differentiable sigmoid function.\label{fig:threshold}}
\end{figure}

To make the \emph{convolve} - \emph{threshold} - \emph{reduce} pipeline differentiable, only the thresholding needs to be modified.
The hard thresholding is therefore approximated using a scaled and shifted differentiable sigmoid function\footnote{We use the same function as in~\cite{seibert_reconstructing_2021-1}.} as shown in Figure~\ref{fig:threshold}.
This introduces errors, because the ensemble to average does not contain only ones and zeros but also intermediate values.
Unlike in~\cite{seibert_reconstructing_2021-1}, where a similar error for~$\tilde{S}$ could be eliminated by deriving a correction step, the difference between~$L$ and~$\tilde{L}$ can not be quantified easily.
Thus, it is clear that~$\tilde{L}$ is only an approximation to~$L$, not a generalization.
This is not problematic if the same descriptor is used for characterization and reconstruction.

To the author's best knowledge, this concludes the first differentiable approximation to the lineal path function.

\section{Underlying technologies}
\label{sec:details}
\emph{MCRpy} is programmed in Python and based on very common packages like \emph{Numpy}, \emph{Scipy}, \emph{Matplotlib} and \emph{TensorFlow}.
While simulated annealing is implemented from scratch, the gradient-based optimizers are taken from \emph{Scipy} and \emph{TensorFlow}.
Furthermore, \emph{TensorFlow} is used for automatic differentiation of the loss function and the descriptors as well as for just-in-time compilation via \emph{AutoGraph}.
This allows \emph{MCRpy} to run highly optimized code on GPUs despite being written entirely in Python.
As optional dependencies, the \emph{Gooey} is required to run the \emph{MCRpy} GUI and \emph{pyMKS} is used in a descriptor plugin for FFT-based 2-point correlations.
A summary of required and optional dependencies and their versions is given in Table~\ref{tab:dependencies}. 
\begin{table}[h]
	\centering
	\caption{Software dependencies for the current version of \emph{MCRpy}.}
	\label{tab:dependencies}
	\begin{tabular}{l | l | l }
		\toprule
		package & required & version \\
		\midrule
		\emph{numpy} & yes & $\geq 1.20.1$ \\
		\emph{matplotlib} & yes & $\geq 3.3.4$ \\
		\emph{scipy} & yes & $\geq 1.6.2$ \\
		\emph{tensorflow} & yes & $\geq 2.3.1$ \\
		\emph{pymks} & for FFT-based correlations & $\geq 0.4.1$ \\
		\emph{gooey} & for GUI & $\geq 1.0.8.1$ \\
	\end{tabular}
\end{table} 
\\
\\
\\
\\
\\

\bibliographystyle{spphys}
{\small \bibliography{preprint}}

\end{document}